\documentclass[apj,numberedappendix]{emulateapj}
\usepackage{lscape}
\newcommand{\Ha}{\mbox{H$\alpha$}}
\newcommand{\Hb}{\mbox{H$\beta$}}

\newcommand{\Te}{\mbox{$T_{\rm{e}}$}}
\newcommand{\Ne}{\mbox{$N_{\rm{e}}$}}

\newcommand{\VK}{\mbox{$V-K$}}
\newcommand{\VR}{\mbox{$V-R$}}

\newcommand{\Rone}{\mbox{\textsc{r1}}}
\newcommand{\Rtwo}{\mbox{\textsc{r2}}}
\newcommand{\Rthree}{\mbox{\textsc{r3}}}
\newcommand{\Rfour}{\mbox{\textsc{r4}}}
\newcommand{\Rfive}{\mbox{\textsc{r5}}}
\newcommand{\Rsix}{\mbox{\textsc{r6}}}
\newcommand{\Rseven}{\mbox{\textsc{r7}}}

\begin{document}

\slugcomment{Accepted for publication in {\it The Astrophysical Journal}}

\title{Integral Field Spectroscopy of Blue Compact Dwarf Galaxies}

%
%

\author{Bego\~na Garc{\'\i}a-Lorenzo}
\email{bgarcia@iac.es}
\affil{Instituto de Astrof{\'\i}sica de Canarias, E-38200  La Laguna, 
Tenerife, Canary Islands, Spain}

\author{Luz M. Cair\'os} 
\email{luzma@aip.de} 
\affil{ Astrophysikalisches Institut Potsdam, An der Sternwarte 16, 
D-14482 Potsdam, Germany} 

\author{Nicola Caon}
\email{nicola.caon@iac.es}
\affil{Instituto de Astrof{\'\i}sica de Canarias, E-38200  La Laguna, 
Tenerife, Canary Islands, Spain}

\author{Ana Monreal-Ibero}
\email{amonreal@iac.es}
\affil{Instituto de Astrof{\'\i}sica de Canarias, E-38200  La Laguna, 
Tenerife, Canary Islands, Spain}

\author{Carolina Kehrig}
\email{kehrig@iaa.es}
\affil{Instituto de Astrof{\'\i}sica de Andaluc{\'\i}a, C. Bajo de Huetor, 
E-18008 Granada, Spain}


\shortauthors{Garcia-Lorenzo et al.}
\shorttitle{IFS observations of BCDs}

\begin{abstract} 

We present results on integral-field optical spectroscopy of five luminous
Blue Compact Dwarf galaxies. The data were obtained using the fiber system
\textsc{integral} attached at the William Herschel telescope. The galaxies
Mrk~370, Mrk~35, Mrk~297, Mrk~314 and III~Zw~102 were observed. The central
$33\farcs6\times29\farcs4$ regions of the galaxies were mapped with a spatial
resolution of $2\farcs7$~spaxel$^{-1}$, except for Mrk~314, in which we
observed the central $16\arcsec\times12\arcsec$ region with a resolution of
$0\farcs9$~spaxel$^{-1}$. We use high-resolution optical images to isolate the
star-forming knots in the objects; line ratios, electron densities and oxygen
abundances in each of these regions are computed. We build continuum and
emission-line intensity maps as well as maps of the most relevant line ratios:
[\ion{O}{3}]~$\lambda5007$/\Hb, [\ion{N}{2}]~$\lambda6584$/\Ha, and \Ha/\Hb,
which  allow us to obtain spatial information on the ionization structure and
mechanisms. We also derive the gas velocity field from the \Ha\ and
[\ion{O}{3}]~$\lambda5007$ emission lines.

We find that all the five galaxies are in the high end of the metallicity
range of Blue Compact Dwarf galaxies, with oxygen abundances varying from
$Z_{\sun}\simeq0.3$ to $Z_{\sun}\simeq1.5$. The objects show \ion{H}{2}-like
ionization in the whole field of view, except the outer regions of III~Zw~102
whose large [\ion{N}{2}]~$\lambda6584$/\Ha\ values suggest the presence of
shocks. The five galaxies display inhomogeneous extinction patterns, and three
of them have high \Ha/\Hb\ ratios, indicative of a large dust content;  all
galaxies display complex, irregular velocity fields in their inner regions.

\end{abstract}


\section{Introduction}

Blue Compact Dwarf (BCD) galaxies are low-luminosity and compact objects,
which show optical spectra similar to those presented by \ion{H}{2} regions in
spiral galaxies \citep{ThuanMartin81}. They are also low-metallicity systems
\citep{Searle72,Lequeux79,KunthSargent83,Masegosa94} with intense star
formation activity (rates ranging between 0.1 and $1\;M_{\sun}$ yr$^{-1}$;
\citealt{Fanelli88}). In most BCDs the gas consumption time scale is much
shorter than the age of the Universe, which, together with their low metal
content, raised initially the hypothesis that they could be truly young
galaxies, forming their first generation of stars \citep{Searle72,Kunth88}.
Nowadays it is clear that the great majority of them are not young, but rather
old systems (\citealt{Papaderos96a}; \citealt[][=C01a]{Cairos01a}; 
\citealt[][=C01b]{Cairos01b}; \citealt{BergvallOstlin02};
\citealt[][=C03]{Cairos03}; \citealt{Gildepaz03}) undergoing short starbursts
followed by longer quiescent periods. 

An essential requisite to comprehend the formation and evolution of BCDs, and
to elaborate coherent pictures of dwarf galaxy evolution, is to understand the
process of star-formation in these galaxies ---~how their starburst activity
ignites and how it propagates afterward~---, and to derive their star-forming
(SF) histories. Even though both issues have been the subject of a
considerable observational and interpretative effort during the last years,
they are still far from being well understood.

Different scenarios have been invoked to explain the onset of star formation
in these galaxies. Some of them favored internal processes, such as the
Stochastic Self-Propagating Star Formation proposed by \cite{Gerola80}, or the
hypothesis of cyclic gas re-processing, that is, the cyclic expulsion and
later accretion of the Interstellar Medium \citep{Davies88}. Alternatively,
interactions and/or mergers have been proposed as the mechanism responsible
for triggering the star formation in dwarfs. From optical surveys
\citep{Campos93,TellesTerlevich95,TellesMaddox00} it was first concluded that
gravitational interactions with optically bright galaxies are too rare to play
a relevant role.  However, studies at radio wavelengths
\citep{Taylor93,Taylor95,Taylor96} established that \ion{H}{2}/BCDs do have
companions, most of them gas-rich, faint objects, which may go undetected in
the optical. In agreement with these findings, more recent optical studies
that extended the searches at lower masses and luminosities also found that a
substantial fraction of SF dwarfs possess low mass companion galaxies
\citep{Noeske01,Pustilnik01}. \cite{Ostlin01} and \cite{BergvallOstlin02} put
forward the idea that a merger between two galaxies with different
metallicities, one gas-rich and one gas-poor, or infall of intergalactic
clouds, could be the most plausible explanation for the starburst activity, at
least in the most luminous BCDs. Recent studies, focused on individual
objects, have shown that interactions do play a decisive role in the evolution
of these systems \citep{Johnson04,BravoAlfaro04,BravoAlfaro06}. 

Much work must still be done to elucidate the star formation histories of
BCDs. The first step is to derive the properties of their stellar populations,
quite a difficult task in such complex objects. The great majority of them
cannot be resolved into stars, and the only way to study their stellar
populations is by comparing their integrated properties with the predictions
of evolutionary synthesis models. At any location in the galaxy, the emitted
flux is the sum of the emission from the local starburst, the flux produced by
the nebula surrounding the young stars, and the emission from the underlying,
old stellar population, all possibly modulated by dust (\citealt{Cairos00};
\citealt[][=C02]{Cairos02}; \citealt[][=C07]{Cairos07}). 

Another issue that was recently brought into discussion is the possible
heterogeneity of the BCD class. Among the galaxies classified as BCDs we find
a wide variety of objects spanning a large interval in luminosities ($-21.5 <
M_{B} < -14.0$) and morphologies. \cite{LooseThuan86} developed a classification
scheme for BCDs based on the appearance of the starburst and the shape of the
external envelopes, distinguishing four subclasses: \textit{iE} galaxies,
which show a complex inner structure with several SF regions over-imposed on
an outer regular envelope; \textit{nE}, objects with a clearly defined nucleus
and regular isophotes; \textit{iI}, which have irregular outer and inner parts
and, finally, \textit{iO} galaxies, in which an outer envelope is not found.
C01b have grouped the BCDs in four categories according to the position and
morphology of the SF knots: {\em nuclear starburst}, which are galaxies with a
prominent central starburst;  {\em extended starburst}, galaxies with star
formation spread over the entire galaxy, {\em chain starburst}, objects in
which the SF knots are aligned to form a chain and finally, {\em cometary
starburst}, galaxies with a "cometary" appearance, that is the star formation
concentrated in one side of the galaxy.

This finding introduces an additional complication, because it opens the
possibility that different star formation mechanisms may operate in BCDs and
that galaxies classified as BCDs may actually have different star formation
histories and evolutionary paths.

Spectrophotometric studies that put together high quality data in a large
wavelength range are fundamental to get insights into the star-formation
process and history of BCDs. However, very few such studies have been carried
out so far, and all of them focused on individual galaxies
(\citealt{Papaderos99,GildePaz00}; C02;
\citealt{Guseva03a,Guseva03b,Guseva03c}; C07). These studies rely on
conventional observing techniques: they combine good resolution broad-band
frames in different bands (to spatially isolate the different stellar
populations and map the dust distribution), with narrow-band imaging (needed
to map the gas distribution, isolate the  starburst regions and to derive
their physical properties) and spectroscopy data (to derive the internal
extinction, compute physical parameters and chemical abundances of the gas,
and remove the contribution from emission lines); a sequence of long-slit
spectra sweeping the regions of interest is usually taken. Therefore, these
analyses require a great amount of observing time; besides, varying
instrumental and  atmospheric conditions make combining all the data together
quite complicated. Long-slit spectroscopy has the additional problem of the
uncertainty on the exact location of the slit.

We have thus undertaken a project to carry out an Integral
Field Spectroscopy (IFS) mapping of a large sample of BCD galaxies. IFS is the
ideal observational technique to study BCDs: each single exposure contains
both spatial and spectral information in a large area of the galaxy, so just
in one shot we collect information for all the SF regions as well as for the
low surface brightness (LSB) stellar component. Besides, the kinematical
information also allows us to investigate what mechanisms ignite the
star-formation in BCDs. In terms of observing time, IFS observations of BCDs
are an order of magnitude more efficient than traditional observing
techniques, and provide simultaneous data for all spatial resolution elements
under the same instrumental and atmospheric conditions, which guarantees a
greater homogeneity of the dataset.

Here we present the first results from this project: we study five galaxies
representative of the group of BCDs populating the high luminosity and
metallicity range, and having rather complex morphologies. These objects,
often referred to as Luminous Blue Compact galaxies (LBCGs, see
\citealt{Ostlin98,Kunth00}; C01b) have a special significance in the general
scenario of galaxy formation and evolution, as they are claimed to be the
local counterparts of the luminous, compact, SF galaxies detected at higher
redshifts \citep{Garland04}. The proximity of these systems allows us to study
their structure, stellar populations, star formation processes and chemical
abundances with an accuracy and spatial resolution that cannot be achieved at
intermediate/high redshift.

\section{Observations and Data Reduction}

\subsection{The Galaxy Sample}

In this pilot study we focus on five luminous galaxies ($M_{B} \leq -17$) with
complex morphologies and where star-formation takes place in a number of
high surface brightness regions distributed atop an underlying, more regular
host envelope (C01b). The galaxies were selected from the larger sample of 28
BCDs already studied by our group (C01a,b). On one hand, these bright systems
are the most suitable for a pilot study; on the other hand, in such objects, in
which the star formation activity spread over the entire galaxy, an IFS mapping
is essential. 

Following the C01b classification, two galaxies (Mrk~297 and III~Zw~102)
belong to the \textit{extended starburst} group and three (Mrk~370,
Mrk~35 and Mrk~314) to the \textit{chain/aligned starburst} class. 

The basic data for the sample galaxies are given in Table~\ref{Table:data}
\footnote{A collection of color and \Ha\ maps of the galaxies can be found at:
\url{http://www.iac.es/proyect/GEFE/BCDs/}}. Column 5 lists the distances
computed assuming a Hubble flow, with a Hubble constant $H_{0} = 75$ km
s$^{-1}$ Mpc$^{-1}$, after correcting recession velocities relative to the
centroid of the local group for Virgocentric infall. Absolute magnitudes
(column 6) were obtained from the $B$ asymptotic magnitudes (C01b), using the
distances tabulated in column 5. Notice that the asymptotic magnitudes listed
in C01b were corrected for Galactic extinction following \cite{BurstHeil84};
here they are recomputed using the \cite{Schlegel98} extinction values.

\subsection{Observations}

The data analyzed in this paper were obtained in two observing runs. The first
run took place in 1999 August, and the second in 2003 February, both using the
William Herschel Telescope (\textsc{wht}), at the Observatorio del Roque de
los Muchachos (\textsc{orm}) on the island of La Palma.   

The \textsc{wht} was used in combination with the \textsc{integral} fiber
system  \citep{Arribas98} and the \textsc{wyffos} spectrograph
\citep{Bingham94}. \textsc{integral} links the Nasmyth focus of the
\textsc{wht} with the slit of \textsc{wyffos} via three optical fiber bundles.
These three bundles have different spatial configurations on the focal plane
and can be interchanged on-line depending on the scientific program or the
prevailing seeing conditions. At the focal plane, the fibers of each bundle
are arranged in two groups, one forming a rectangle, and the other an outer
ring mapping the sky. The data discussed in this paper were obtained with the
standard bundles 3 and 2 (\textsc{std3} and \textsc{std2} hereafter).
\textsc{std3} consists of 135 fibers, each $2\farcs7$ in diameter on the sky,
115 fibers forming a central rectangle and covering an area of 
$33\farcs6\times29\farcs4$ on the sky, and the other 20 fibers forming a ring 
$90\arcsec$ in diameter. \textsc{std2} has 219 fibers (189 science
and 30 sky fibers), each $0\farcs9$ in diameter and covering a field of
view of $16\arcsec\times12\arcsec$.

In 1999 August the \textsc{wyffos} spectrograph was equipped with a 600 groove
mm$^{-1}$ grating and a Tek CCD array of $1124 \times 1124$ pixels 24 $\mu$m
in size. The night was non-photometric and the seeing about $1\farcs3$. In
February 2003 observations of Mrk 35 were done using the 1200 groove mm$^{-1}$
grating and the same Tek CCD. We observed again in non-photometric conditions,
with a seeing of about $1\farcs5$.

A complete log of the observations is provided in Table~\ref{Table:logobs}.

Complementary broad and narrow-band (\Ha) images are also available. These
images were taken in different observing runs, using various telescopes and
instrumental configurations. Details about the observations and the data
reduction process can be found in C01a,b and C07.

\subsection{Data Reduction and Analysis}
\label{Sect:datared}

The data reduction consists of two main steps: i) basic reduction of the
spectra (i.e. bias, flatfielding, wavelength calibration, etc.) and ii)
generation of spectral feature maps (e.g. line intensities, radial velocity, 
etc).

The first step was performed in \textsc{iraf}, following the standard
procedures used in fiber data processing (see e.g. \citealp{GarciaLorenzo99}).
We obtained typical wavelength calibration errors of 0.1 \AA, which give
velocity uncertainties of $\pm 6$ and $\pm 4.5$  km s$^{-1}$ for
[\ion{O}{3}]~$\lambda5007$ and \Ha, respectively.  We corrected the spectra
for differential atmospheric refraction effects following the method proposed
by \cite{Arribas99}; the differential atmospheric refraction was estimated
according to the model given by \cite{Allen73}. In those cases in which the
air mass was close to one, such correction was not necessary.

For the generation of the spectral features maps, a two-dimensional
interpolation was applied using the {\sc ida} package \citep{GarciaLorenzo02}.
In particular we transformed an ASCII file with the actual position of the
fibers and the spectral feature corresponding to each fiber into a regularly
spaced rectangular grid. In this way we built up images of $70\times 70$
pixels with a scale $0\farcs5$ pixel$^{-1}$ for data from \textsc{std3}, and
$67\times 67$ pixels with a scale $0\farcs3$ pixel$^{-1}$ for data from
\textsc{std2}. These images can then be treated with standard astronomical
software packages. Maps generated in this way are presented in section 3 and 
beyond.

As mentioned before, observations were done under non-photometric conditions,
and no spectrophotometric stars were observed, so no absolute flux calibration
could be done. However, a relative calibration, aimed at computing the correct
line flux ratios, was carried out by means of the spectral data published by
\cite{MoustakasKennicutt06}, which lists line fluxes integrated over a large
area of the galaxies (all of our 5 objects are included in their sample).

We proceeded as follows. First, for each object we obtained a "total" spectrum
by adding up the spectra of all the science fibers, and computed the ratios,
in counts, between the \Hb, [\ion{O}{3}]~$\lambda5007$,
[\ion{N}{2}]~$\lambda6584$, [\ion{S}{2}]~$\lambda\lambda6717,\;6731$ lines and
\Ha. We then compared these values to the line ratios derived from the
\cite{MoustakasKennicutt06} data, and determined, for each such ratio, and for
each object, the factor to convert count ratios into flux ratios. For each
line ratio, the final conversion factor was obtained by averaging out the
corresponding values for each object, and the associated uncertainty estimated
from its scatter, after taking into account measurement errors (see notes in
Table~\ref{Table:emissionlines}). This method relies on the assumption that,
while the field of view of {\sc integral} is a fraction of the area covered by
\cite{MoustakasKennicutt06} spectra, it includes most of the galaxy light, and
thus area-related differences in line ratios are negligible.

\section{Results}

\subsection{The spectra}

As a first step in the spectroscopic analysis, we identify and select a number
of spatial regions for each galaxy by using high resolution continuum and \Ha\
images already analyzed by our group (see C01b and C07). The selected regions
include the galaxy nucleus (defined as the peak in the continuum emission),
and the most luminous SF knots; for all the galaxies, we also produce what we
call the LSB spectrum, obtained by adding up all the fibers spectra outside
the nucleus and the brighter knots; this spectrum, free from strong gas
emission, is more suitable for searching for absorption features.

\begin{figure*} 
\centerline{\hbox{
\includegraphics[width=.5\textwidth]{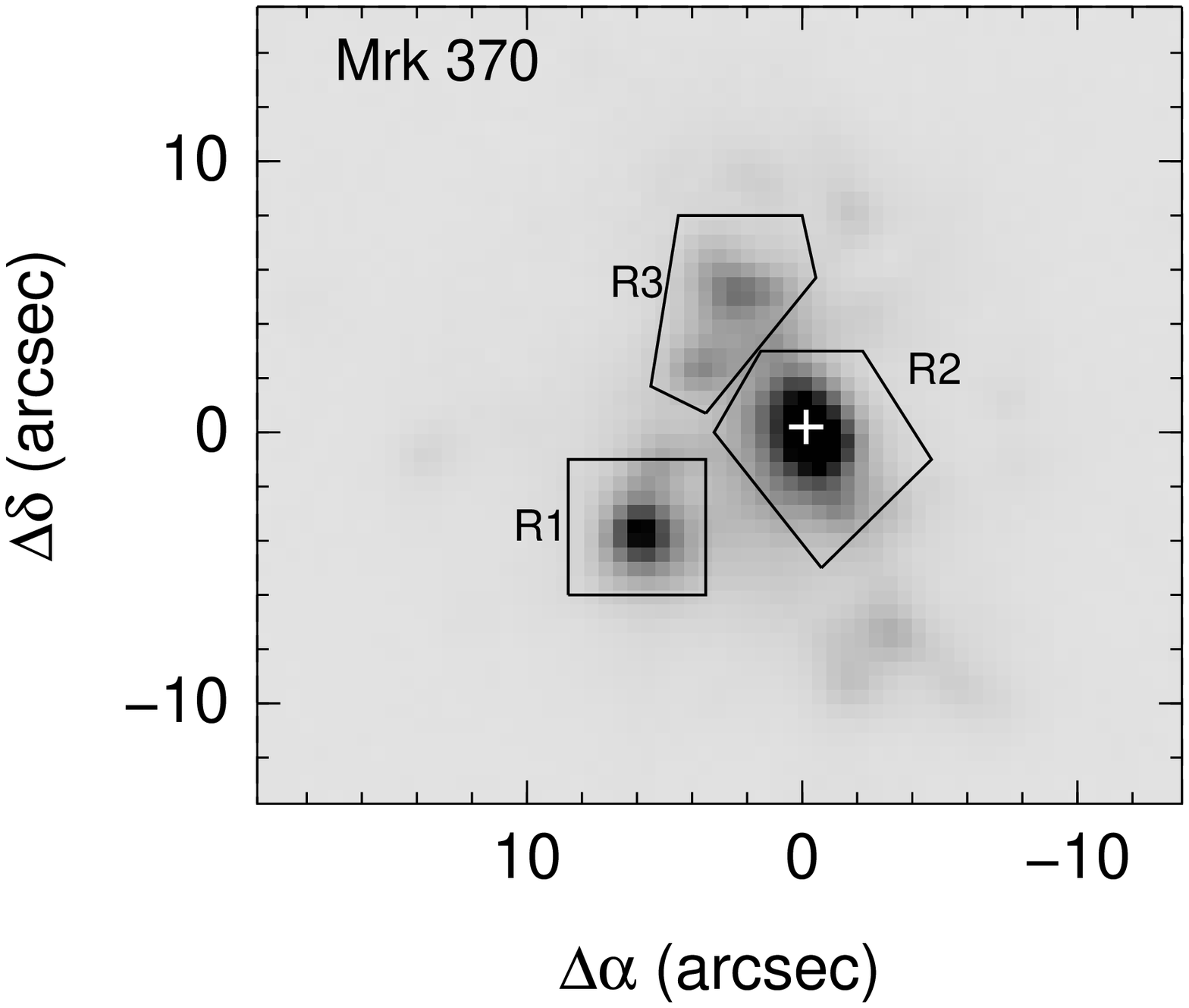}
\includegraphics[width=.5\textwidth]{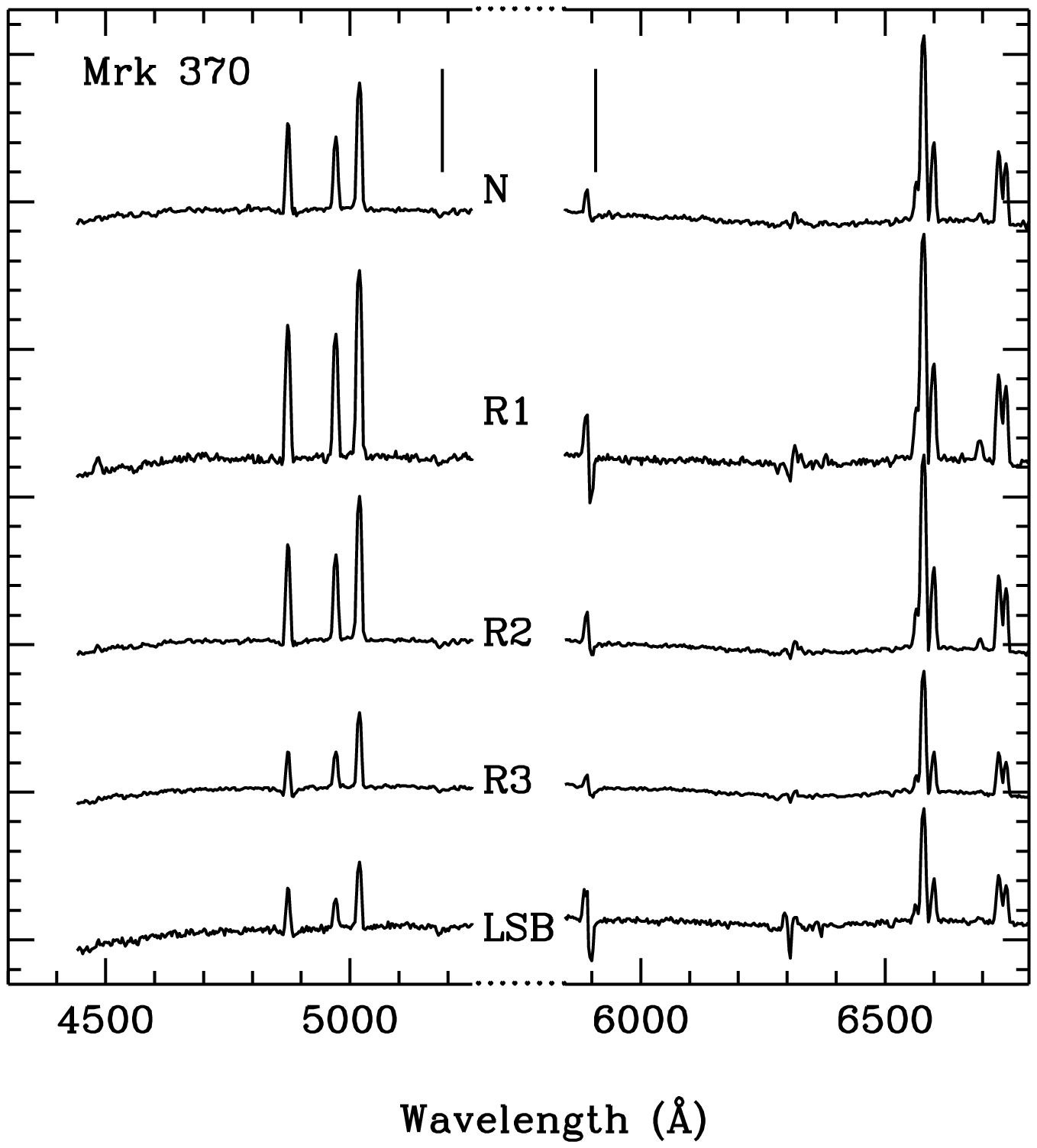}
}}
\caption{
\textit{Left:} Continuum subtracted \Ha\ image of Mrk~370 from C01b; the 
different regions selected in the galaxy are labeled. The white cross marks
the galaxy nucleus.
The area shown is the \textsc{integral} field of view. 
\textit{Right:} Representative spectra of the regions selected for analysis. 
The horizontal axis does not represent a continuous range in wavelength,
but two sub-intervals around \Hb\ and [\ion{O}{3}]~$\lambda\lambda4959,\;5007$ 
and around \Ha+[\ion{N}{2}]~$\lambda\lambda6548,\;6584$ and
[\ion{S}{2}]~$\lambda\lambda6717,\;6731$.
For sake of clarity the spectra are shown in logarithmic scale and arbitrarily
shifted on the Y axis. The step between small tickmarks is 0.2 dex. 
Two vertical lines mark the position of the \ion{Mg}{1} and \ion{Na}{1} D 
absorption lines. N is the spectrum of the nucleus.}
\label{Fig:mrk370spectra}
\end{figure*}

\begin{figure*}[h]   
\centerline{\hbox{
\includegraphics[width=.5\textwidth]{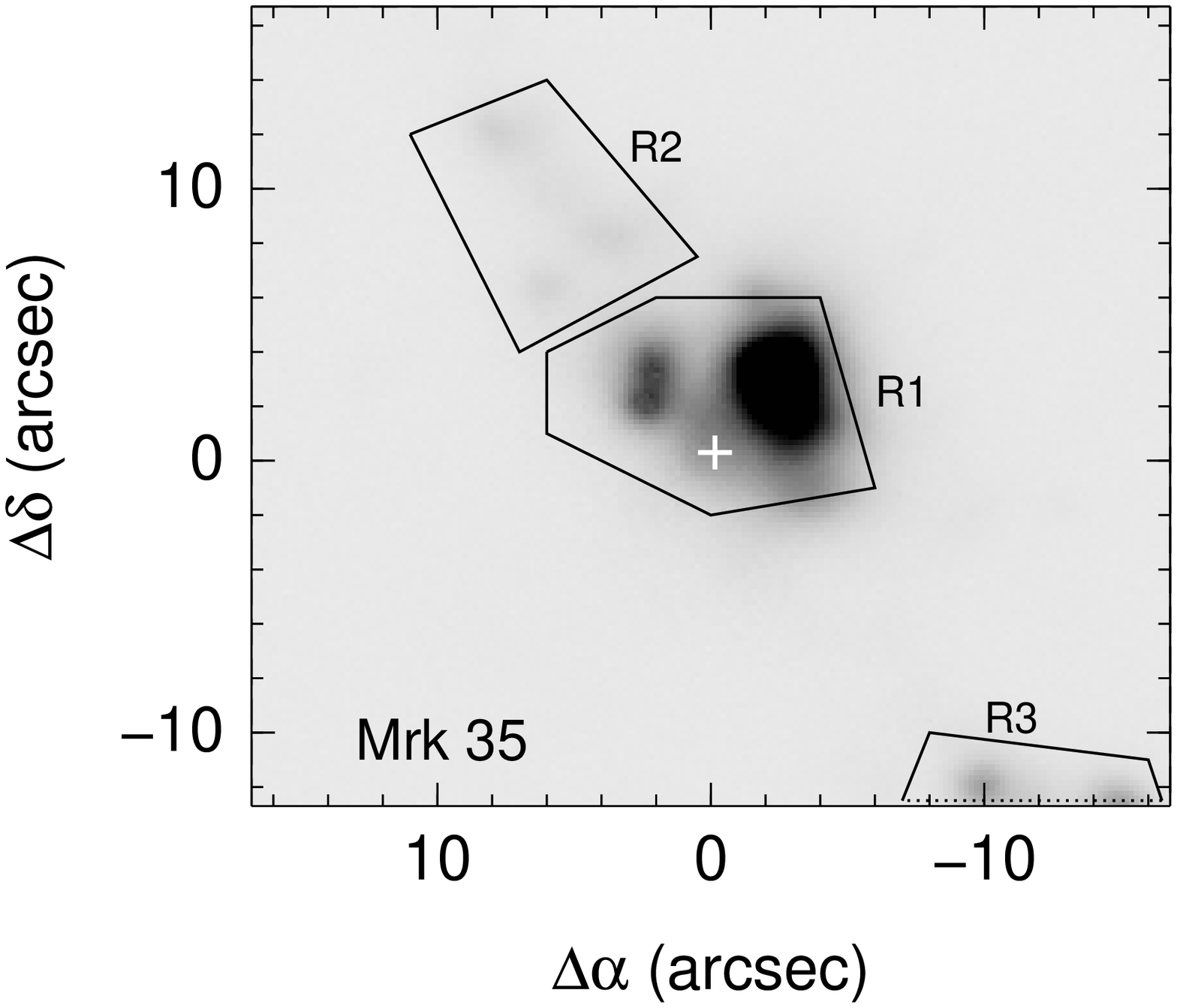}
\includegraphics[width=.5\textwidth]{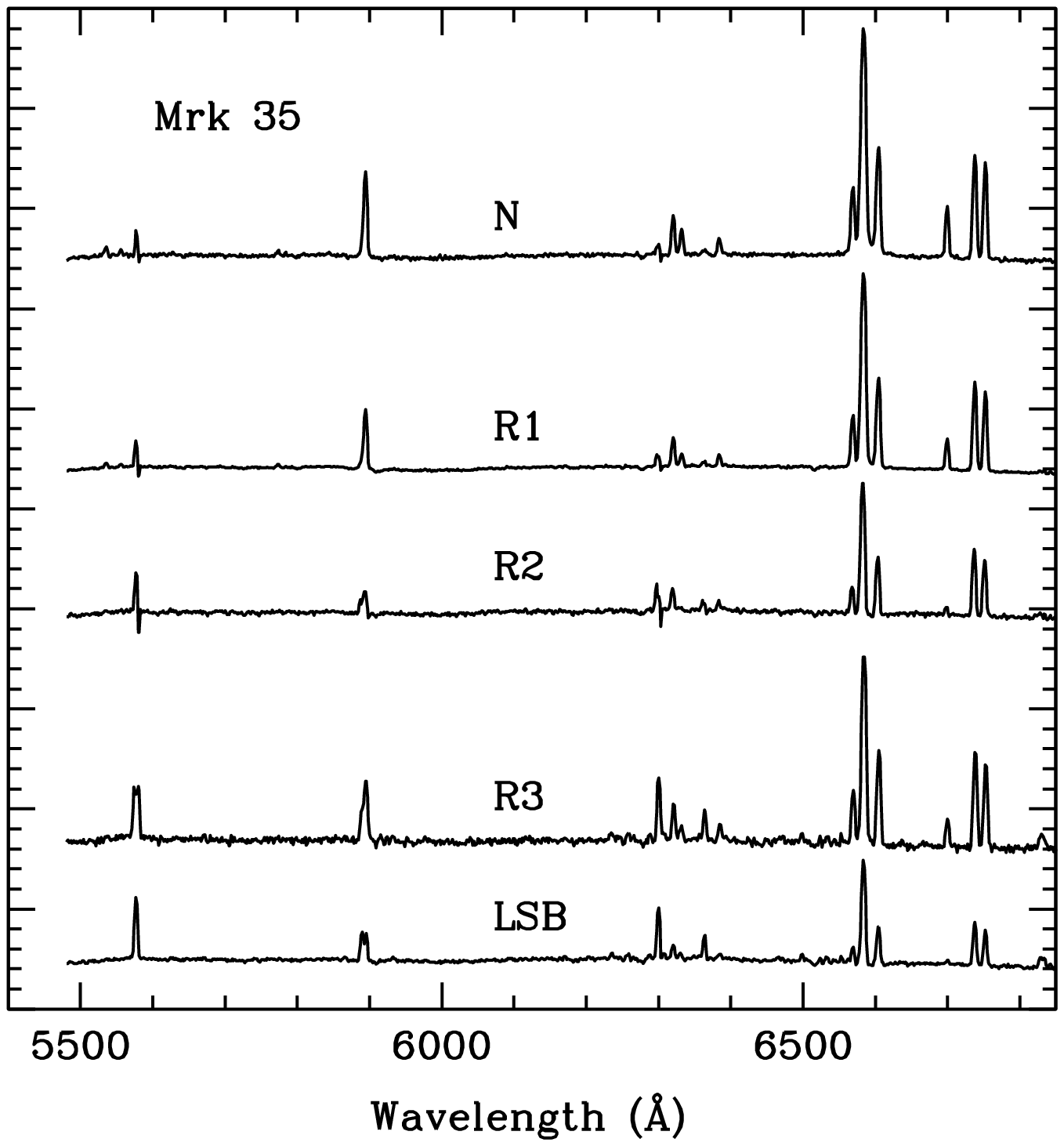}
}}
\caption{
Same as in Figure~\ref{Fig:mrk370spectra}, for Mrk~35
(Continuum subtracted \Ha\ image from C07)}
\label{Fig:mrk35spectra}
\end{figure*}

\begin{figure*}[h]   
\centerline{\hbox{
\includegraphics[width=.5\textwidth]{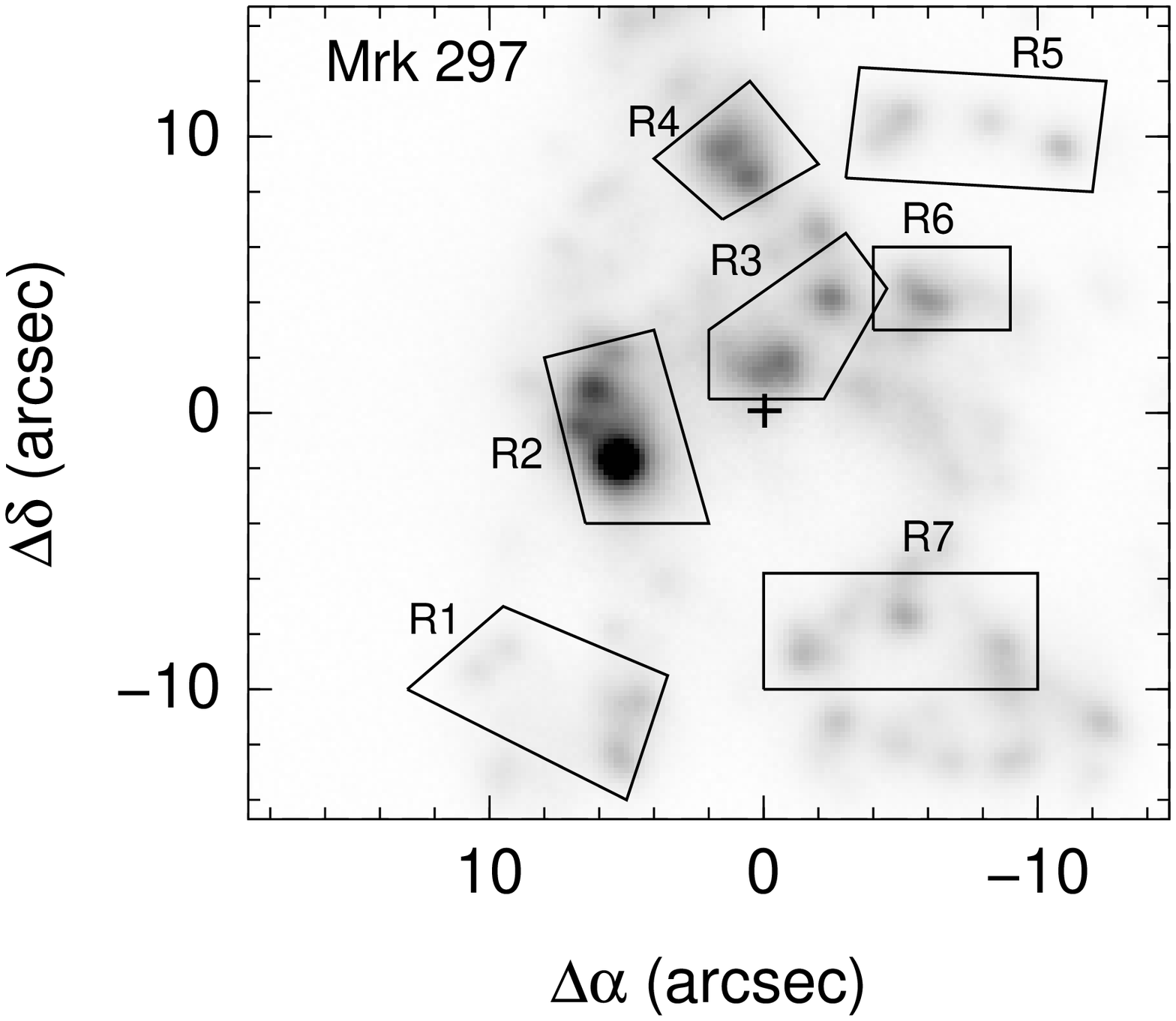}
\includegraphics[width=.5\textwidth]{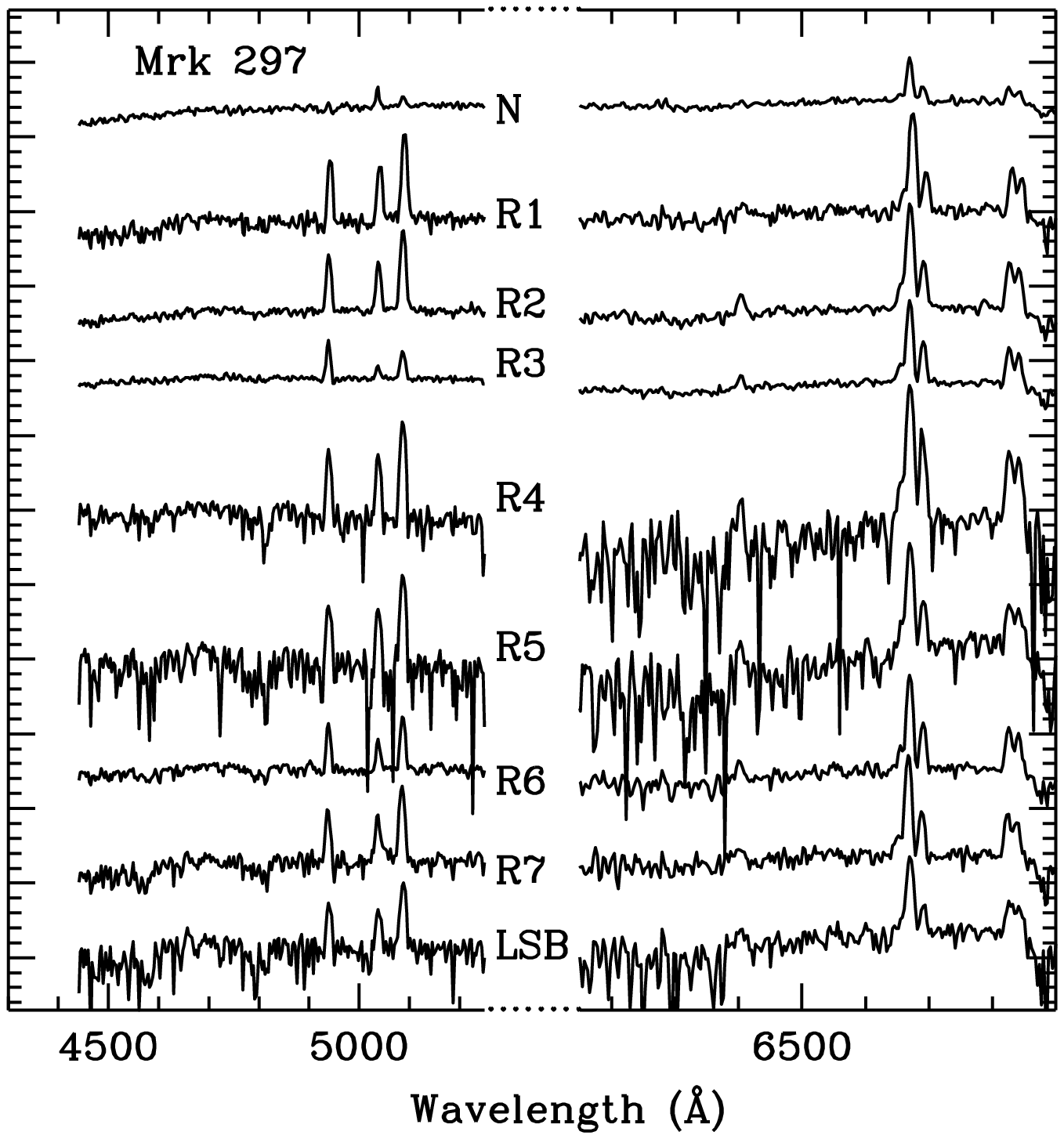}
}}
\caption{
Same as in Figure~\ref{Fig:mrk370spectra}, for Mrk~297.
}
\label{Fig:mrk297spectra}
\end{figure*}

\begin{figure*}[h]   
\centerline{\hbox{
\includegraphics[width=.5\textwidth]{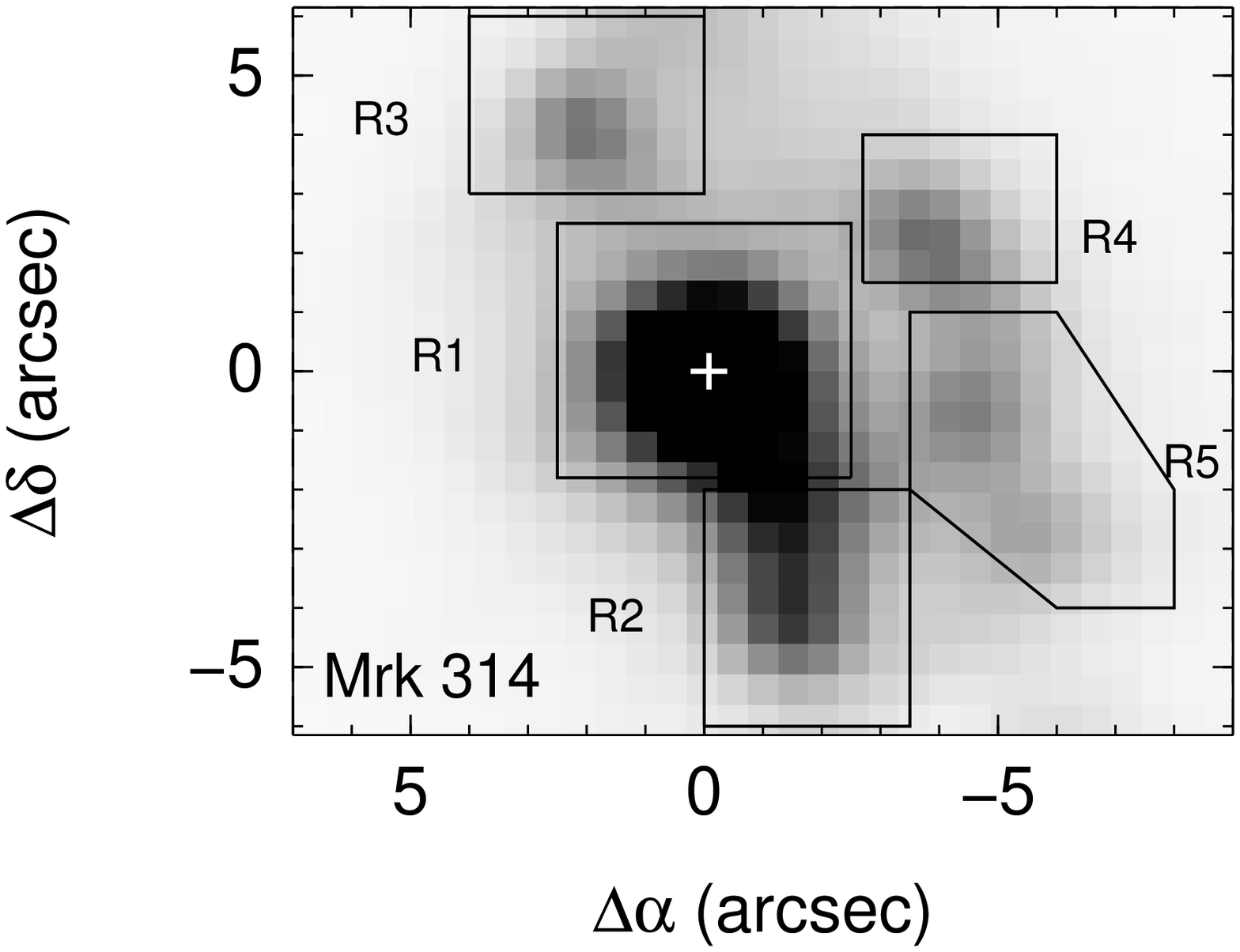}
\includegraphics[width=.5\textwidth]{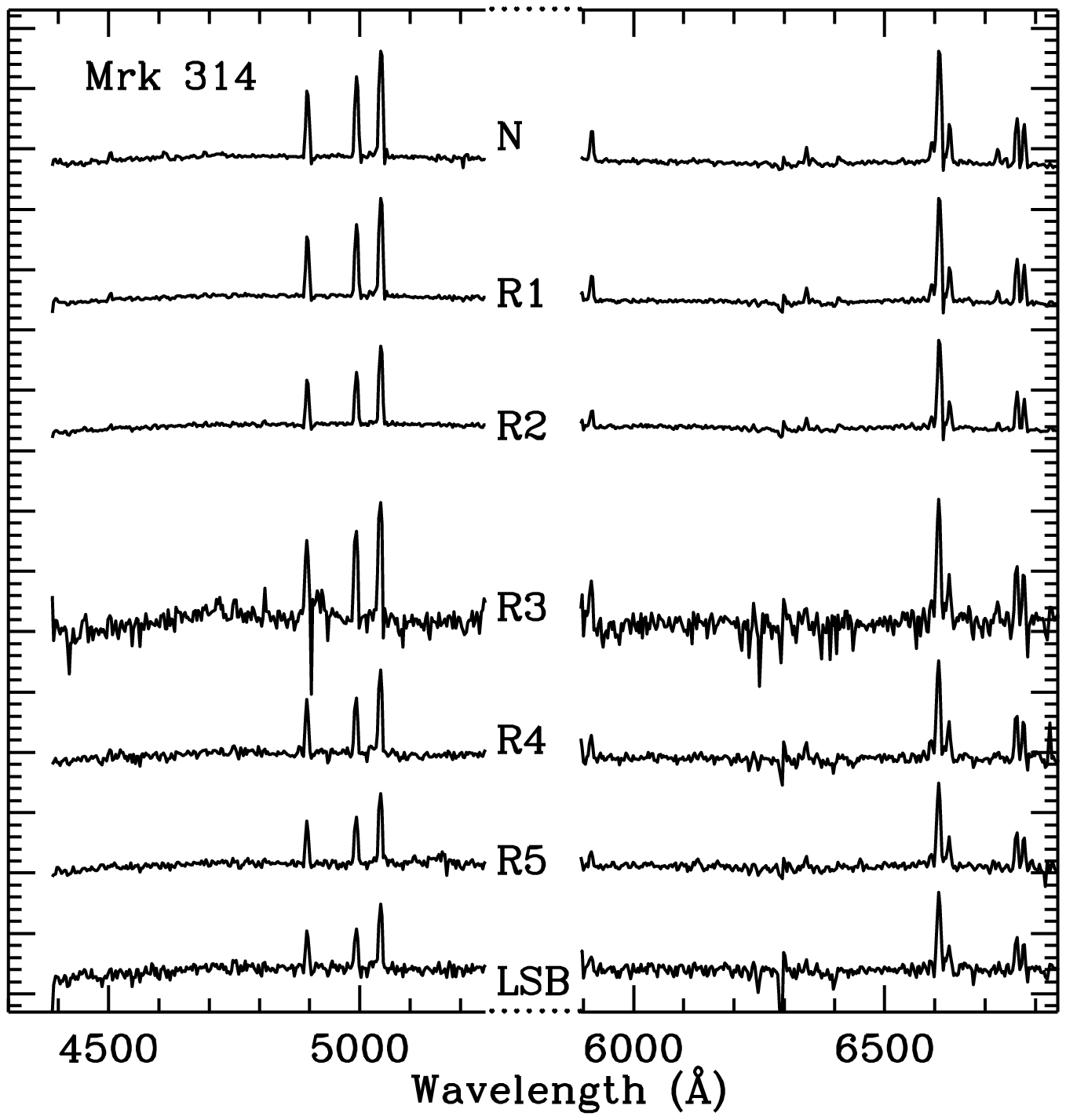}
}}
\caption{
Same as in Figure~\ref{Fig:mrk370spectra}, for Mrk~314.
}
\label{Fig:mrk314spectra}
\end{figure*}

\begin{figure*}[h]   
\centerline{\hbox{
\includegraphics[width=.5\textwidth]{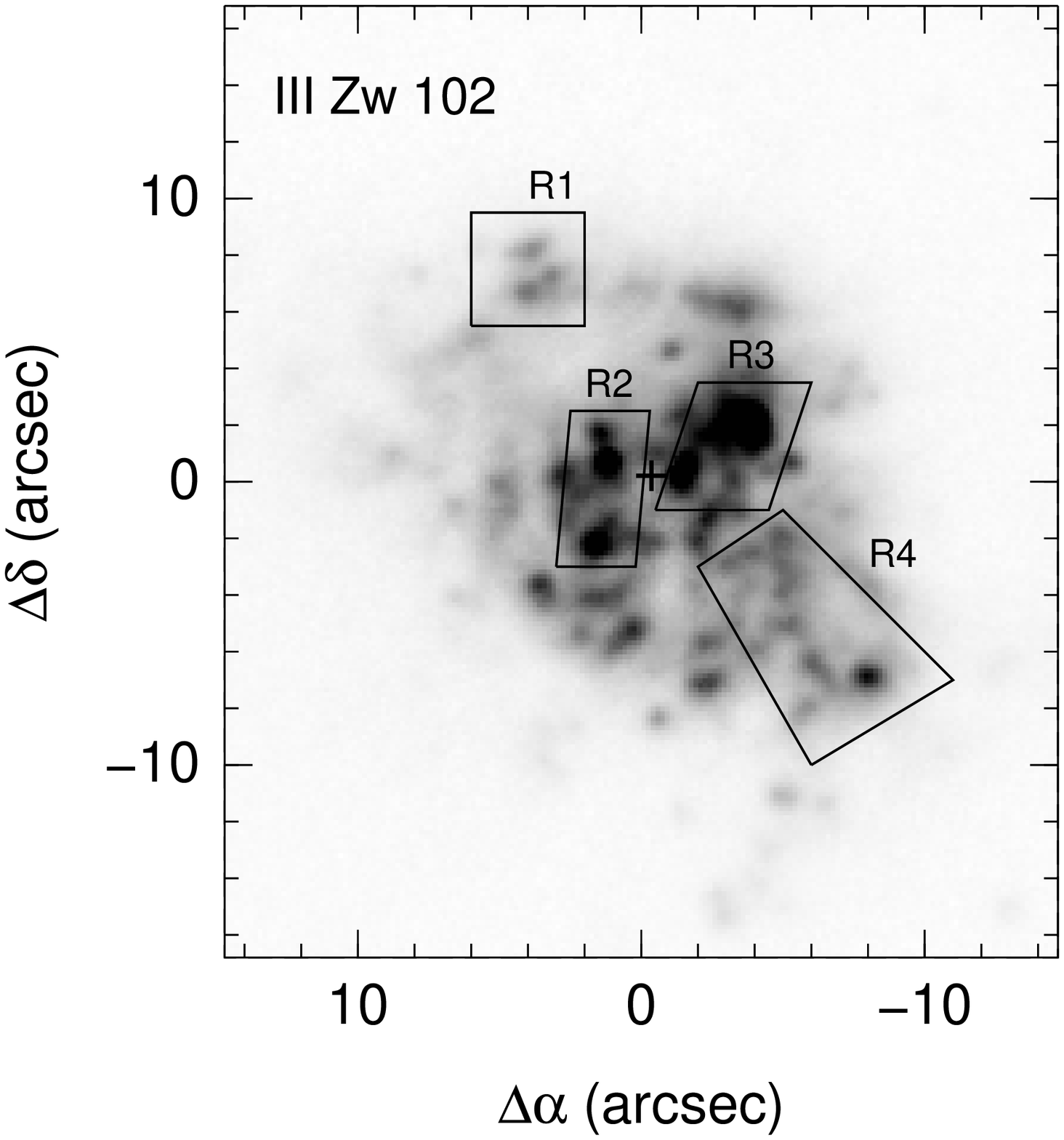}
\includegraphics[width=.5\textwidth]{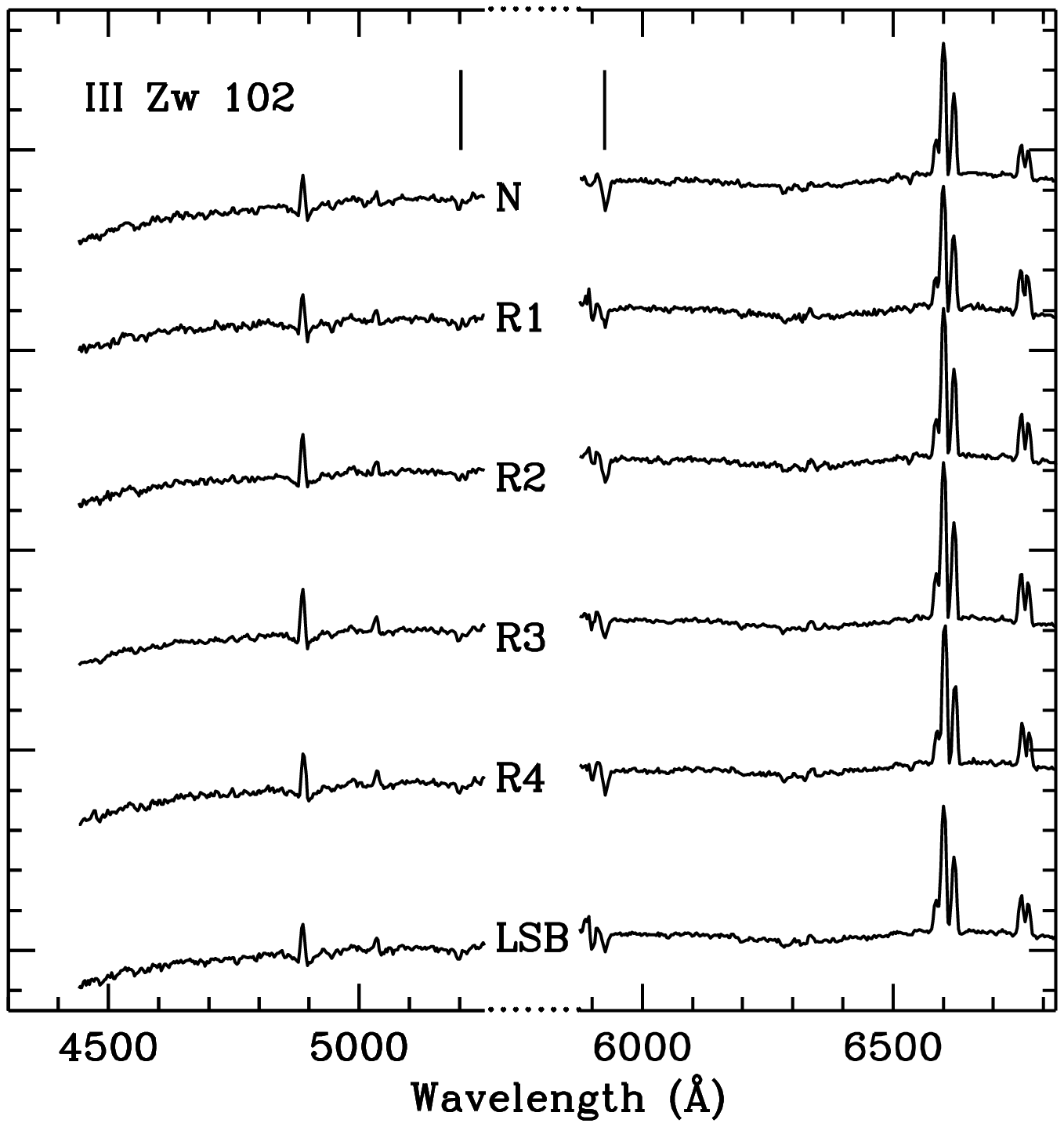}
}}
\caption{
Same as in Figure~\ref{Fig:mrk370spectra}, for III~Zw~102.
}
\label{Fig:iiizw102spectra}
\end{figure*}

By direct comparison of our reference images and \textsc{integral} maps, we
identify those fibers that cover the area of interest and create the final
spectrum for each region by summing them. Figures~\ref{Fig:mrk370spectra} to
\ref{Fig:iiizw102spectra} display the selected regions in each galaxy and their
spectra.

The dominating features in the final spectra are the emission lines
(\Hb, [\ion{O}{3}]~$\lambda5007$, \Ha, [\ion{N}{2}]~$\lambda6584$ and 
[\ion{S}{2}]~$\lambda\lambda6717,\;6731$), but significant differences appear
both among the objects, and among different regions in a same galaxy. In
particular, several regions show a very high continuum, with strong
absorption features (indicating the presence of a substantial contribution of
old stars) whereas other regions display a flat spectrum, with prominent
emission lines and no absorption features, characteristic of a dominant OB
population. The shape of the continuum, the strength of the emission and
absorption features or the presence of several lines strongly vary across the
objects. A more detailed description of the spectra of each galaxy is
provided below:

\begin{itemize}

\item {\em Mrk~370 --} We define five spatial regions (see
Figure~\ref{Fig:mrk370spectra}): the nucleus, the two major \Ha\ knots 
(\Rone\ and \Rtwo), the smaller \Ha\ knots at the north
(\Rthree) and the LSB component. The spectra of the nuclear region and
\Rtwo\ show a high continuum and strong absorption wings around H$\gamma$
and \Hb, indicating the presence of a substantial fraction of older stars; the
absorption lines \ion{Mg}{1}~$\lambda\lambda5167,\;5173$ and
\ion{Na}{1}~D~$\lambda5892$ are detected. In \Rthree, with weaker gas
emission, the absorption features are more evident. \Rone, on the
contrary, has a flat continuum, with no absorption features, characteristic of
a population of young stars. \ion{Mg}{1}~$\lambda\lambda5167,\;5173$ and
\ion{Na}{1}~D~$\lambda5892$ are visible in the LSB component spectrum.

\item {\em Mrk~35 --} This galaxy was observed with a different configuration,
which provides higher spectral resolution in a smaller wavelength range. Five
regions have been defined (see Figure~\ref{Fig:mrk35spectra}).
\textsc{integral} does not allow to resolve the central knots, which we lump
together within the aperture \Rone; the nucleus (continuum peak) does
not coincide with any \Ha\ peak. Both the nucleus and \Rone\ have flat
spectra with prominent emission lines, in which
[\ion{O}{1}]~$\lambda\lambda6300,\;6363$, \ion{He}{1}~$\lambda5876$,
\ion{He}{1}~$\lambda6678$ and
[\ion{S}{3}]~$\lambda6312$ are also visible. \Rtwo, which includes several 
smaller SF regions towards the NE, and \Rthree, which corresponds with
the two knots detached from the main body of the galaxy, both display similar
flat spectra. No absorption lines are visible in the spectrum of the LSB
component.

\item {\em Mrk~297 --} We study seven SF regions, plus the nucleus and the LSB
emission (see Figure~\ref{Fig:mrk297spectra}). The characteristics of the
spectra in the selected regions differs significantly. The nucleus, which does
not coincide with any \Ha\ peak, shows a very high continuum, and displays
strong H$\gamma$ and \Hb\ absorption lines; \Rone\ and \Rtwo\ have a flat
continuum, with no evident absorption features; both spectra show
\ion{He}{1}~$\lambda5969$, but only in \Rtwo\ [\ion{O}{1}]~$\lambda6300$ is
apparent. \Rthree, located in the central region of the galaxy, has a higher
stellar continuum, significant absorption features, and strong
\ion{He}{1}~$\lambda5969$ and [\ion{O}{1}]~$\lambda6300$ lines. \Rfour,
\Rfive, \Rsix\ and \Rseven, show flat spectra; [\ion{O}{1}]~$\lambda6300$ is
detected in all these regions except \textsc{r7}.  The signal-to-noise in the
LSB spectra is not large enough to search for absorption features.

\item {\em Mrk~314 --} This galaxy was observed with the \textsc{std2} of
\textsc{integral}, which provides higher spatial resolution in a smaller field
of view. Our \textsc{ifu} data allow us to resolve the five larger knots
detected in the \Ha\ frames (Figure~\ref{Fig:mrk314spectra}). The nucleus,
which coincides with the \Ha\ knot \Rone, displays a nebular spectrum
dominated by emission lines; [\ion{O}{1}]~$\lambda\lambda6300,\;6363$, 
\ion{He}{1}~$\lambda5876$ and \ion{He}{1}~$\lambda6678$ are visible and
\ion{Mg}{1}~$\lambda5167$ is detected in absorption. The five selected regions
show flat spectra, typical of \ion{H}{2} regions. No absorption features are
detected in the spectrum of the LSB component.

\item {\em III~Zw~102 --} We study four spatial regions besides the nucleus and
the LSB component (Figure~\ref{Fig:iiizw102spectra}). The nucleus does not
coincide with any \Ha\ peak and has a very high continuum, with prominent
absorption features (e.g. \ion{Mg}{1}~$\lambda5167$,
\ion{Na}{1}~D~$\lambda5892$ and absorption wings around \Hb); interestingly,
there is very weak [\ion{O}{3}]~$\lambda\lambda4959,\;5007$ in emission. The
spectra of the four selected regions are quite similar: all are dominated by a
high continuum, have faint [\ion{O}{3}]~$\lambda\lambda4959,\;5007$ lines and
display pronounced absorption features. Indeed, the shape of the spectra
resembles those typical of spiral galaxies, rather than of BCDs. 
\end{itemize}

In  Appendix we present the individual spectra corresponding to each fiber 
in a selected spectral interval centered around \Hb\ (in the case of Mrk~35, 
around \Ha).

\subsection{Emission line ratios}
\label{Sect:emissionlineratios}

We measured the intensity of the lines in the spectrum of each region defined
in the previous section. The emission line profiles were fitted by a single
Gaussian function using the {\sc dipso} package inside the {\sc starlink}
environment\footnote{http://www.starlink.rl.ac.uk/}. In order to obtain
accurate values of the Balmer emission line fluxes, we took the underlying
stellar absorption into account by fitting two Gaussians in those cases where
absorption wings are visible. The 
\Ha+[\ion{N}{2}]~$\lambda\lambda6548,\;6584$ lines were fitted simultaneously,
imposing that the three lines have the same velocity shift and the same width,
and fixing to 3 the ratio between the two nitrogen lines. Also the
[\ion{O}{3}]~$\lambda\lambda4959,\;5007$ lines were fitted simultaneously,
again fixing the line ratio to 3, using the same width for both lines, and
forcing them to have the same velocity shift. The same constraints (except the
one on the flux ratio) were applied to the
[\ion{S}{2}]~$\lambda\lambda6717,\;6731$ doublet. The continuum was fitted
with a linear or quadratic function.

The most relevant line ratios for each region are listed in
Table~\ref{Table:emissionlines}.

In virtually all the regions, the \Ha/\Hb\ line ratio is considerably larger
than the nominal values of 2.86 \citep{Osterbrock06}, indicating that they are
significantly affected by dust obscuration. Particularly high values are found
for Mrk~297 and III~Zw~102. This result is in agreement with previous
findings: the color maps published in C01b and C03 already pointed out to the
presence of dust in both objects; the presence of strong dust lanes in
III~Zw~102 has been long known \citep{Demoulin69,Brosch91}.

\subsubsection{Ionization Mechanisms}
\label{Sect:ionizationmechanisms}

The ionization mechanisms for the individual SF regions can be studied by
means of various diagnostic diagrams (see the classical paper by
\citealp{Baldwin81}). Here we follow the scheme proposed by \cite{Veilleux87}
in which the [\ion{O}{3}]~$\lambda5007$/\Hb, [\ion{O}{1}]~$\lambda6300$/\Ha,
[\ion{N}{2}]~$\lambda6584$/\Ha\ and
[\ion{S}{2}]~$\lambda\lambda6717,\;6731$/\Ha\ line ratios are used to 
distinguish among different ionization mechanisms (radiation from young stars,
shocks and AGNs). 

\begin{figure*}  
\includegraphics[angle=270,width=\textwidth]{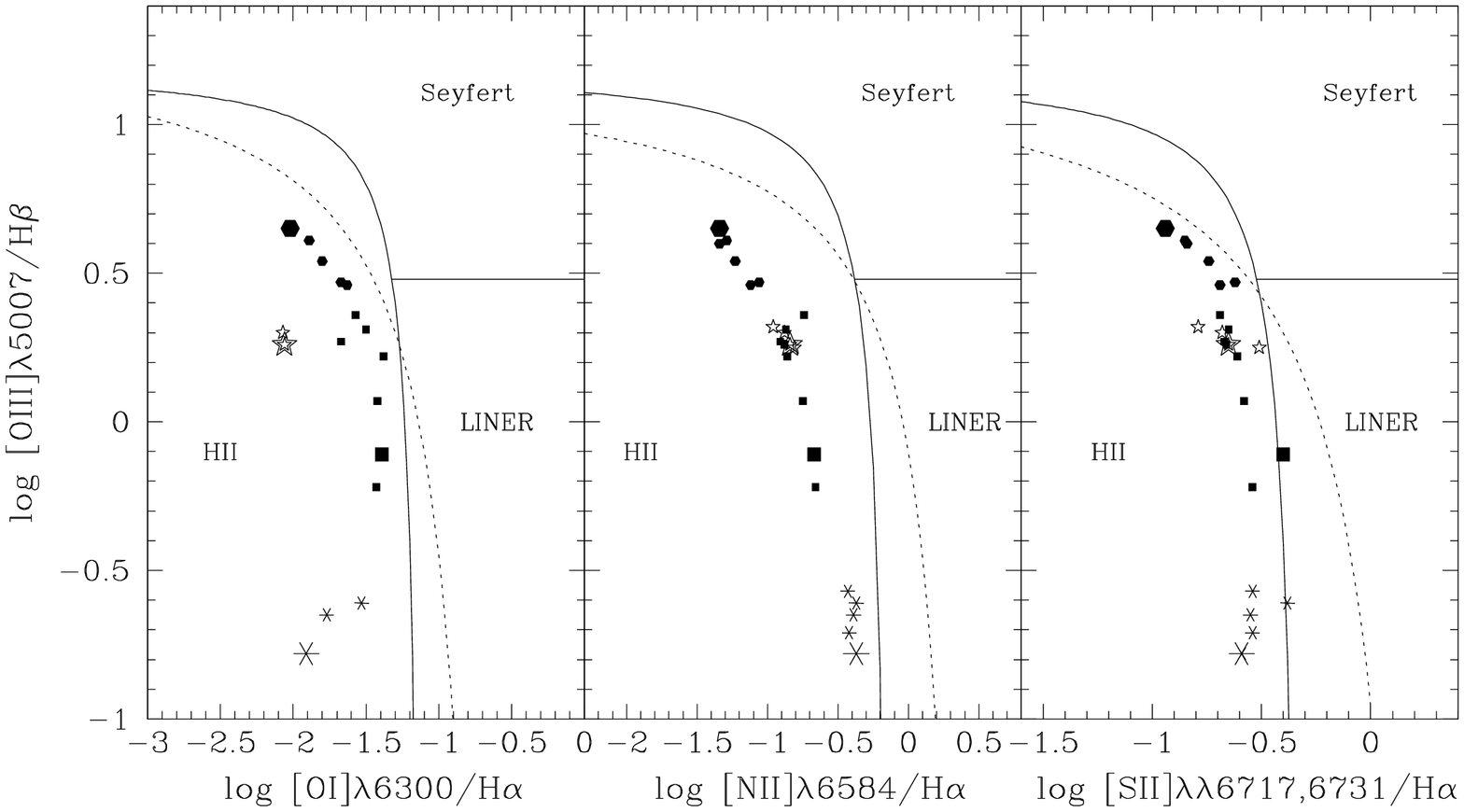}
\caption{Optical emission line diagnostic diagrams for the selected regions in
the sample galaxies.
The curves separate Seyferts and LINERs from \ion{H}{2} region-like
objects;
{\em solid lines} are the empirical borders from \cite{Veilleux87}, while
{\em dotted lines} represent the theoretical borders from \cite{Kewley01}.
Different galaxies are shown with different symbols: 
stars, Mrk~370; squares, Mrk~297; hexagon, Mrk~314; asterisks, III~Zw~102. The
nuclei are shown with a larger symbol.}
\label{Fig:diagnostic}
\end{figure*}

The position of the different regions under study in these diagrams is shown
in Figure~\ref{Fig:diagnostic} (Mrk~35 is not included because in this galaxy
[\ion{O}{3}]~$\lambda5007$ and \Hb\ are outside the observed wavelength
interval). The empirical boundaries between the different zones (from
\citealp{Veilleux87}) as well as the theoretical boundaries proposed by
\cite{Kewley01} are also plotted. The main result emerging from this plot is
that radiation from young stars is the dominant ionization mechanism in all 
the galaxies.
Also, we found that all the different SF regions of a same galaxy lie very
close to each other in the diagram: the SF knots identified in Mrk~314 occupy
the upper zone (with [\ion{O}{3}]~$\lambda5007$/\Hb\ $\geq 3$), while the
III~Zw~102 regions have lower excitation parameters
([\ion{O}{3}]~$\lambda5007$/\Hb\ $\leq 1$). In three galaxies, the nucleus
unequivocally has a \ion{H}{2}-like ionization. Only the nucleus of Mrk~297
falls on the boundary between the \ion{H}{2} and the LINER regions in the
[\ion{S}{2}]~$\lambda\lambda6717,\;6731$/\Ha\ diagram.


\subsubsection{Electron Densities and Chemical Abundances}
\label{Sect:electronicdensities}

From the measured emission line fluxes, we can derive electron densities
and oxygen abundances in the selected galaxy regions.  

As we did not detect the [\ion{O}{3}]~$\lambda$4363 emission line in any of
the spectra, we could not calculate the electron temperature (the alternative
empirical calibrator proposed by \cite{Pilyugin01} to estimate \Te\ relies on
the flux of the [\ion{O}{2}]~$\lambda3727$ line, which is outside the observed
spectral range). Therefore, in order to compute electron densities, we assumed
an electron temperature of 10000 K, a good approximation for most \ion{H}{2}
regions \citep{Osterbrock06}. 

We determined the electron density of the ionized gas from the 
[\ion{S}{2}]~$\lambda$6717/$\lambda$6731 ratio, using the task
\textsc{temden}, based on the \textsc{fivel} program \citep{ShawDufour95},
which is included in the {\sc iraf} package \textsc{nebular}. We find densities in
between $\leq100$ and 360 cm$^{-3}$, typical of classical \ion{H}{2} regions
\citep{Copetti00}. In all galaxies except Mrk~370, we find that the electron
density significantly varies across the galaxy.

Abundances were estimated using the empirical calibrator {\sc N2}
\citep{Storchi94}, following the calibration made by \cite{Denicolo02}. The
abundances range from $\simeq 0.3 Z_{\sun}$ to $\simeq 1.5 Z_{\sun}$ (the
solar value is taken as $12+\log\mbox{(O/H)}=8.69$, \citealp{Lodders03}).

Electron densities and oxygen abundance values are listed in
Table~\ref{Table:physicalpara}.

\subsection {Intensity maps}
\label{Sect:3dmaps}

The procedure that we follow to generate the maps is described in 
Section~\ref{Sect:datared}. All maps are corrected for galactic
extinction.

\subsubsection {Continuum and Line Emission Maps}
\label{Sect:continuummaps}

In order to study the properties of the stellar component, we build continuum
maps within selected wavelength intervals free from emission lines, which
trace the stellar properties (``pure  continua''), e.g. 5600--5800 \AA\ or 
7000--7200 \AA. 

\begin{figure*}[h]   
\mbox{
\centerline{
\hspace*{0.0cm}\includegraphics[width=4.5cm]{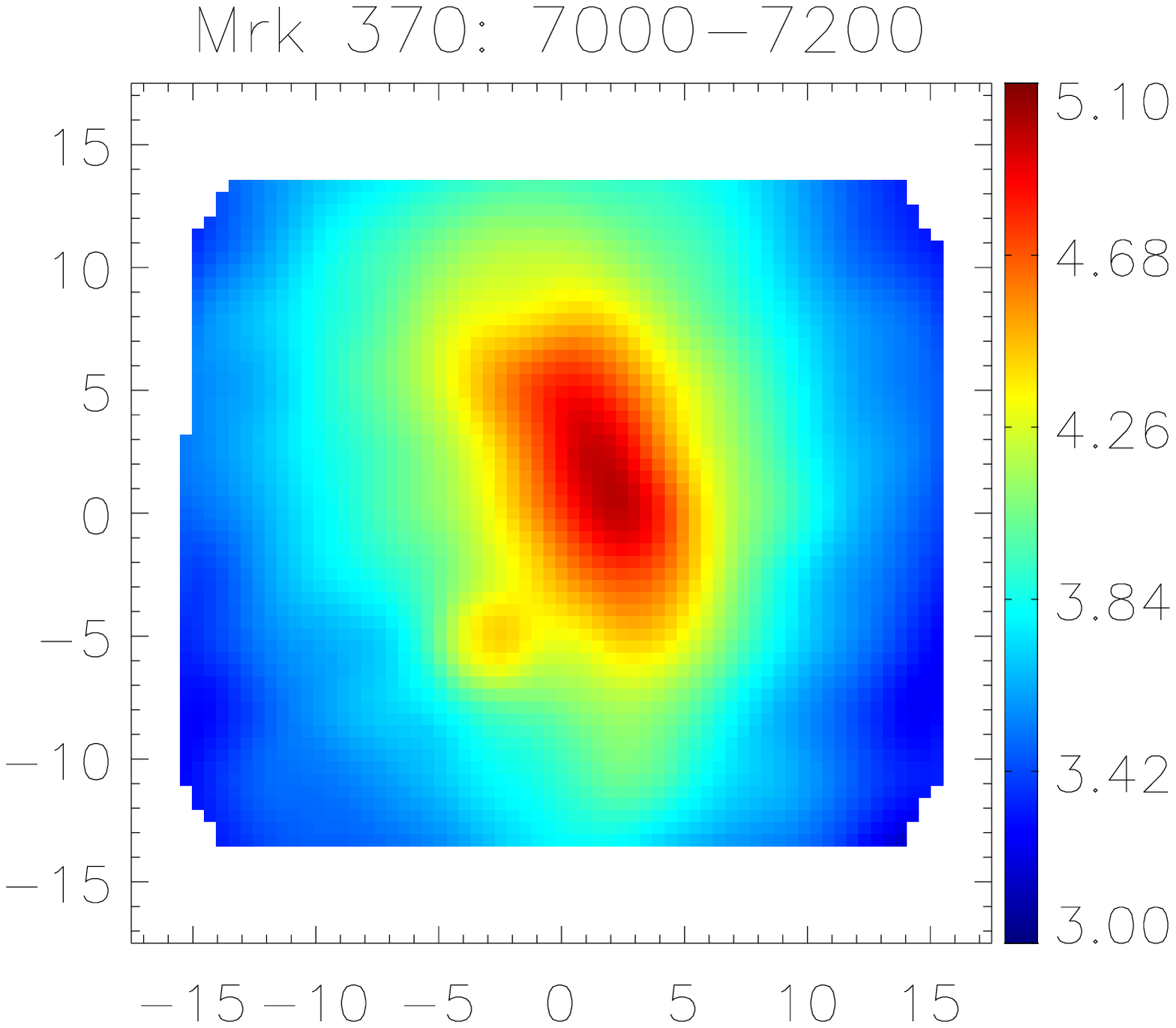}
\hspace*{0.0cm}\includegraphics[width=4.5cm]{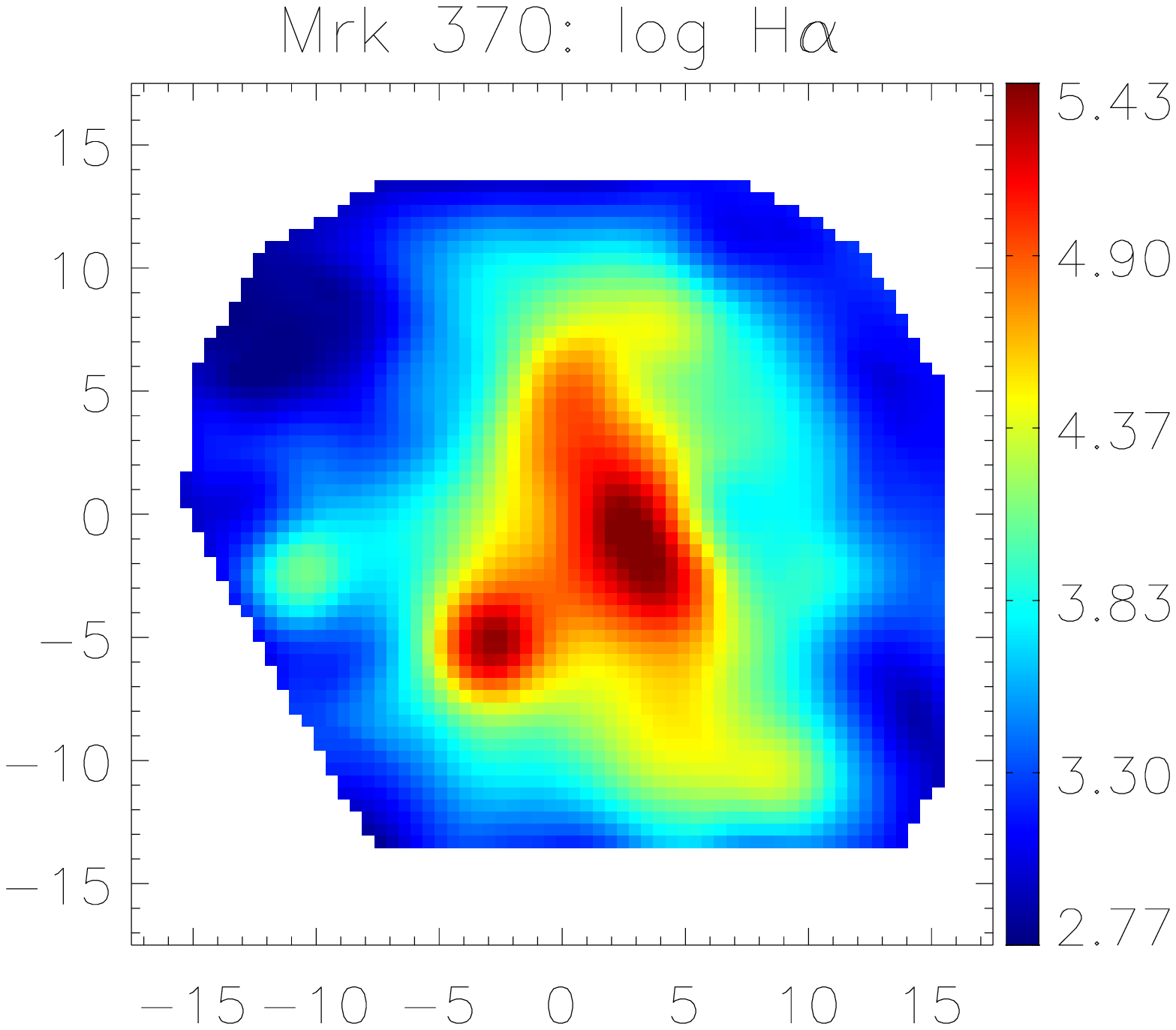}
\hspace*{0.0cm}\includegraphics[width=4.5cm]{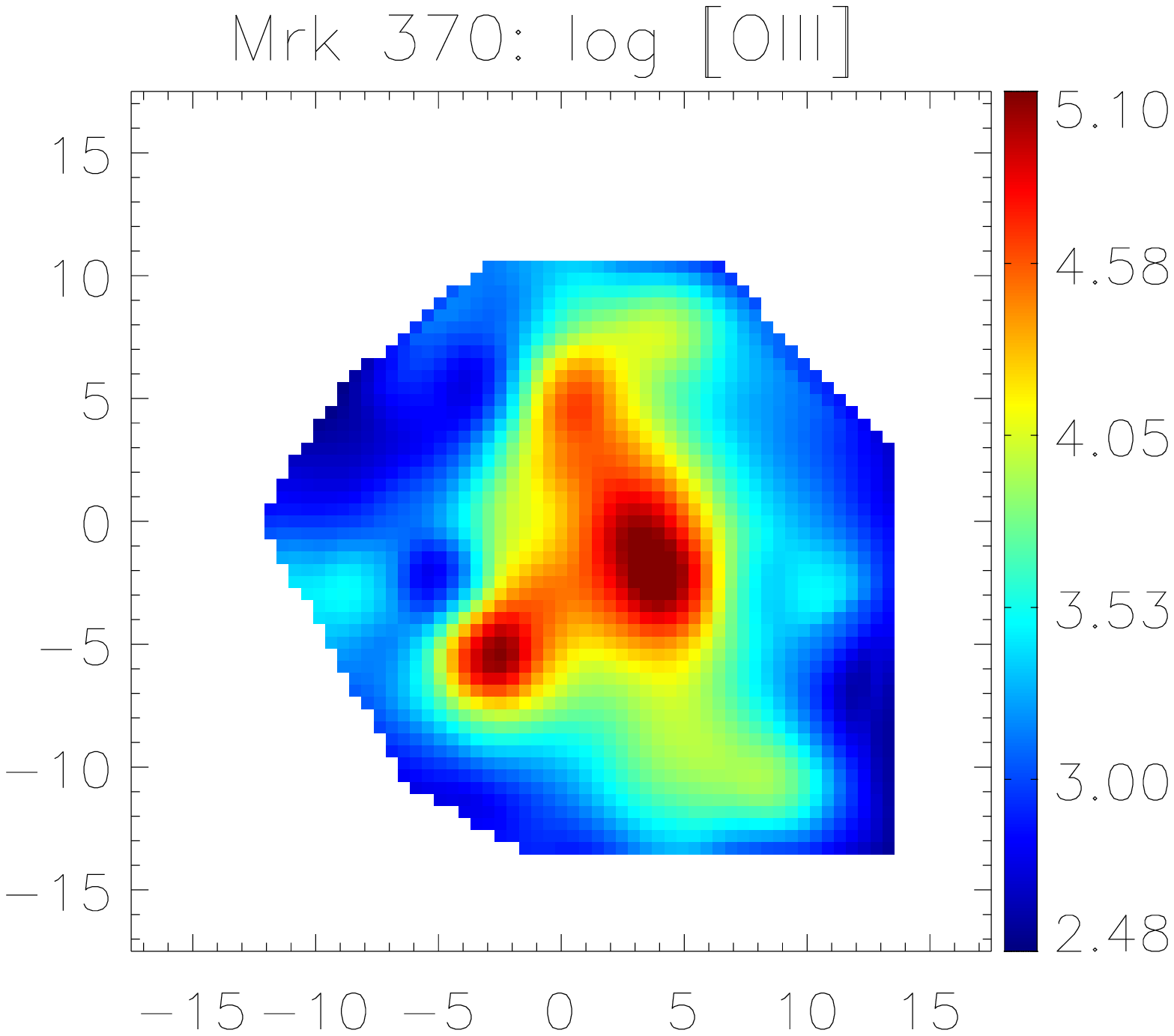}
}}
\mbox{
\centerline{
\hspace*{0.0cm}\includegraphics[width=4.5cm]{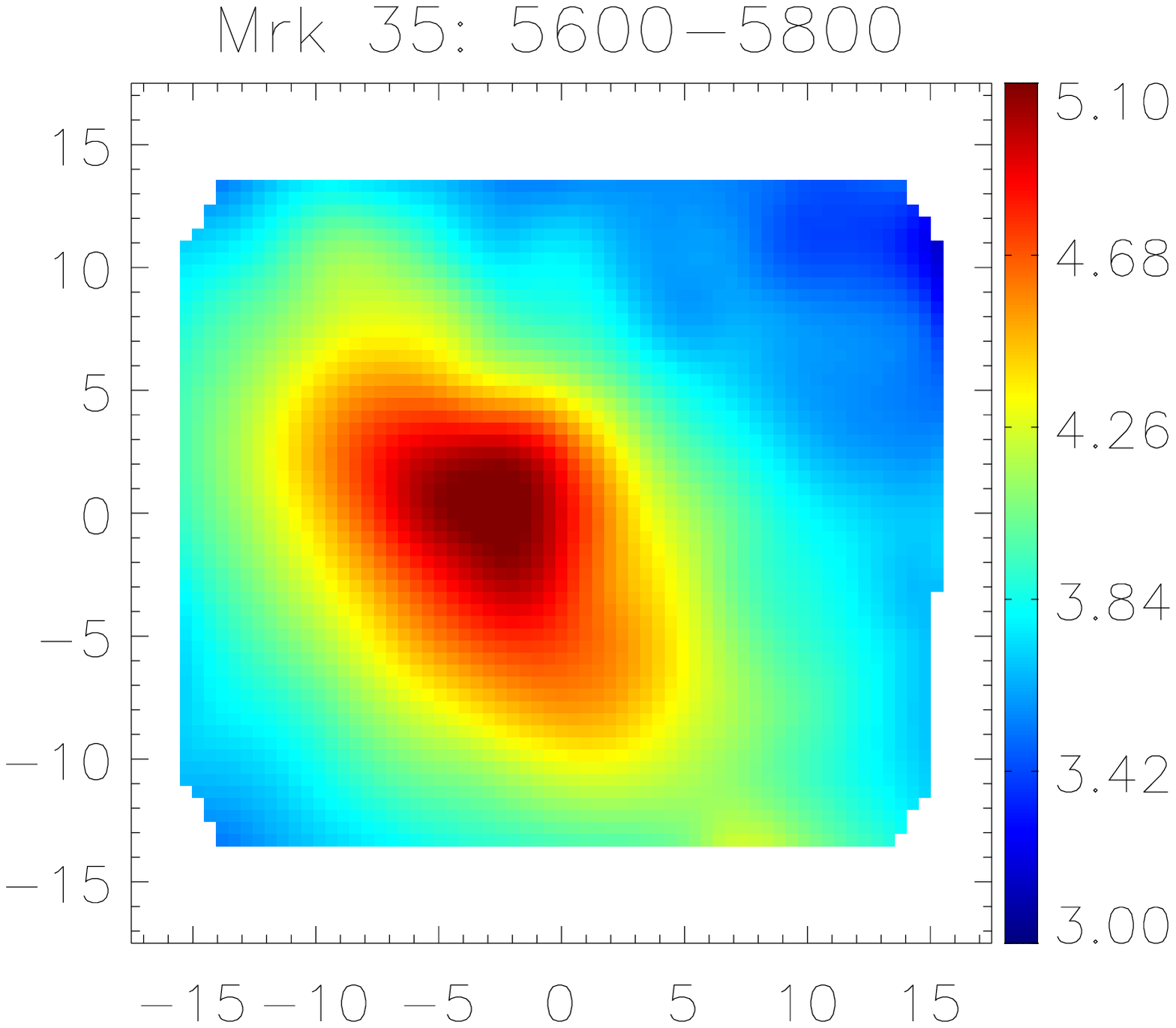}
\hspace*{0.0cm}\includegraphics[width=4.5cm]{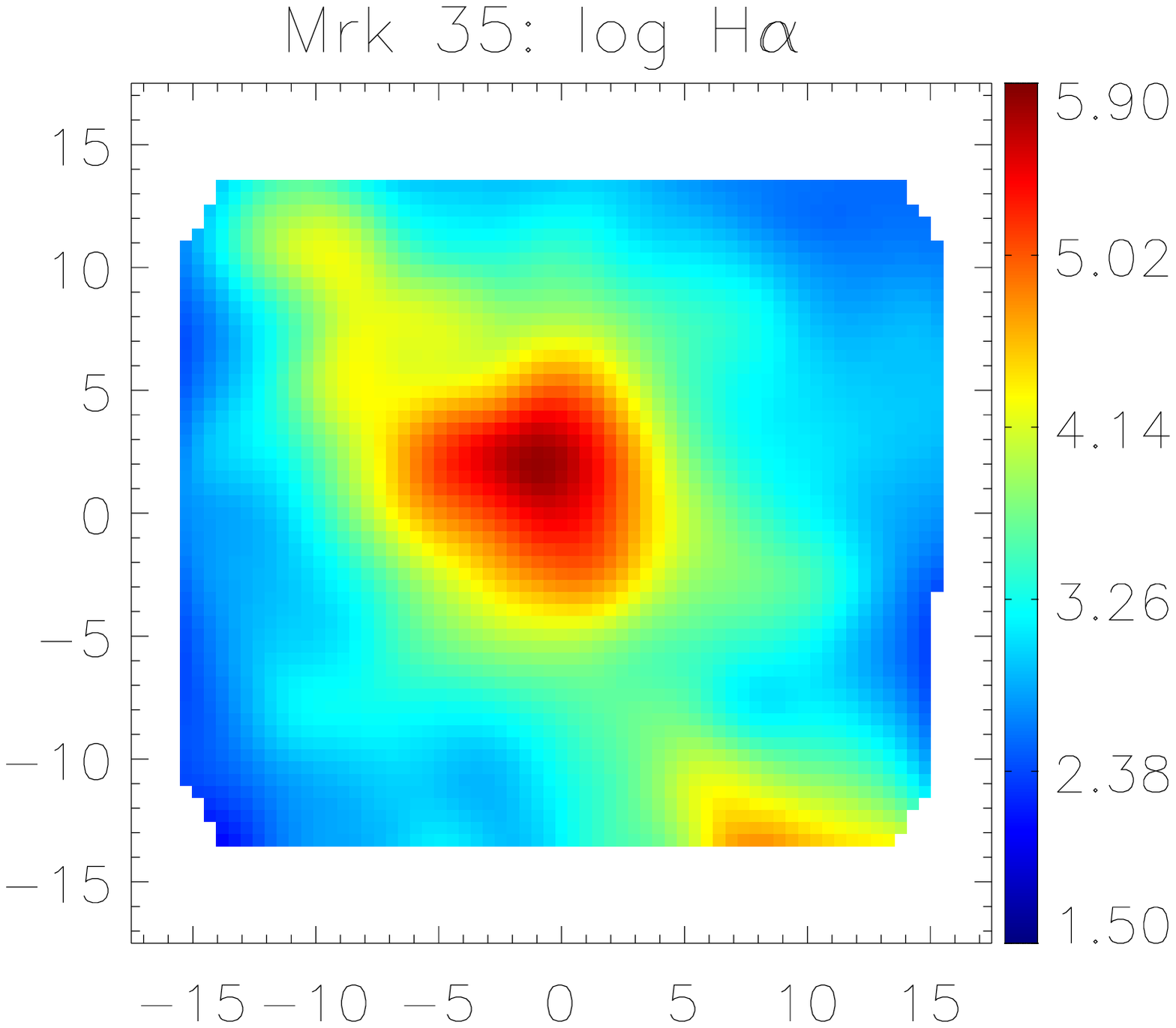}
\hspace*{4.5cm}
}}
\mbox{
\centerline{
\hspace*{0.0cm}\includegraphics[width=4.5cm]{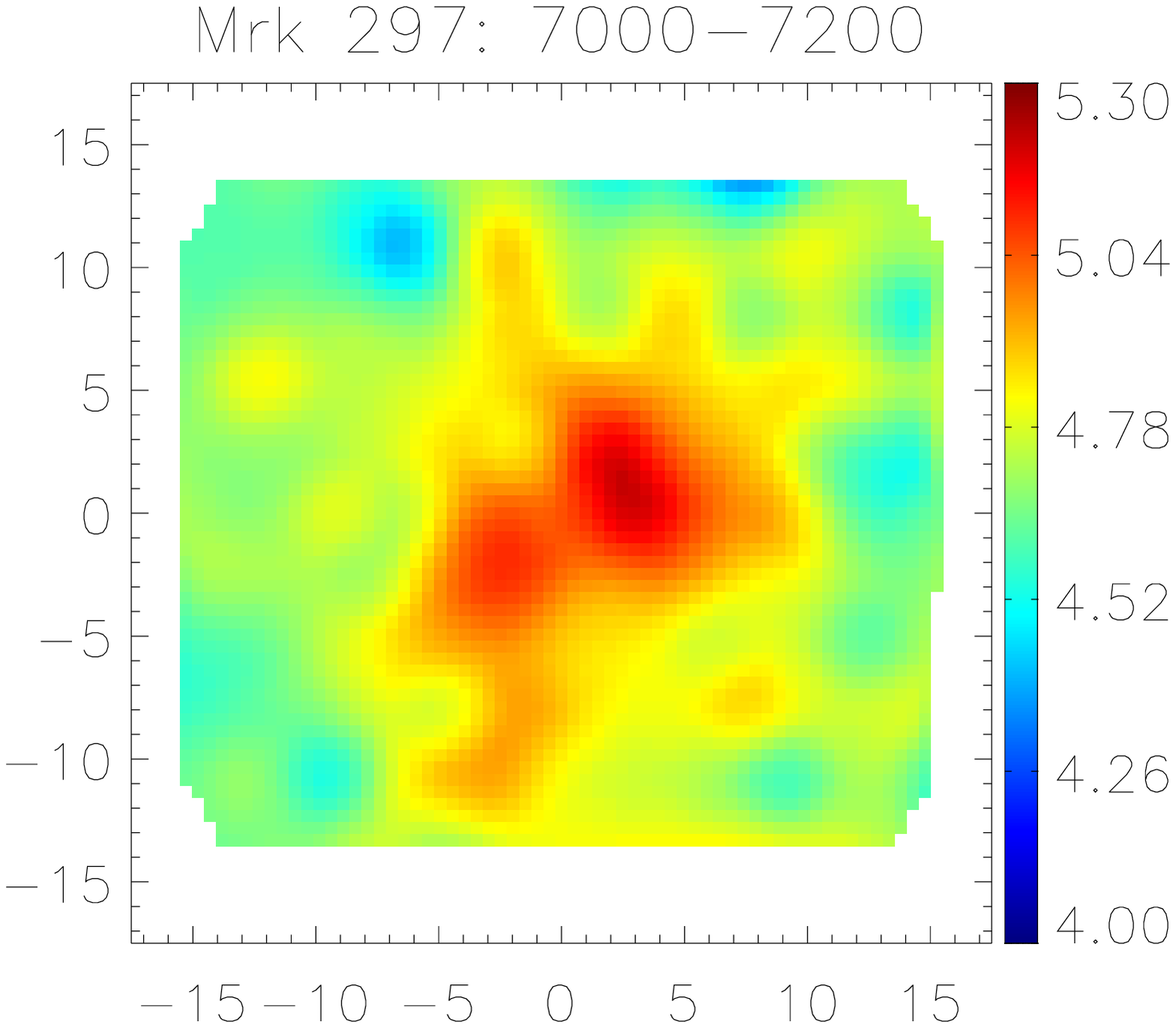}
\hspace*{0.0cm}\includegraphics[width=4.5cm]{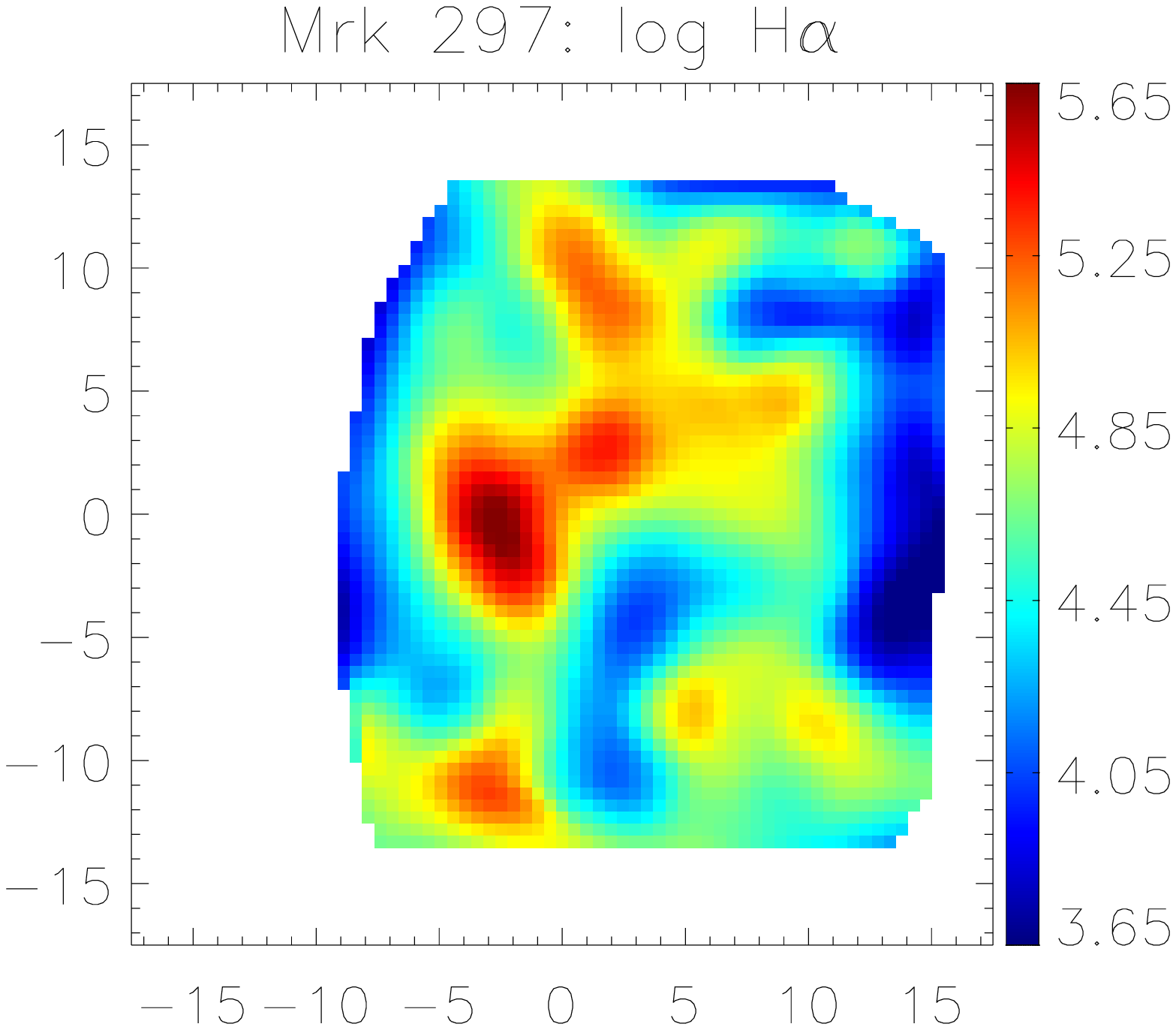}
\hspace*{0.0cm}\includegraphics[width=4.5cm]{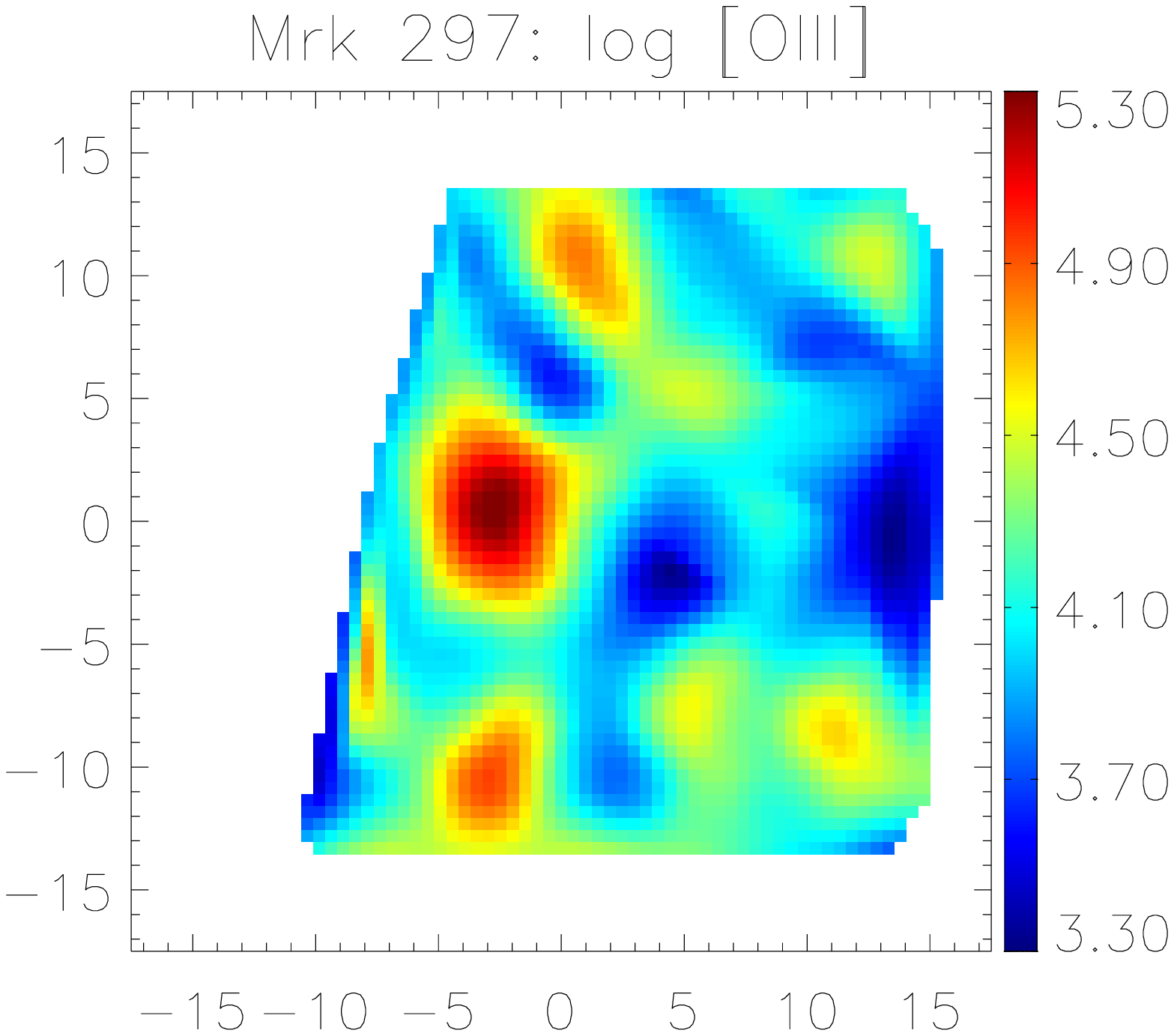}
}}
\mbox{
\centerline{
\hspace*{0.0cm}\includegraphics[width=4.5cm]{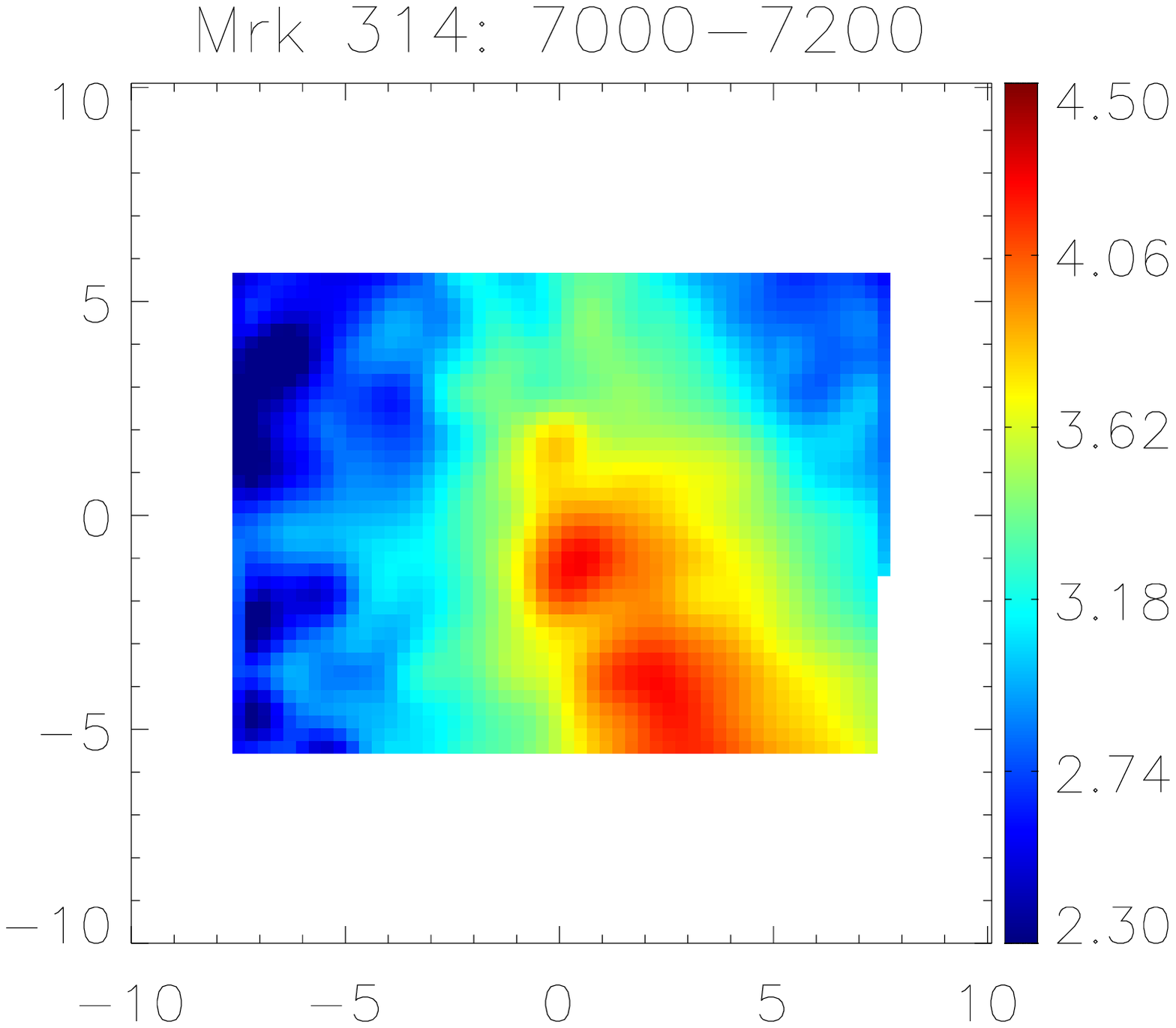}
\hspace*{0.0cm}\includegraphics[width=4.5cm]{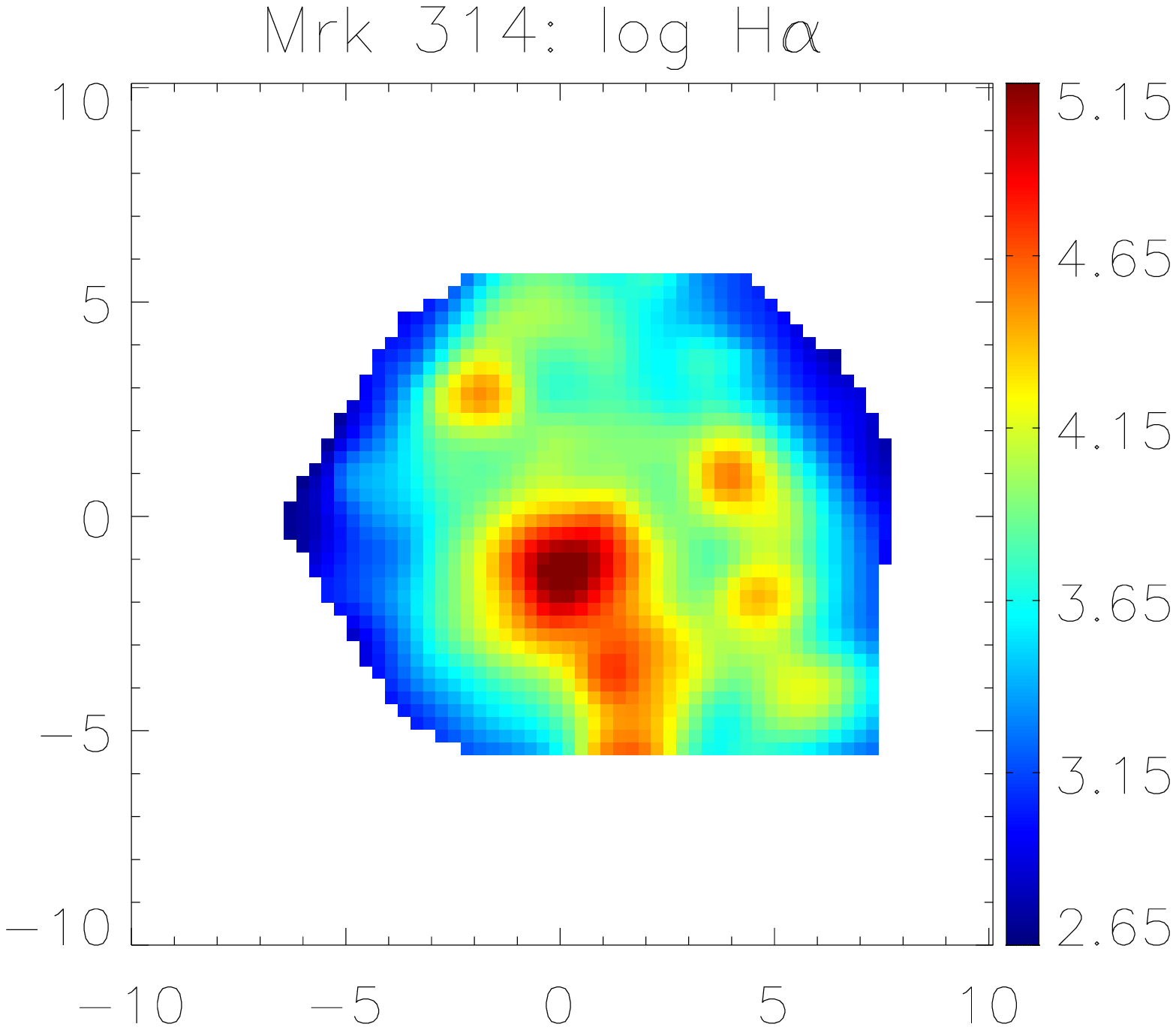}
\hspace*{0.0cm}\includegraphics[width=4.5cm]{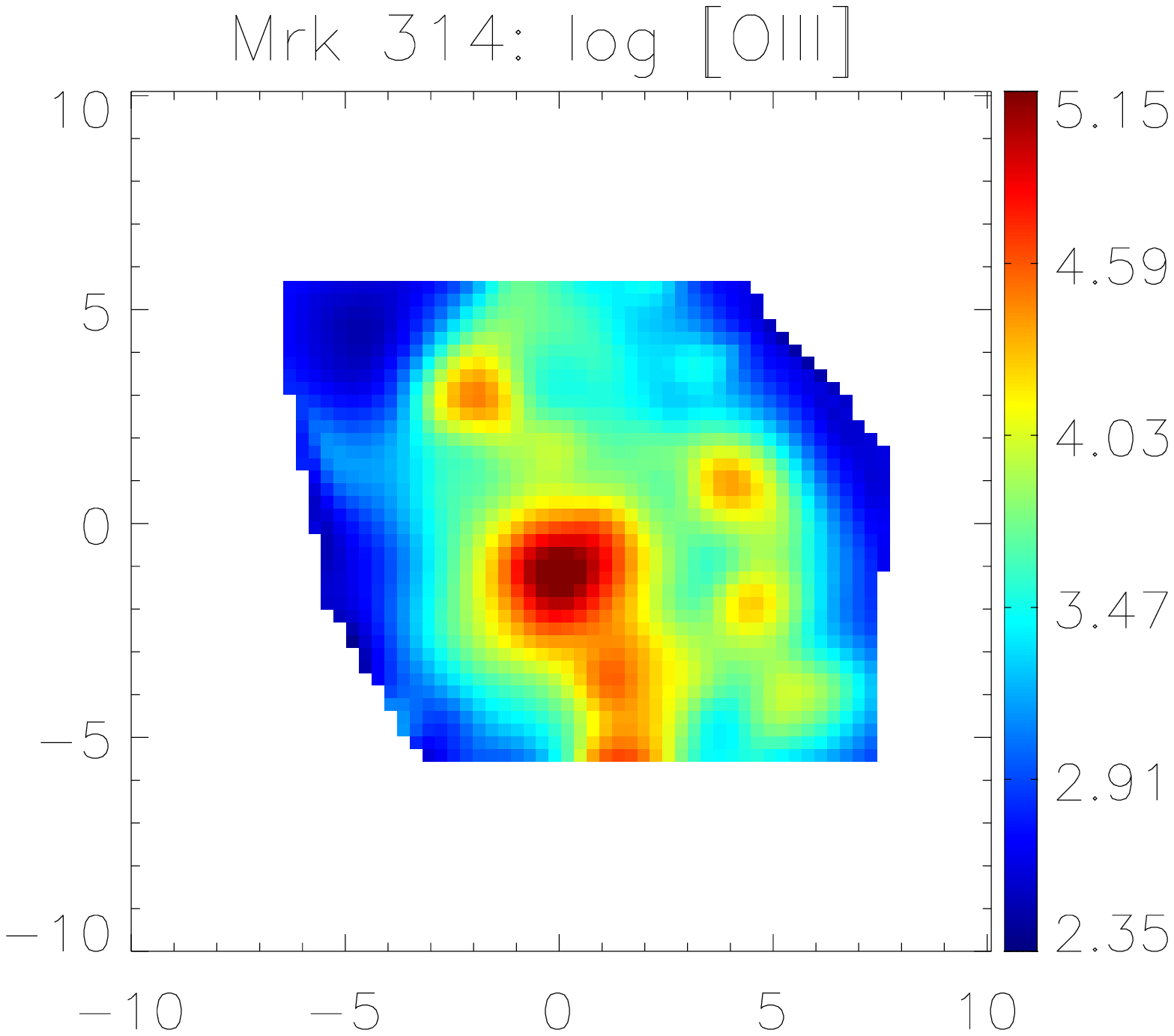}
}}
\mbox{
\centerline{
\hspace*{0.0cm}\includegraphics[width=4.5cm]{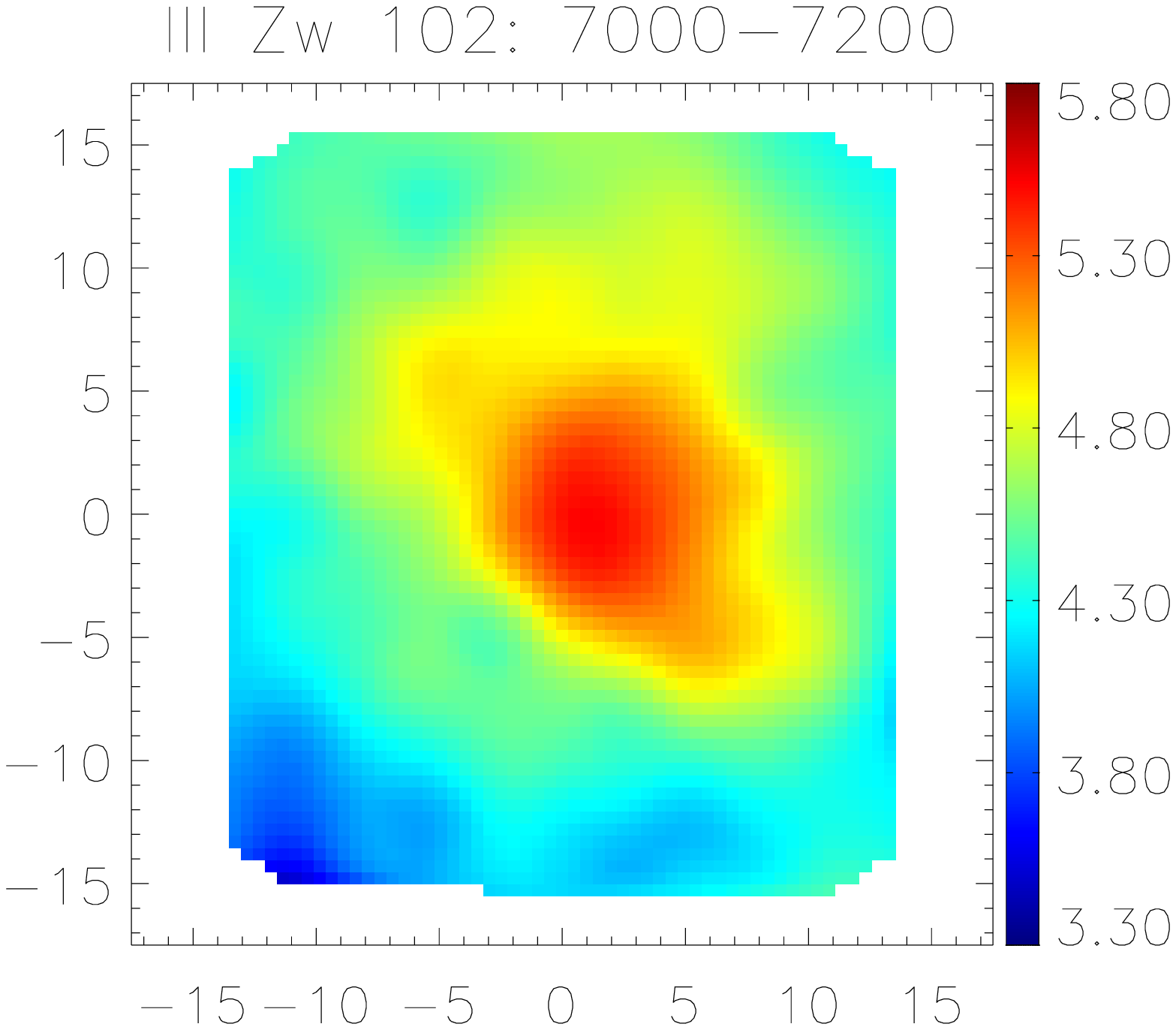}
\hspace*{0.0cm}\includegraphics[width=4.5cm]{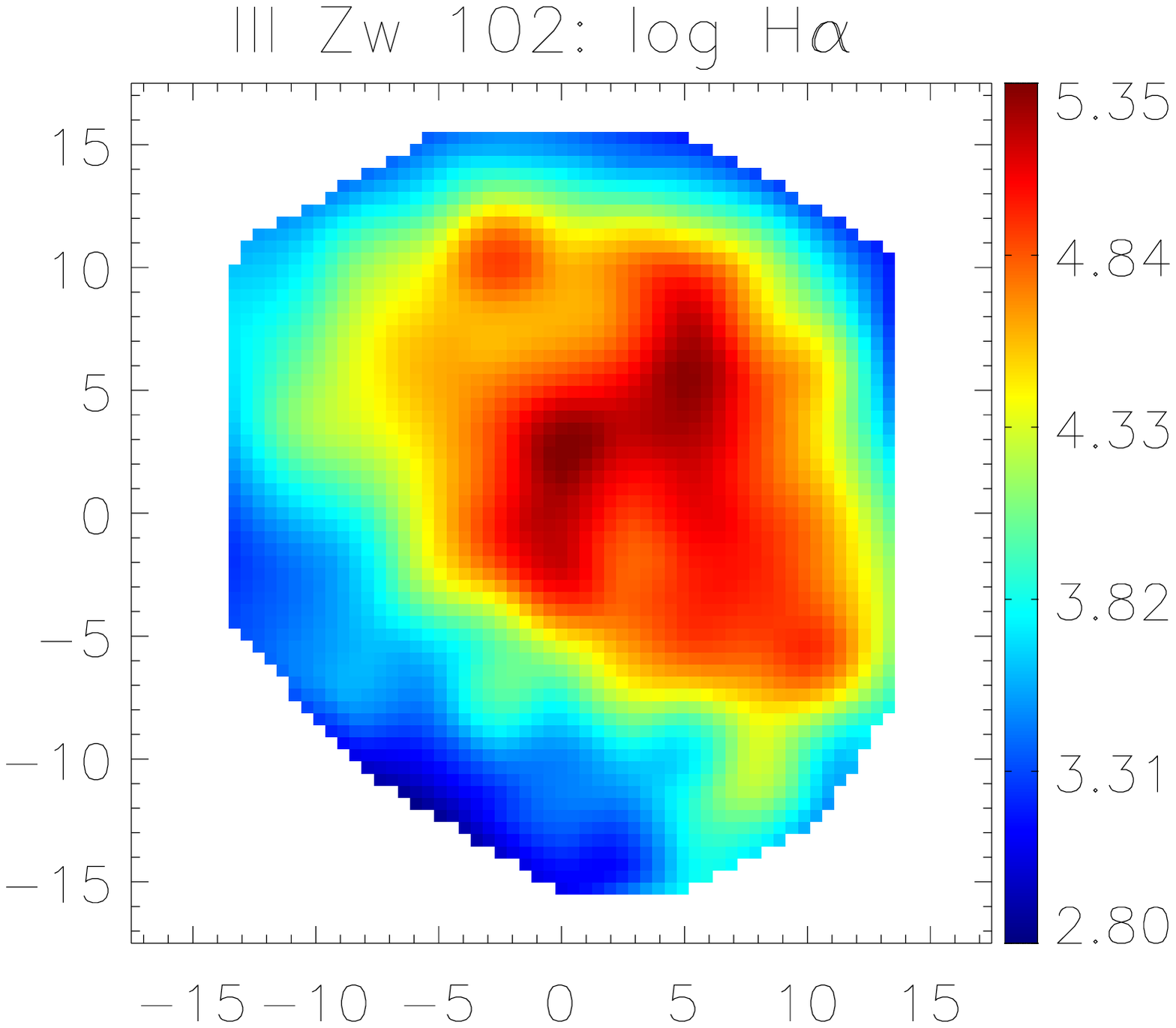}
\hspace*{0.0cm}\includegraphics[width=4.5cm]{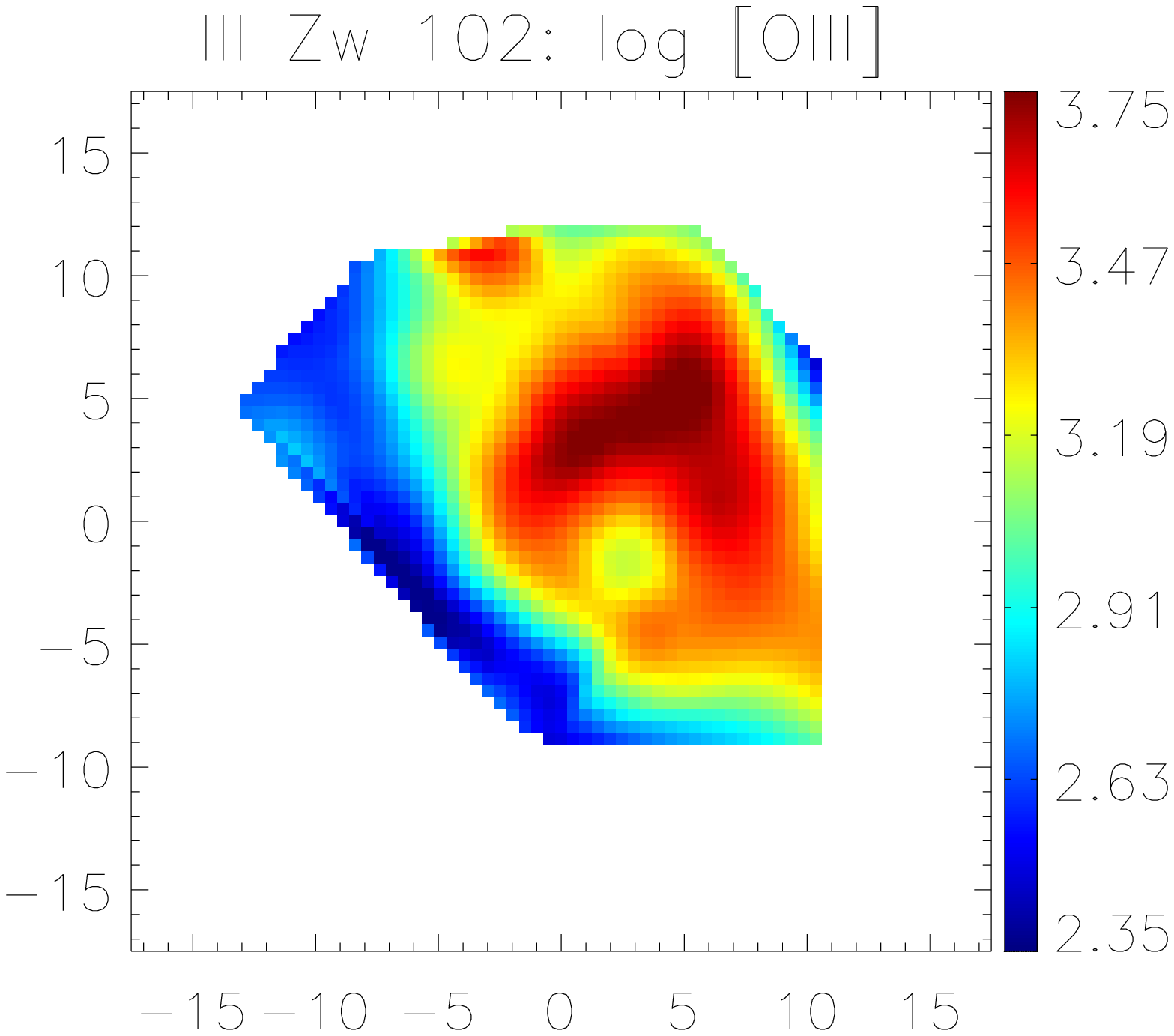}
}}
\caption{Two-dimensional distribution of the continuum emission derived 
by integrating the signal in the indicated spectral ranges free from emission 
line contribution (``pure continua''), and integrated line intensity maps  
derived by a Gaussian fit. Maps are shown on a logarithmic scale to 
bring up the fainter regions. Axis units are arcseconds; north is up, east to 
the left.}
\label{Fig:mapascontinuo}
\end{figure*}

Figure~\ref{Fig:mapascontinuo} shows these continuum maps  in the spectral
interval 7000--7200 \AA\ for all galaxies except Mrk~35, for which we display
the continuum map in the interval 5600--5800 \AA. The location of the peak of
the ``pure continuum '' maps does not depend on what specific "gas-free"
spectral range is used; therefore, we can safely define the continuum peaks as
the ``optical nucleus'' of the galaxies. Mrk~370, Mrk~297 and Mrk~314 have a
second continuum maximum, which could be a secondary nucleus (see the
discussion in Section~\ref{Sect:individual}).


\Ha\ and [\ion{O}{3}]~$\lambda5007$ line intensity maps are also displayed in
Figure~\ref{Fig:mapascontinuo}. For all galaxies, both \Ha\ and
[\ion{O}{3}]~$\lambda5007$ maps have a similar morphology, as expected in
objects ionized by stars, significantly different from the continuum maps. The
emission line maps show several peaks distributed across the whole field of
view, whereas the continuum distribution is more regular. Only in Mrk~314 the
peak in the emission line maps spatially coincides with the peak in the
continuum. In Mrk~297 and III~Zw~102, the current star-formation activity is
spread all over the mapped region.

\subsubsection {Line ratio maps}
\label{Sect:lineratiomaps}

\begin{figure*}                         
\mbox{
\centerline{
\hspace*{0.0cm}\includegraphics[width=4.5cm]{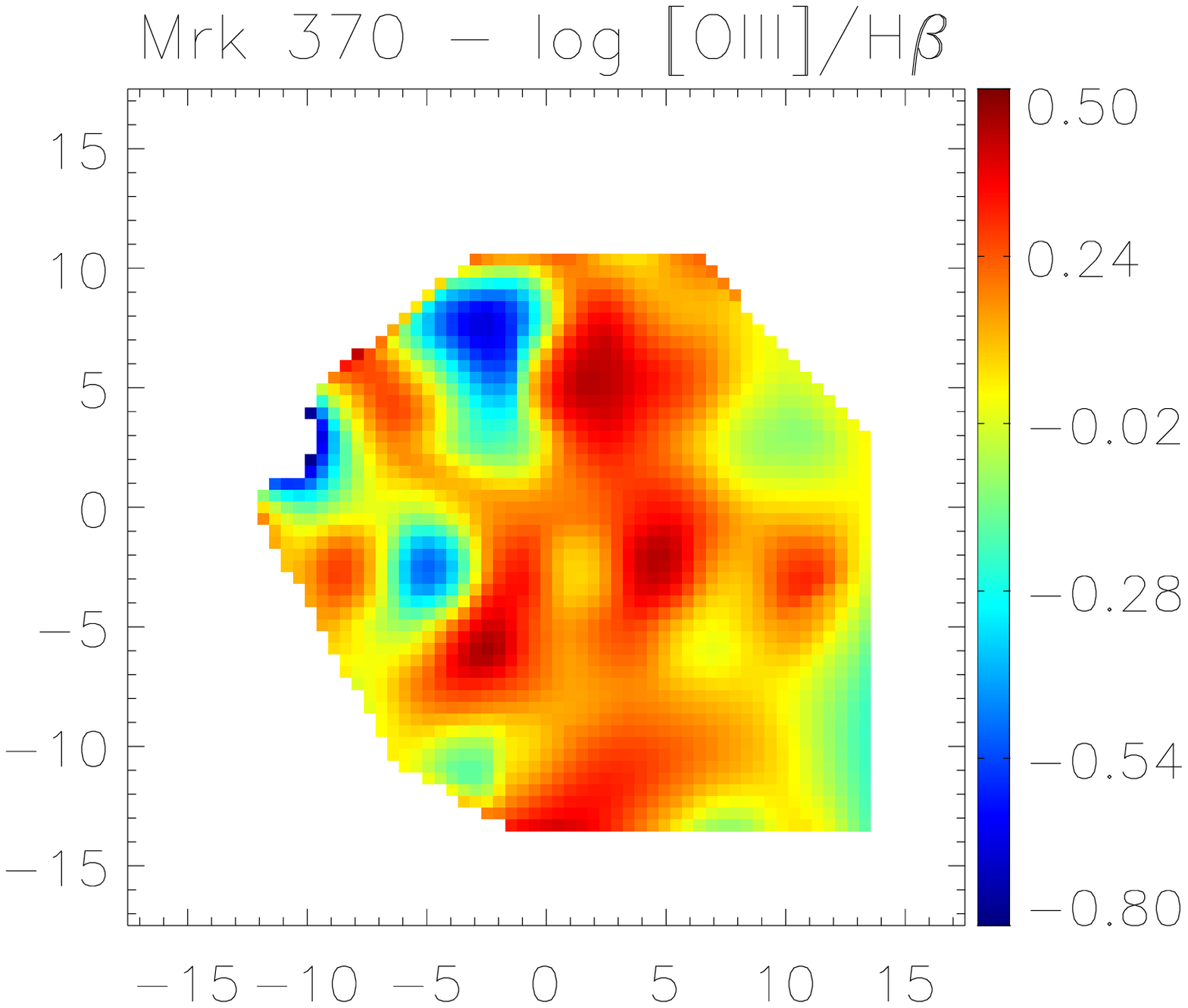}
\hspace*{0.0cm}\includegraphics[width=4.5cm]{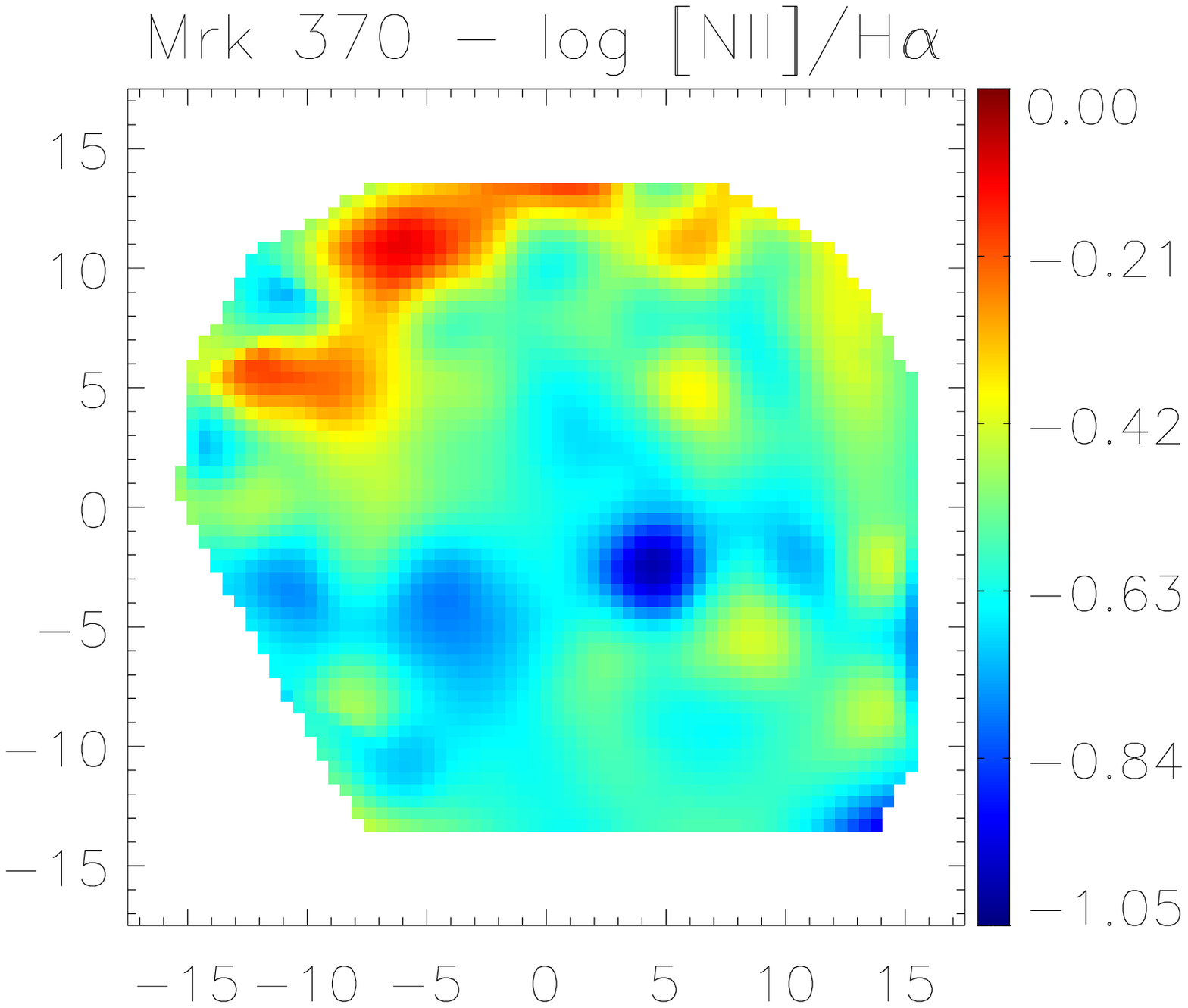}
\hspace*{0.0cm}\includegraphics[width=4.5cm]{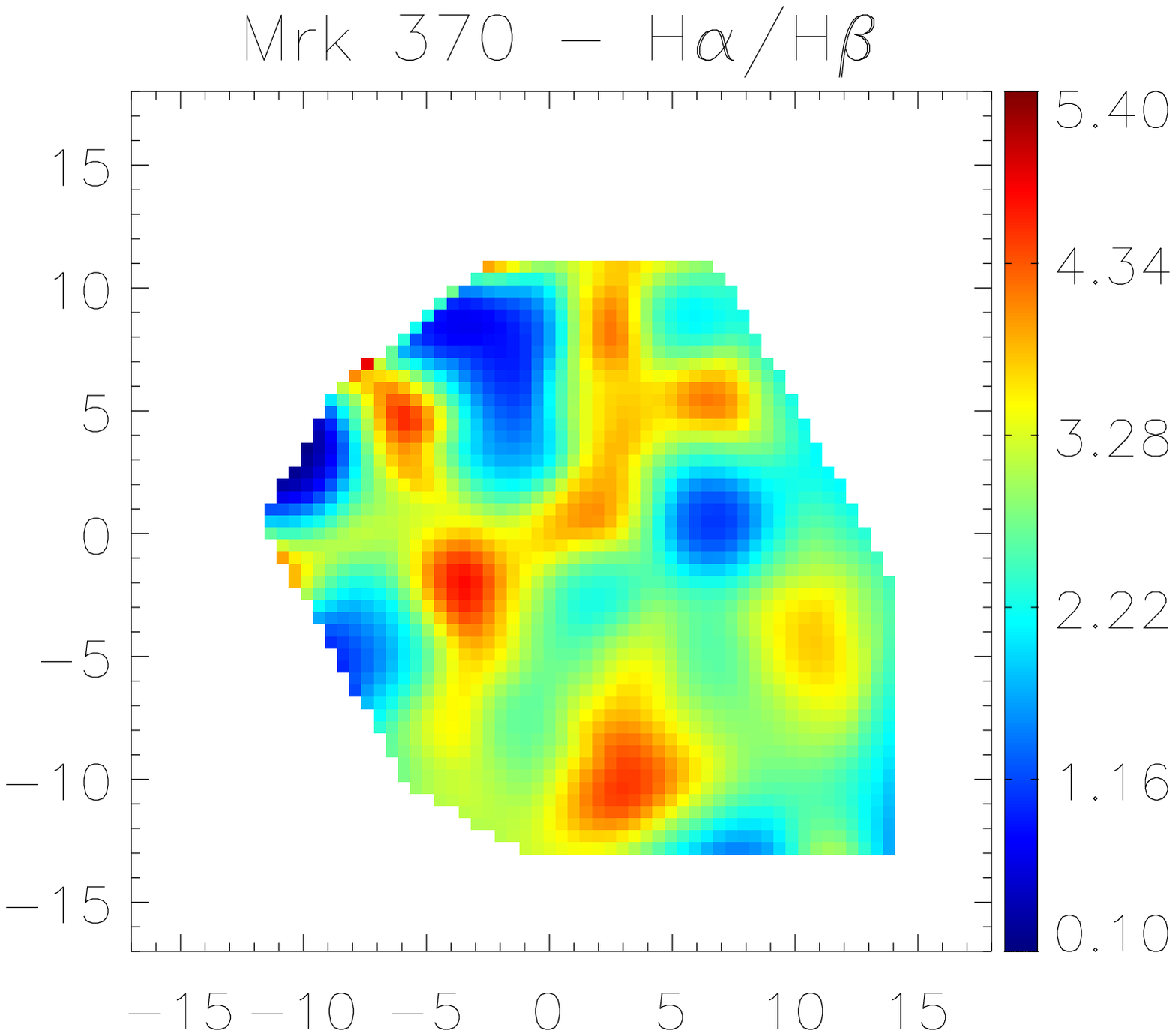}
}}
\mbox{
\centerline{
\hspace*{0.0cm}\includegraphics[width=4.5cm]{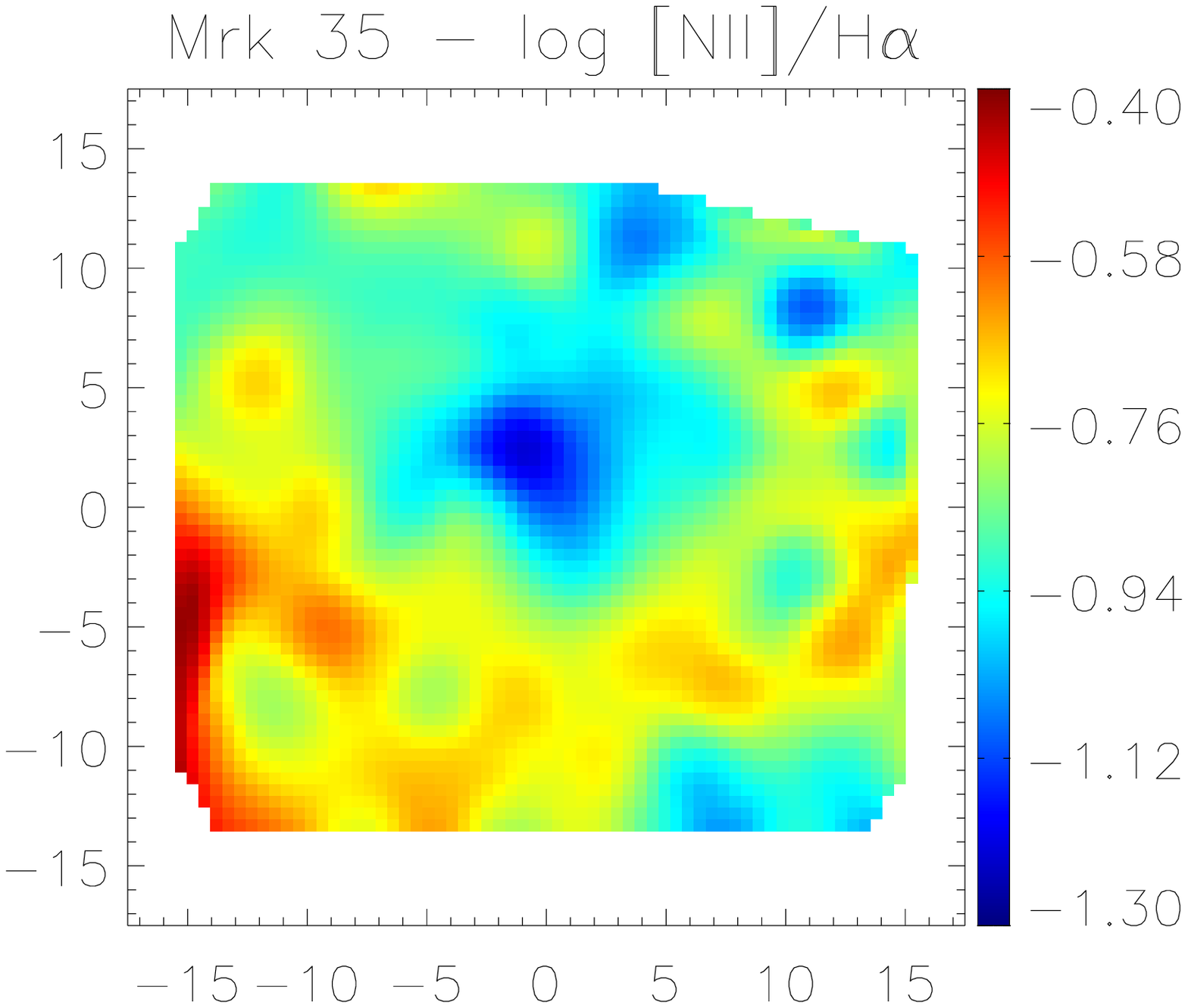}
}}
\mbox{
\centerline{
\hspace*{0.0cm}\includegraphics[width=4.5cm]{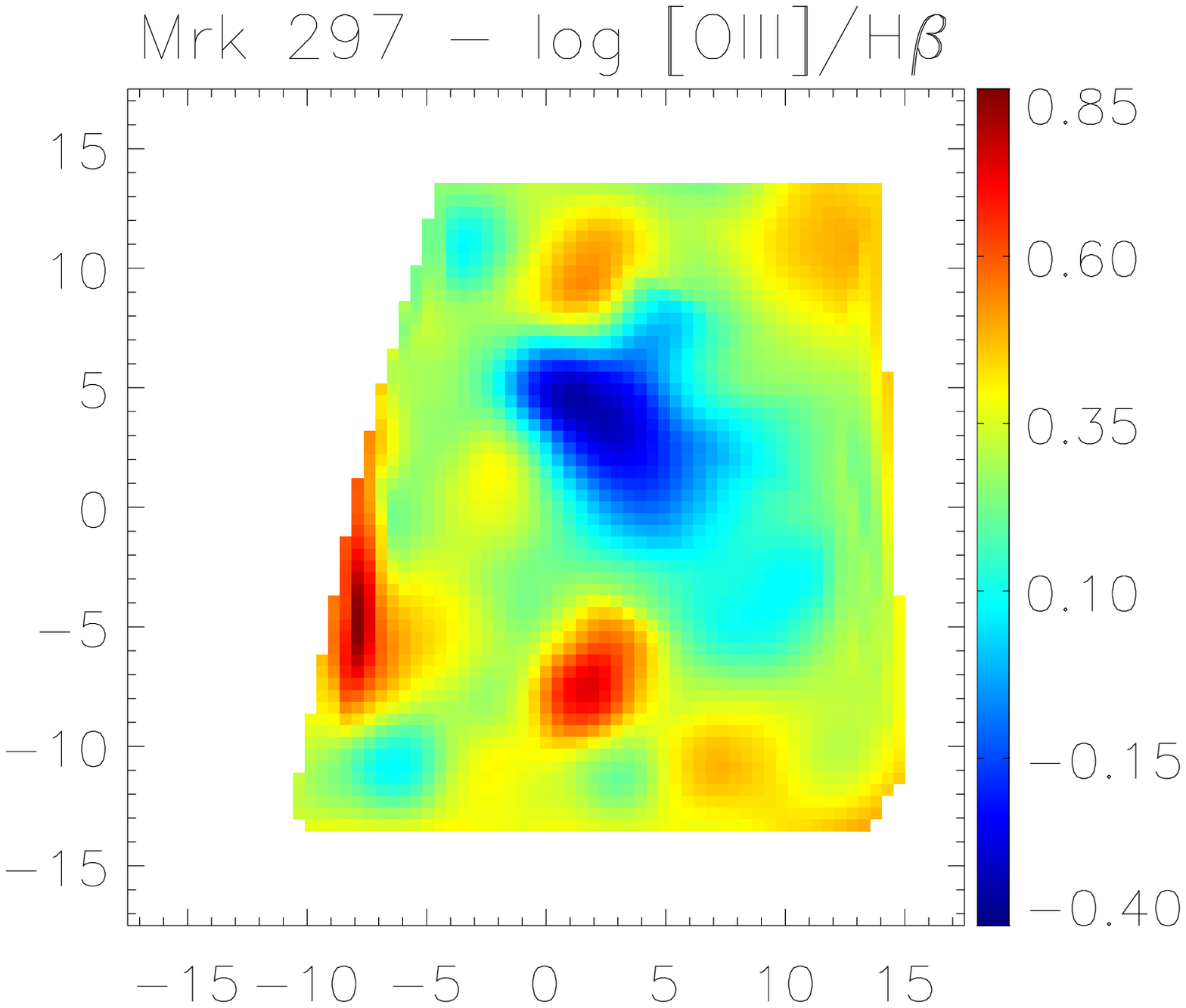}
\hspace*{0.0cm}\includegraphics[width=4.5cm]{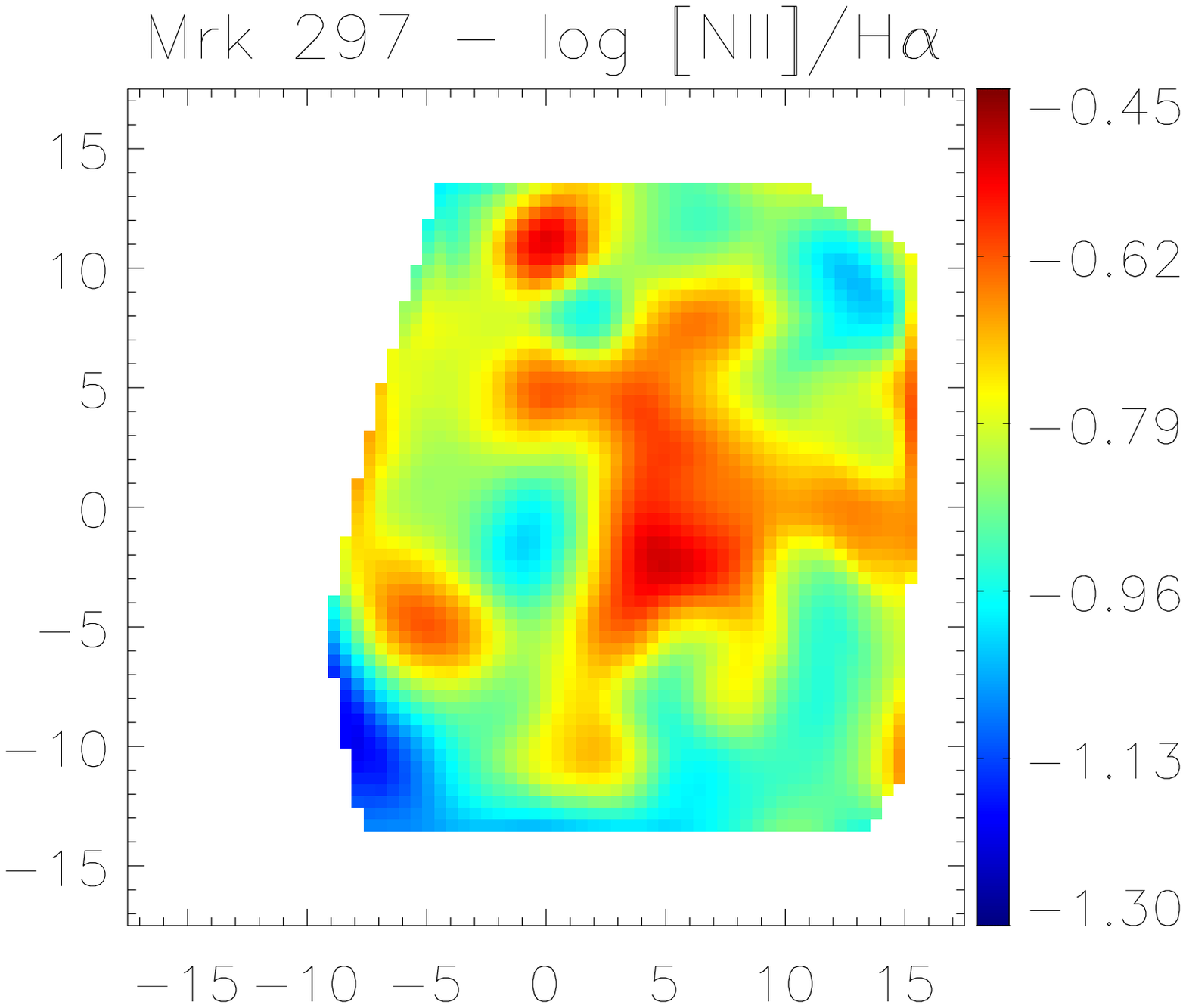}
\hspace*{0.0cm}\includegraphics[width=4.5cm]{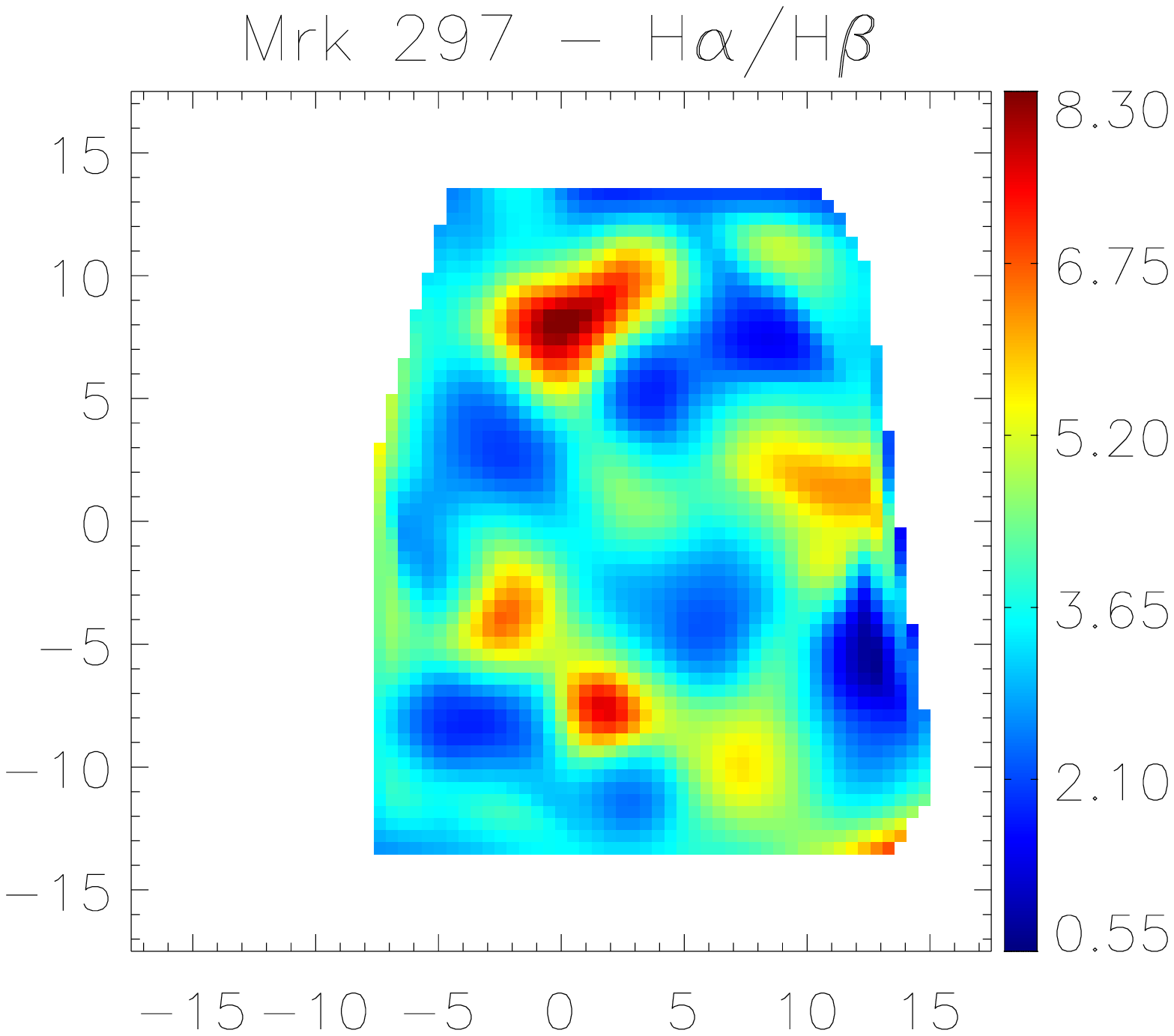}
}}
\mbox{
\centerline{
\hspace*{0.0cm}\includegraphics[width=4.5cm]{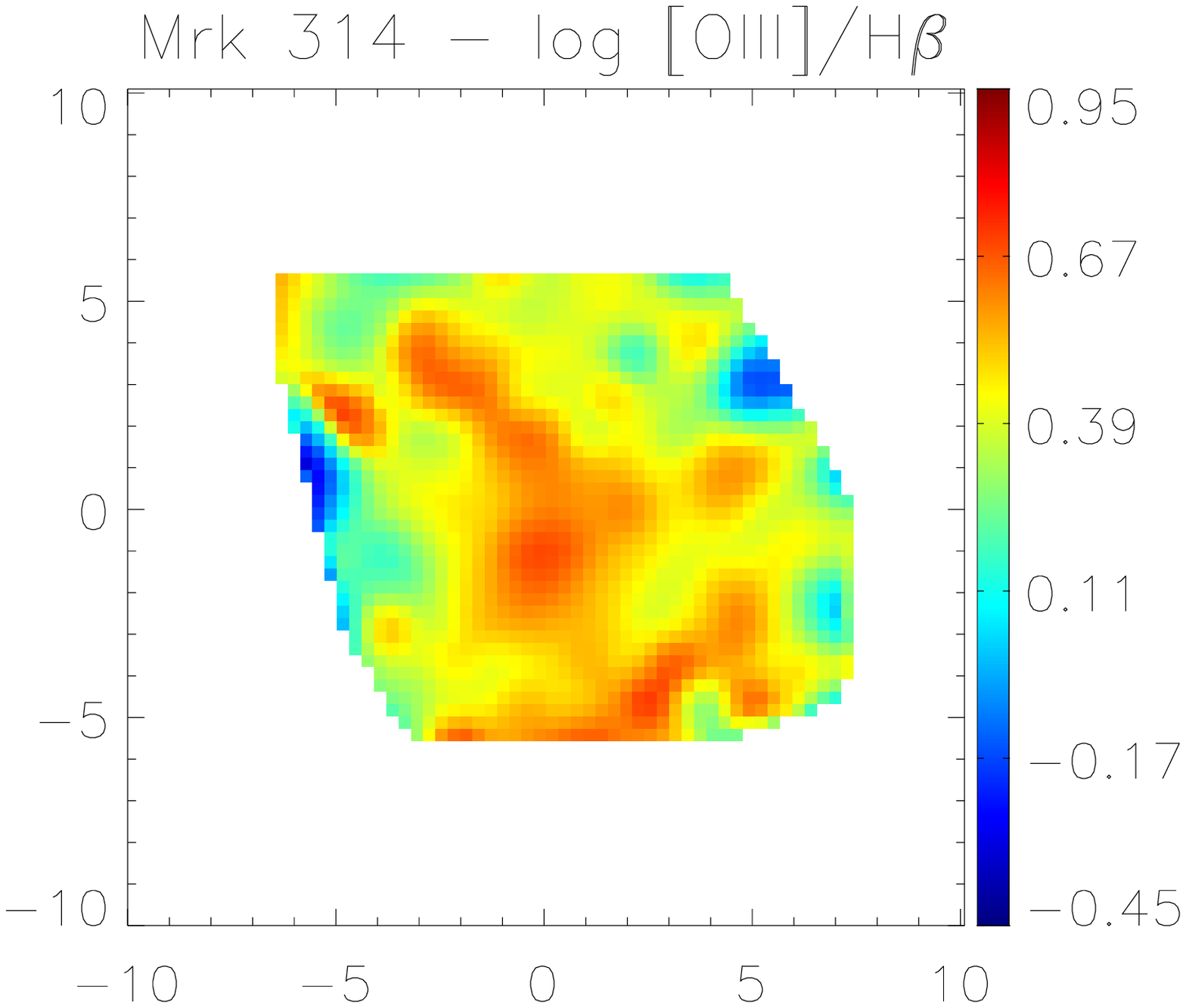}
\hspace*{0.0cm}\includegraphics[width=4.5cm]{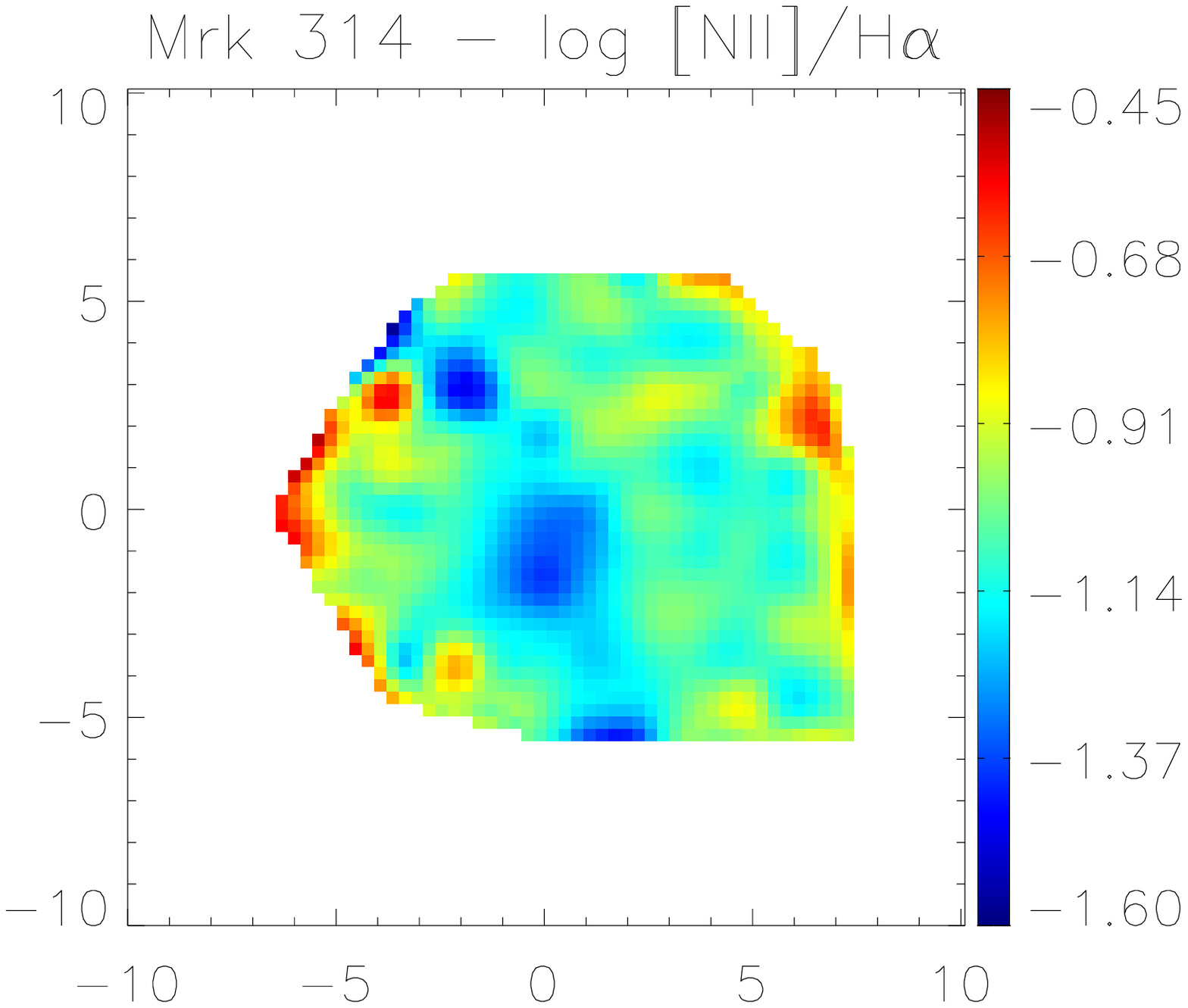}
\hspace*{0.0cm}\includegraphics[width=4.5cm]{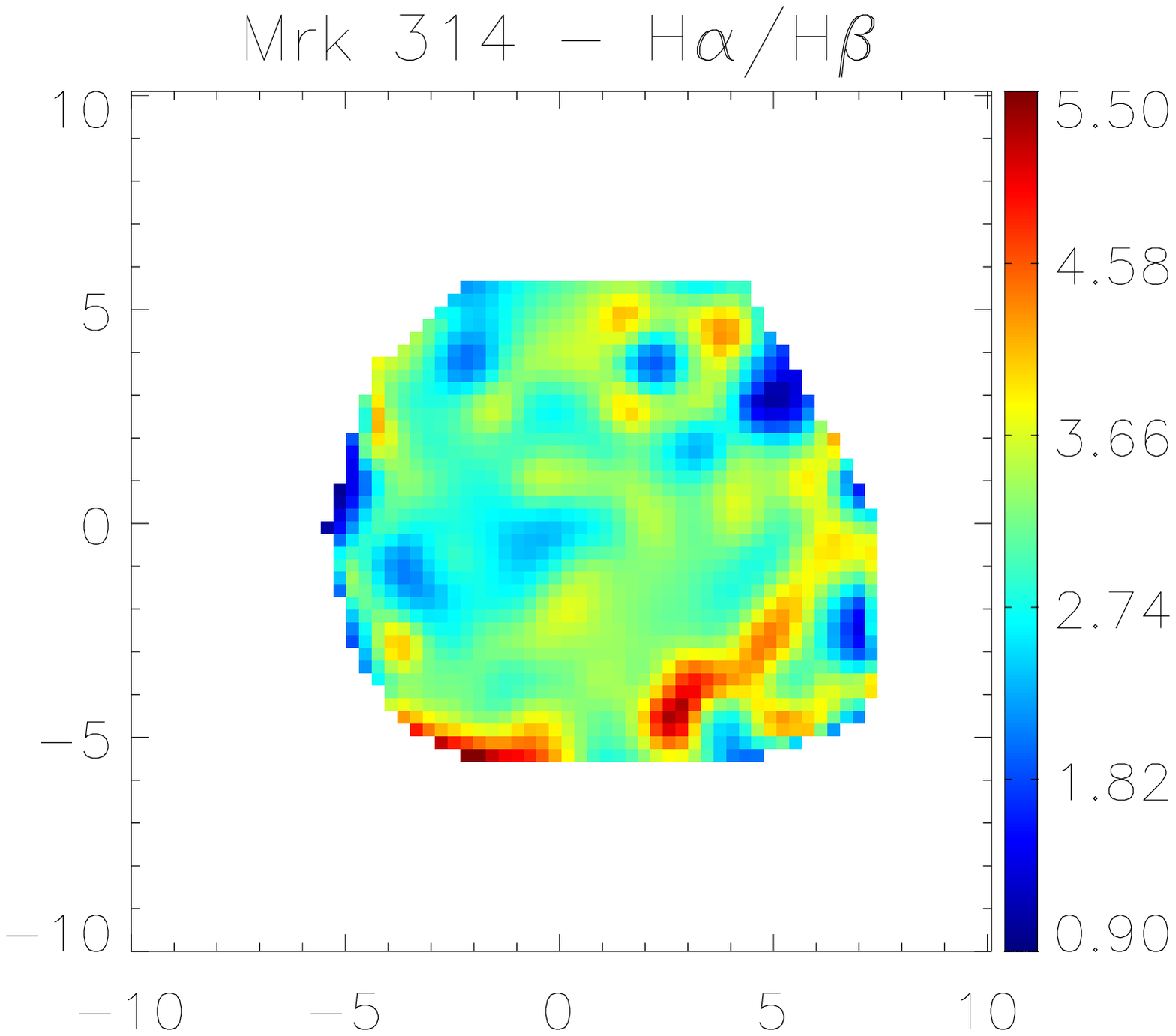}
}}
\mbox{
\centerline{
\hspace*{0.0cm}\includegraphics[width=4.5cm]{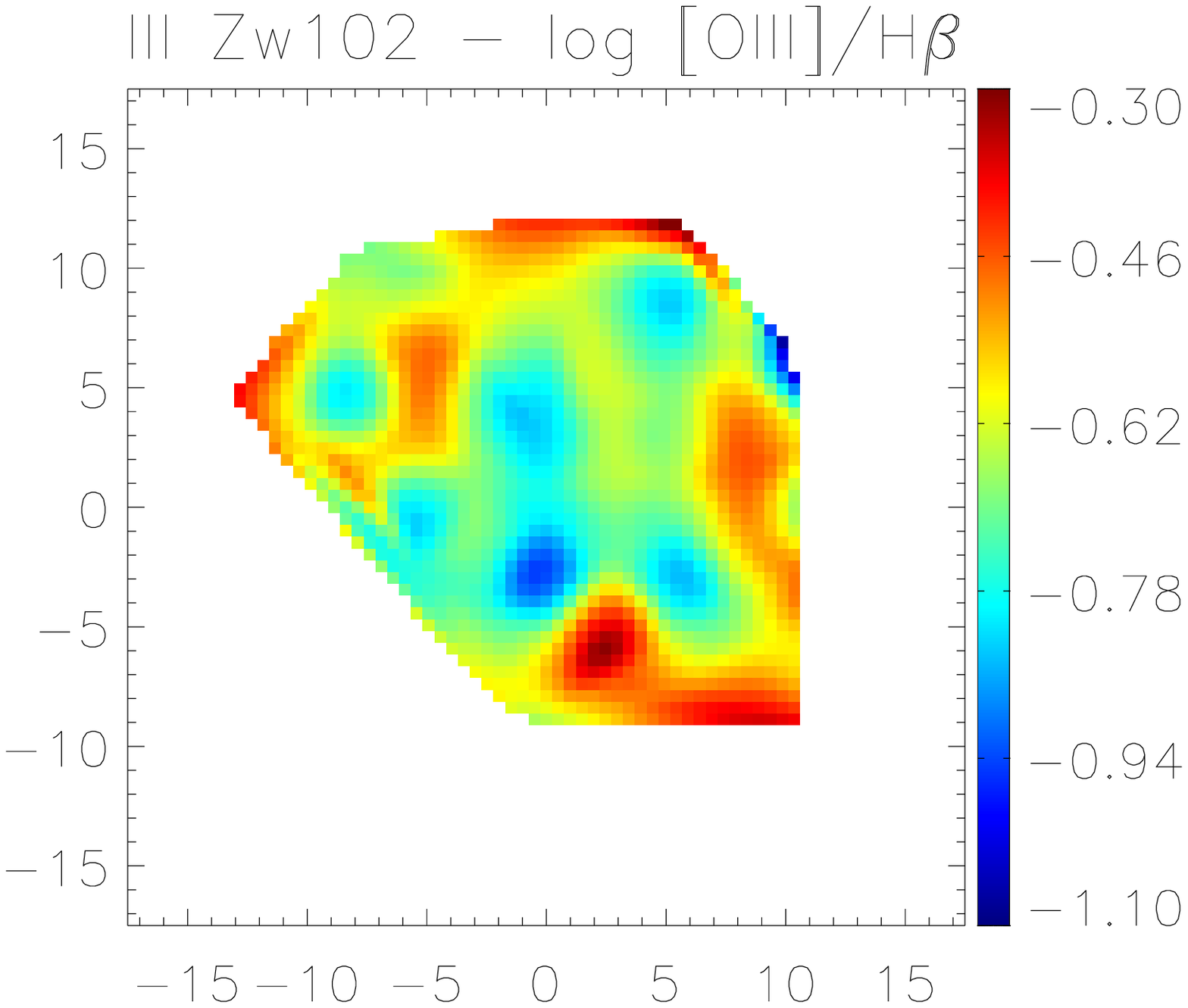}
\hspace*{0.0cm}\includegraphics[width=4.5cm]{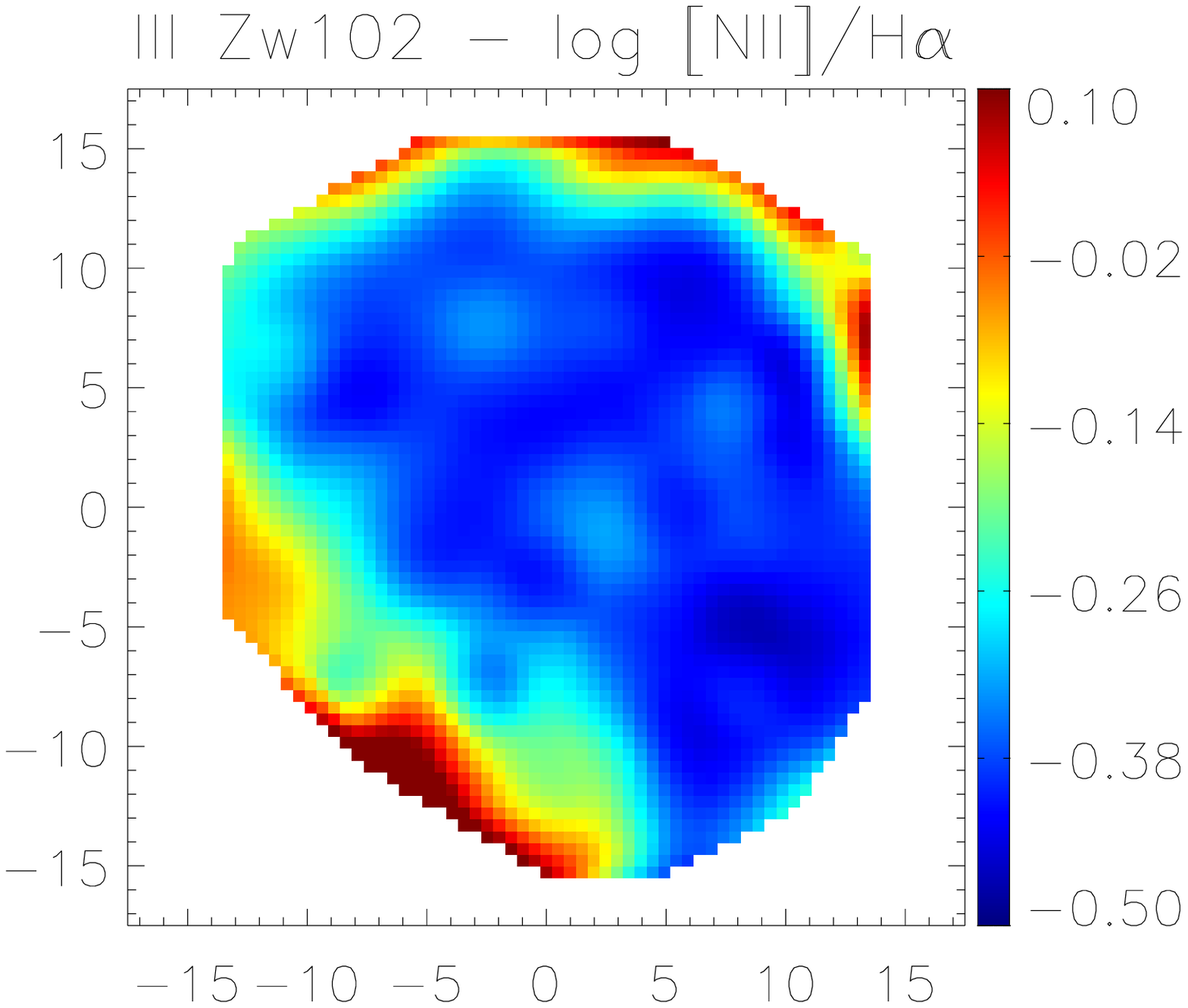}
\hspace*{0.0cm}\includegraphics[width=4.5cm]{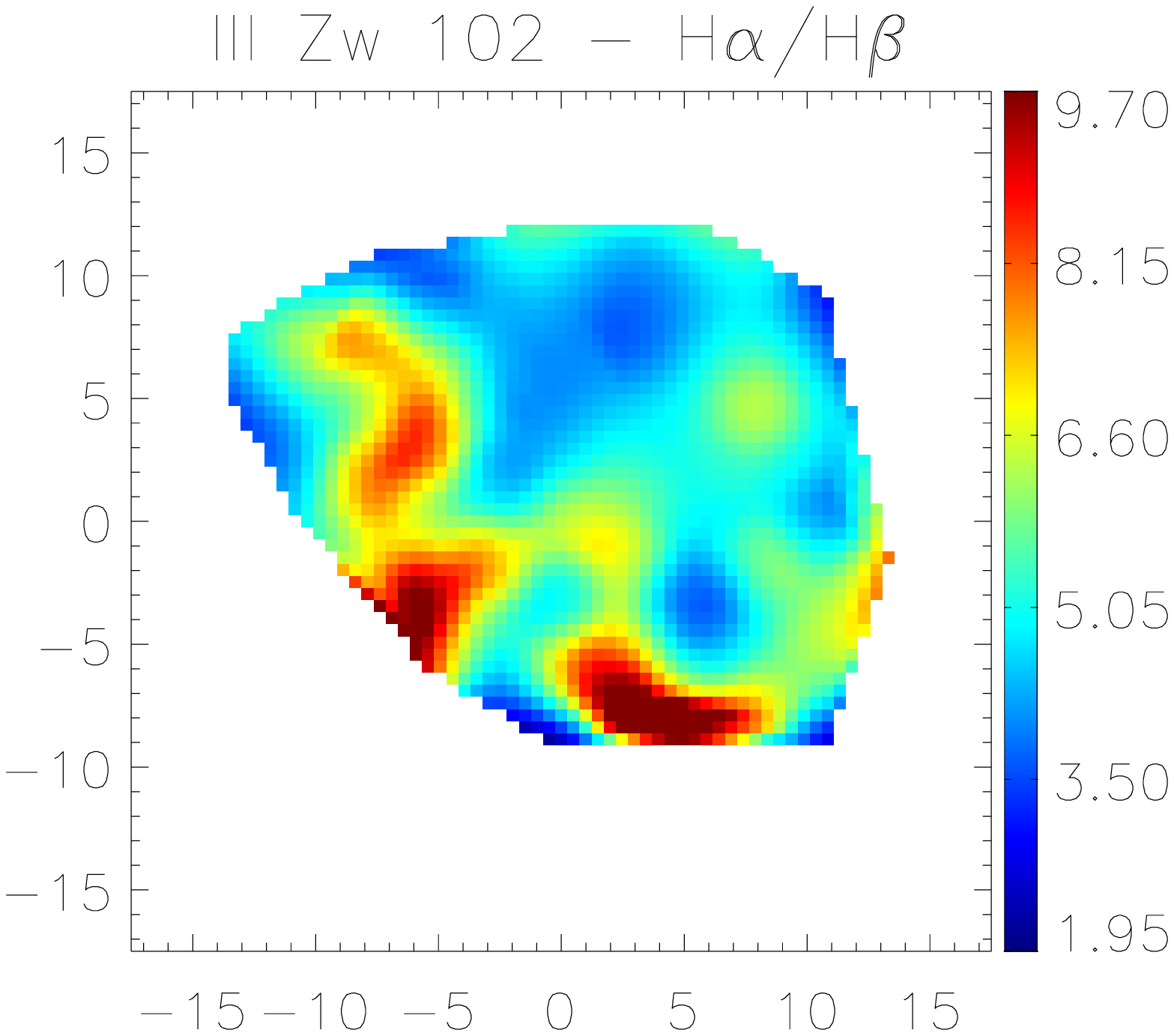}
}}
\caption{Ionization ratios [\ion{O}{3}]/\Hb\ and
[\ion{N}{2}]~$\lambda6584$/\Ha, and extinction maps (\Ha/\Hb) for the observed 
galaxies.}
\label{Fig:ratios}
\end{figure*}

Maps of the [\ion{O}{3}]~$\lambda5007$/\Hb, [\ion{N}{2}]~$\lambda6584$/\Ha, and
\Ha/\Hb\ ratios are presented in Figure~\ref{Fig:ratios}. 

The [\ion{O}{3}]~$\lambda5007$/\Hb\ ratio (Figure~\ref{Fig:ratios}, column~1),
commonly used as an indicator of the excitation degree \citep{McCall85}, shows
a complex pattern in all the objects; the highest
[\ion{O}{3}]~$\lambda5007$/\Hb\ values ($\geq 3$) usually trace the line
emission peaks, except in Mrk~297. The five galaxies have the excitation
levels expected for \ion{H}{2}-like regions in the whole field of view.

The morphology of the [\ion{N}{2}]~$\lambda6584$/\Ha\ maps is similar to the
[\ion{O}{3}]~$\lambda5007$/\Hb\ maps but, as expected, it follows the opposite
trend: the lower [\ion{N}{2}]~$\lambda6584$/\Ha\ values are reached in the
center of the SF knots (where the \Ha\ intensity peaks), while higher ratios
are found in their surroundings. 
(This behavior towards higher excitation degree when the intensity is larger
can be related to the decreasing distance from the young ionizing cluster.)

The outer regions of III~Zw~102 have [\ion{N}{2}]~$\lambda6584$/\Ha\ values
larger than those expected in \ion{H}{2}-regions
([\ion{N}{2}]~$\lambda6584$/\Ha\ $>0.6$), which may suggest the presence
of shock-like excitation mechanisms in the circum-knots environment.


The \Ha/\Hb\ maps all show a highly inhomogeneous structure, with strong
variations of the extinction across the galaxy. Just assuming a single,
spatially constant value for the extinction (for instance obtained from
long-slit spectra and thus dominated by the nuclear/brighter knot emission)
can lead to large errors in the derivation of the magnitudes and fluxes of the
knots, and hence in the determination of the star-formation rate and the
properties of the stellar populations (see also \citealt{Kehrig07}).

\subsection{Kinematics of the Ionized Gas}
\label{Sect:gaskinematics}

The central wavelength of the Gaussians fit to the lines in the individual
spectra gives us the radial velocity of the ionized gas at each observed
position. (The spectral resolution of our data, about 9.5 \AA\ FWHM, prevents
us from deriving reliable velocity dispersions).

\begin{figure*} 
\mbox{ 
\centerline{
\hspace*{0.0cm}\includegraphics[width=4.5cm]{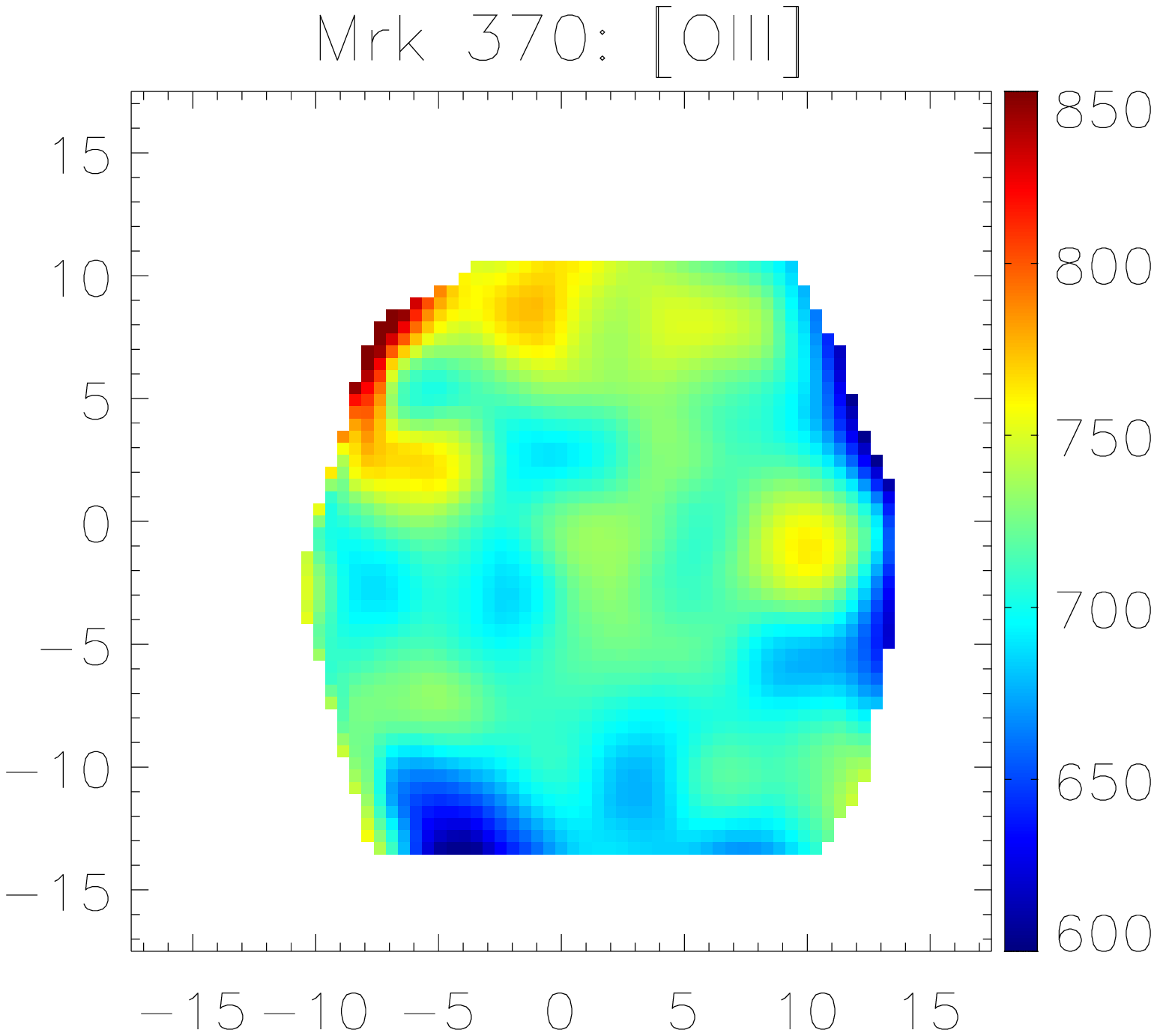}
\hspace*{0.0cm}\includegraphics[width=4.5cm]{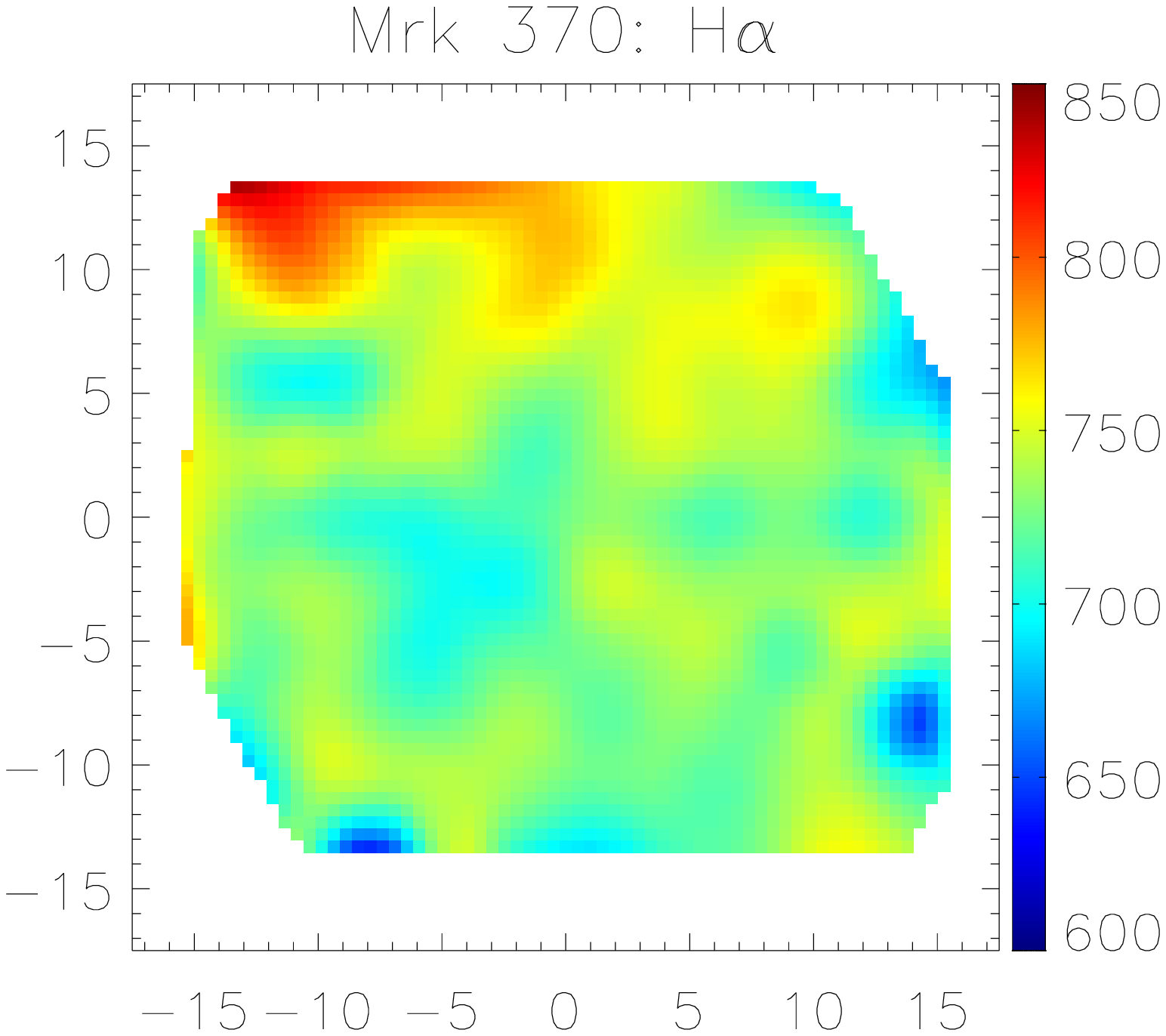} 
}}    
\mbox{
\centerline{ 
\hspace*{0.0cm}\includegraphics[width=4.5cm]{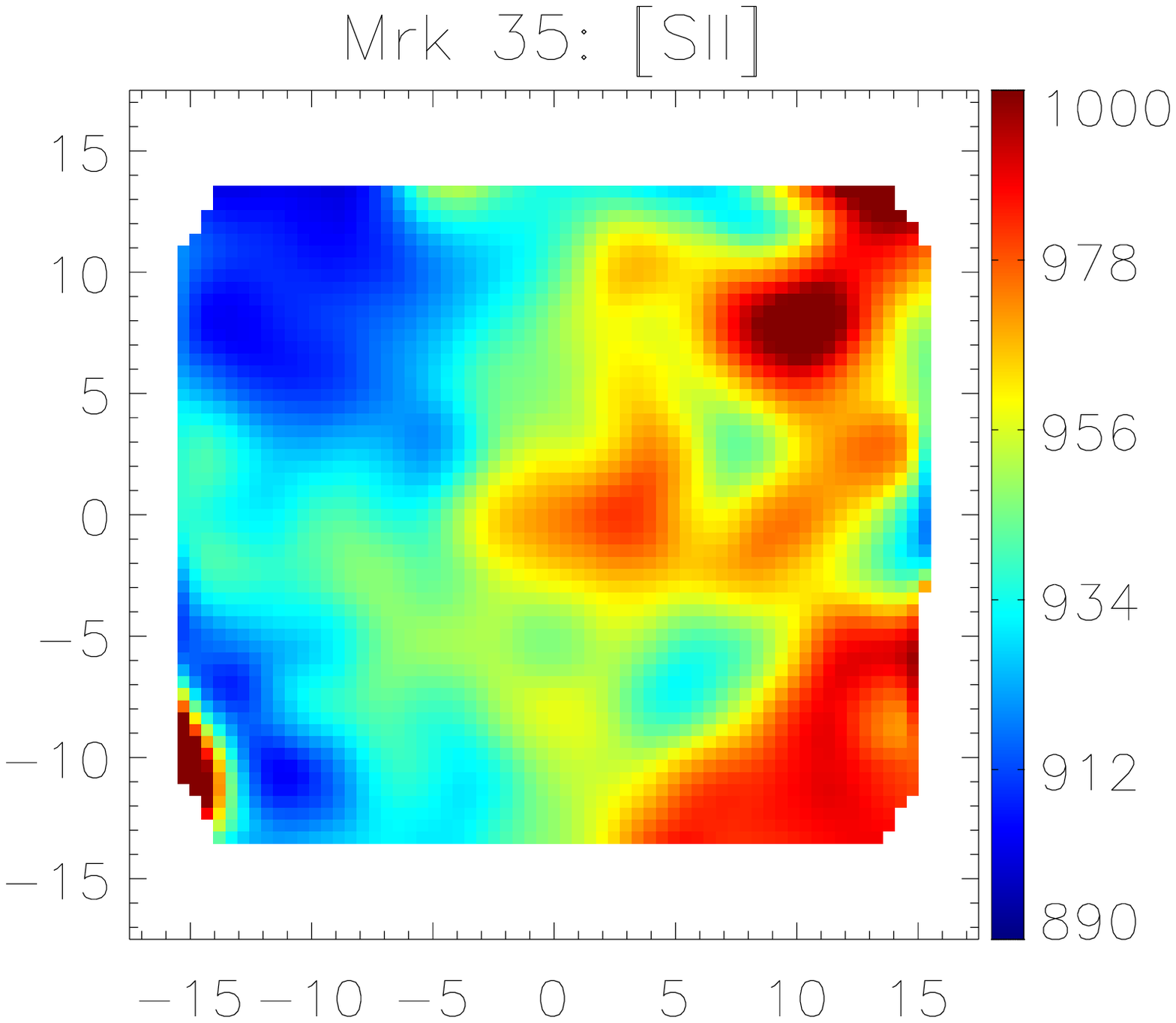}
\hspace*{0.0cm}\includegraphics[width=4.5cm]{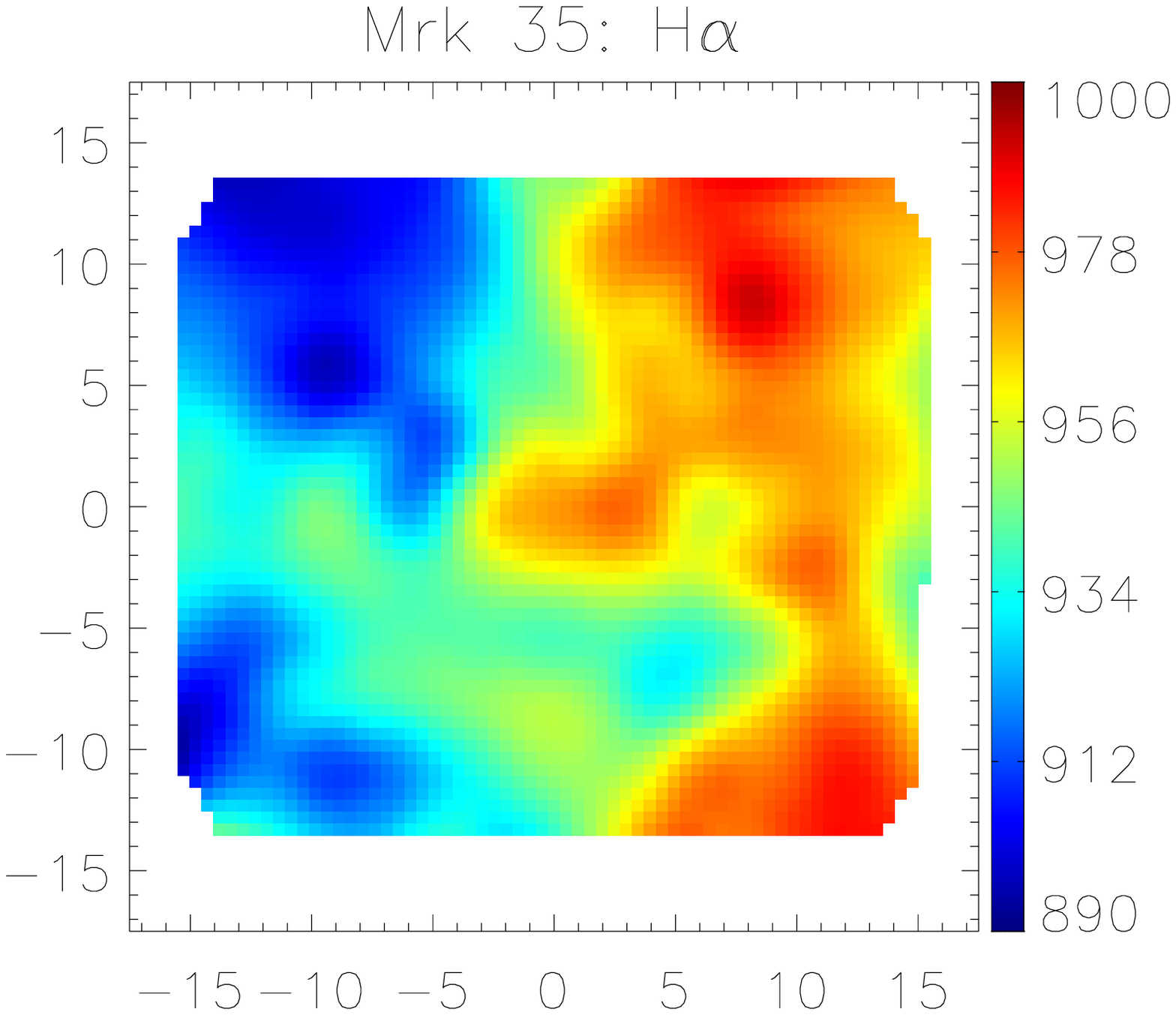} 
}}    
\mbox{
\centerline{ 
\hspace*{0.0cm}\includegraphics[width=4.5cm]{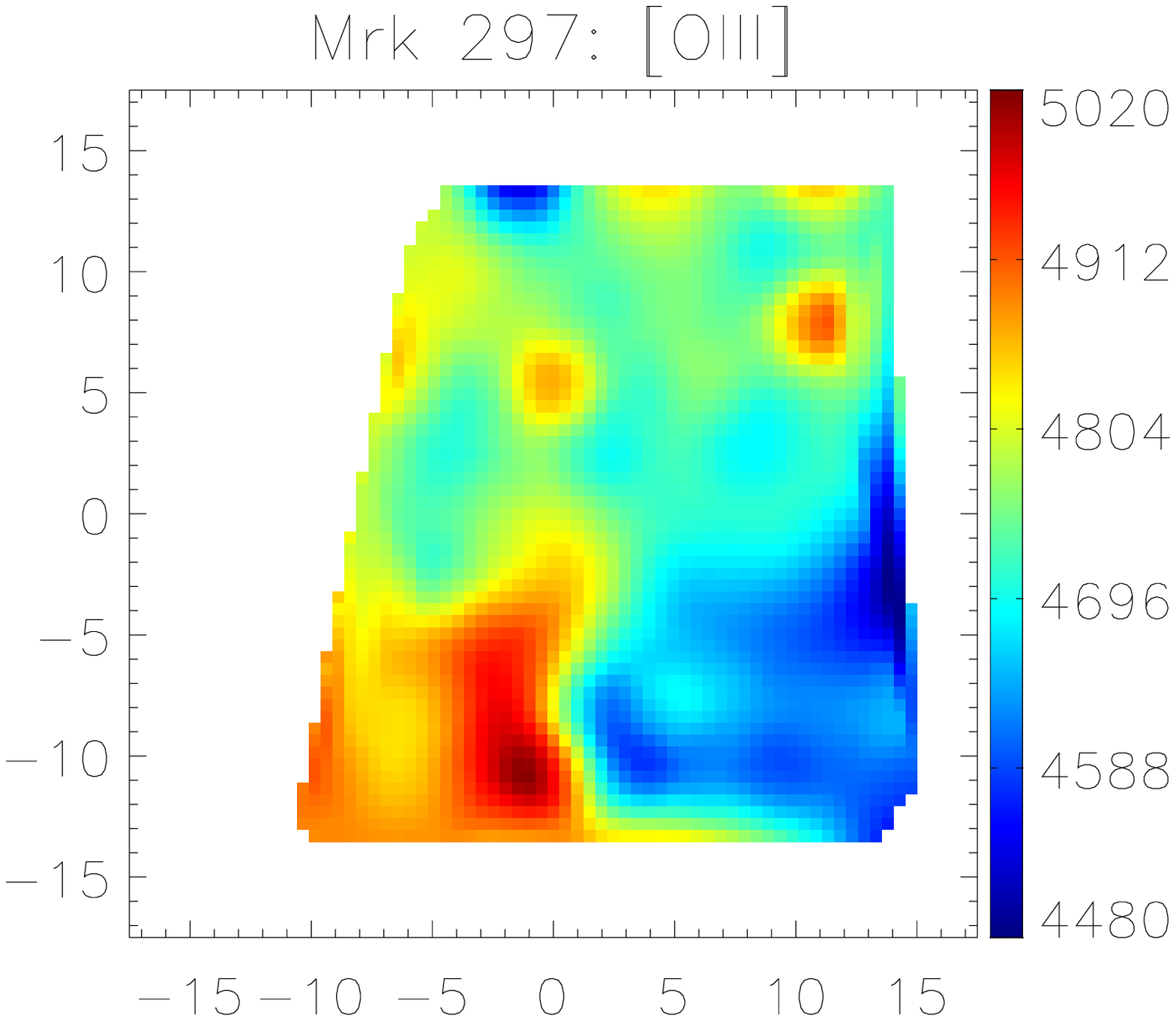}
\hspace*{0.0cm}\includegraphics[width=4.5cm]{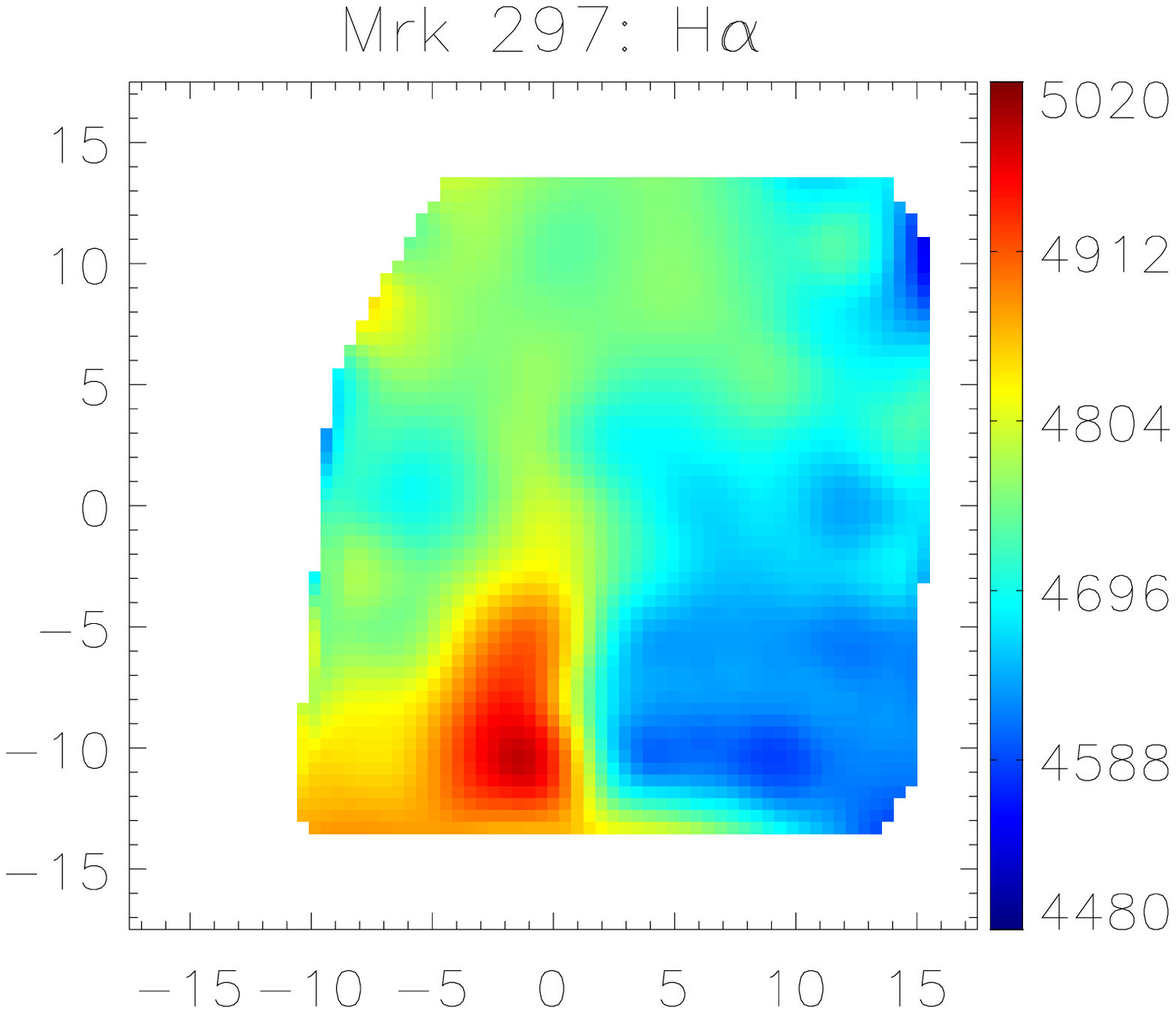} 
}}    
\mbox{
\centerline{ 
\hspace*{0.0cm}\includegraphics[width=4.5cm]{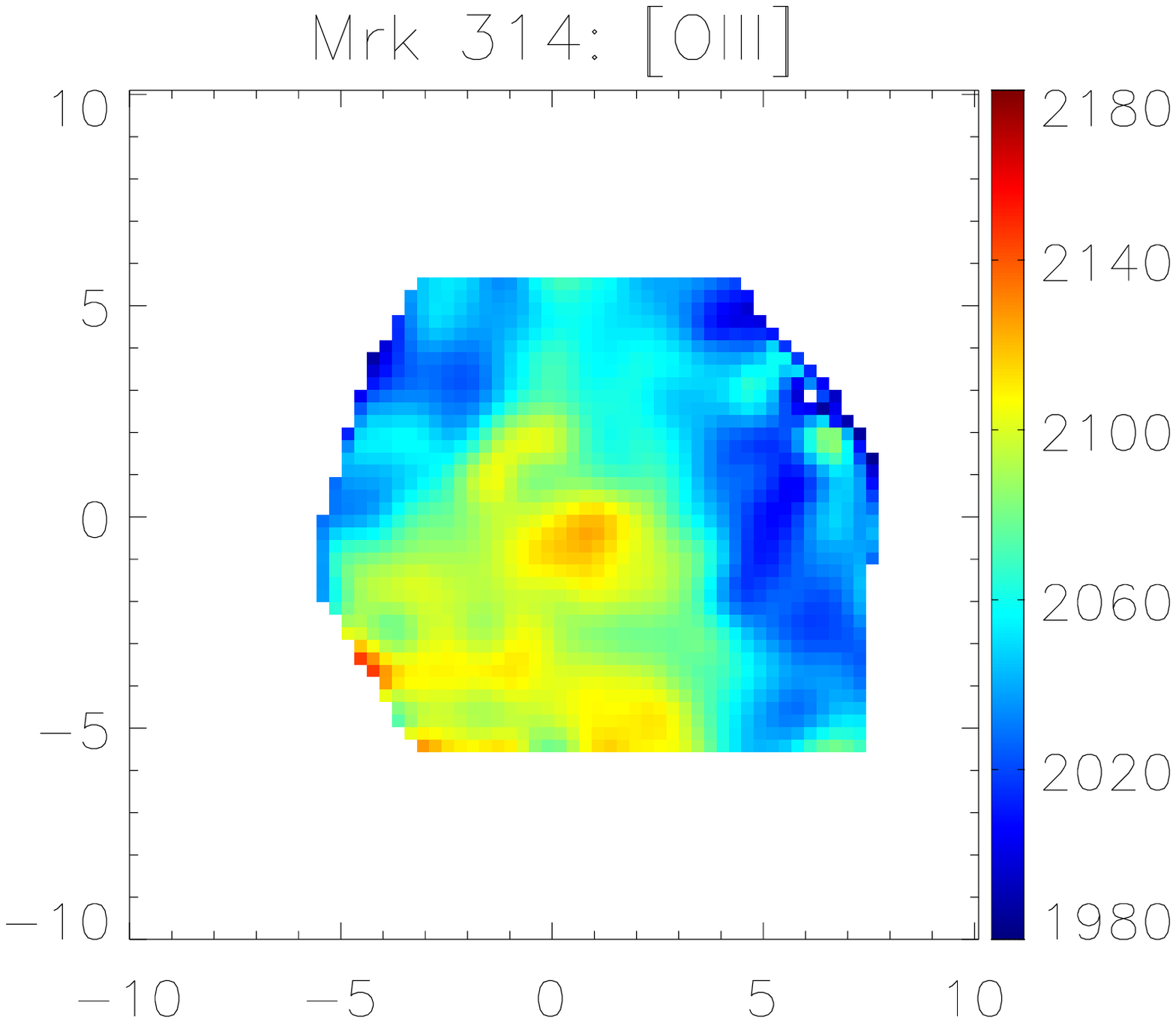}
\hspace*{0.0cm}\includegraphics[width=4.5cm]{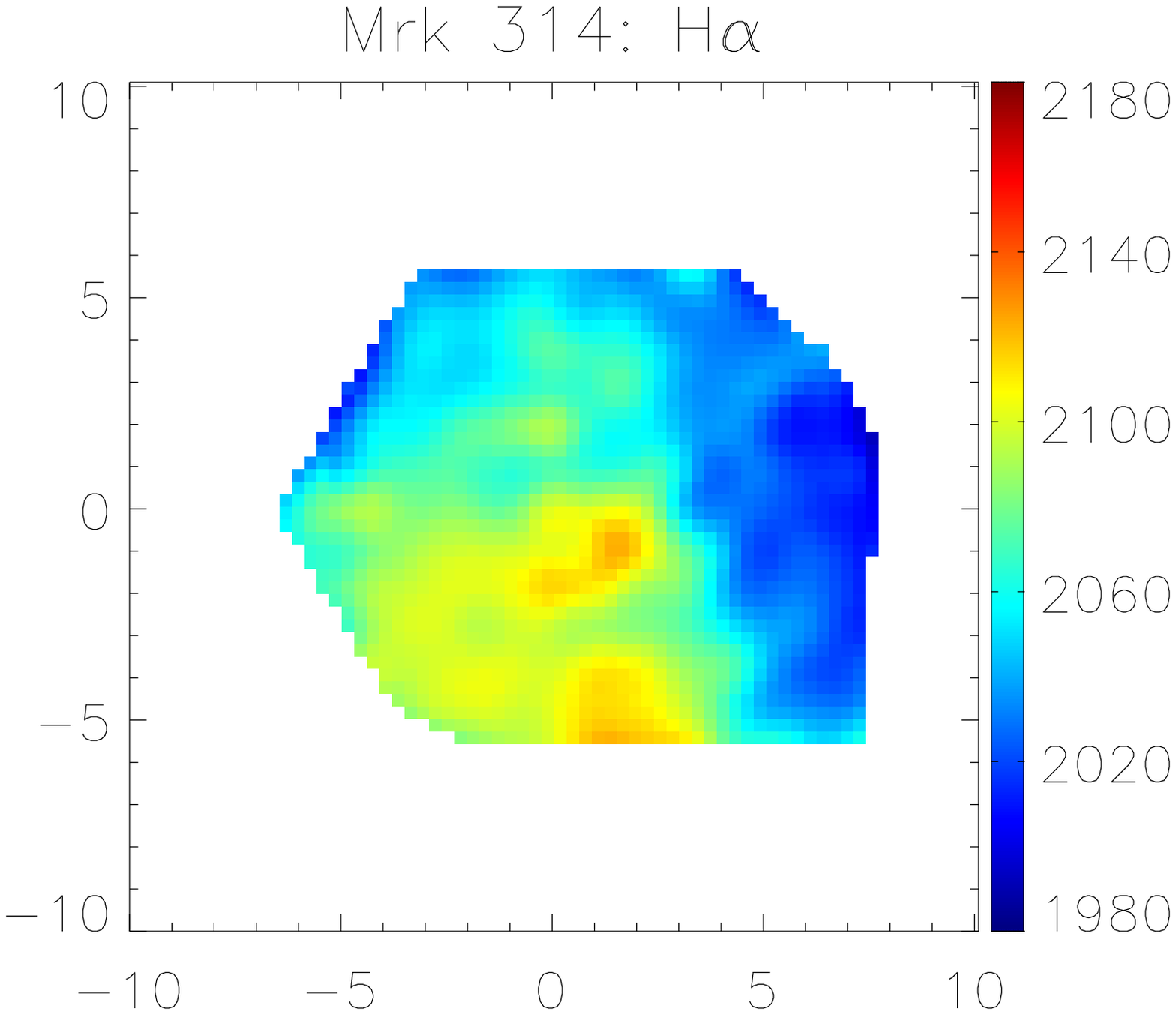} 
}}    
\mbox{
\centerline{ 
\hspace*{0.0cm}\includegraphics[width=4.5cm]{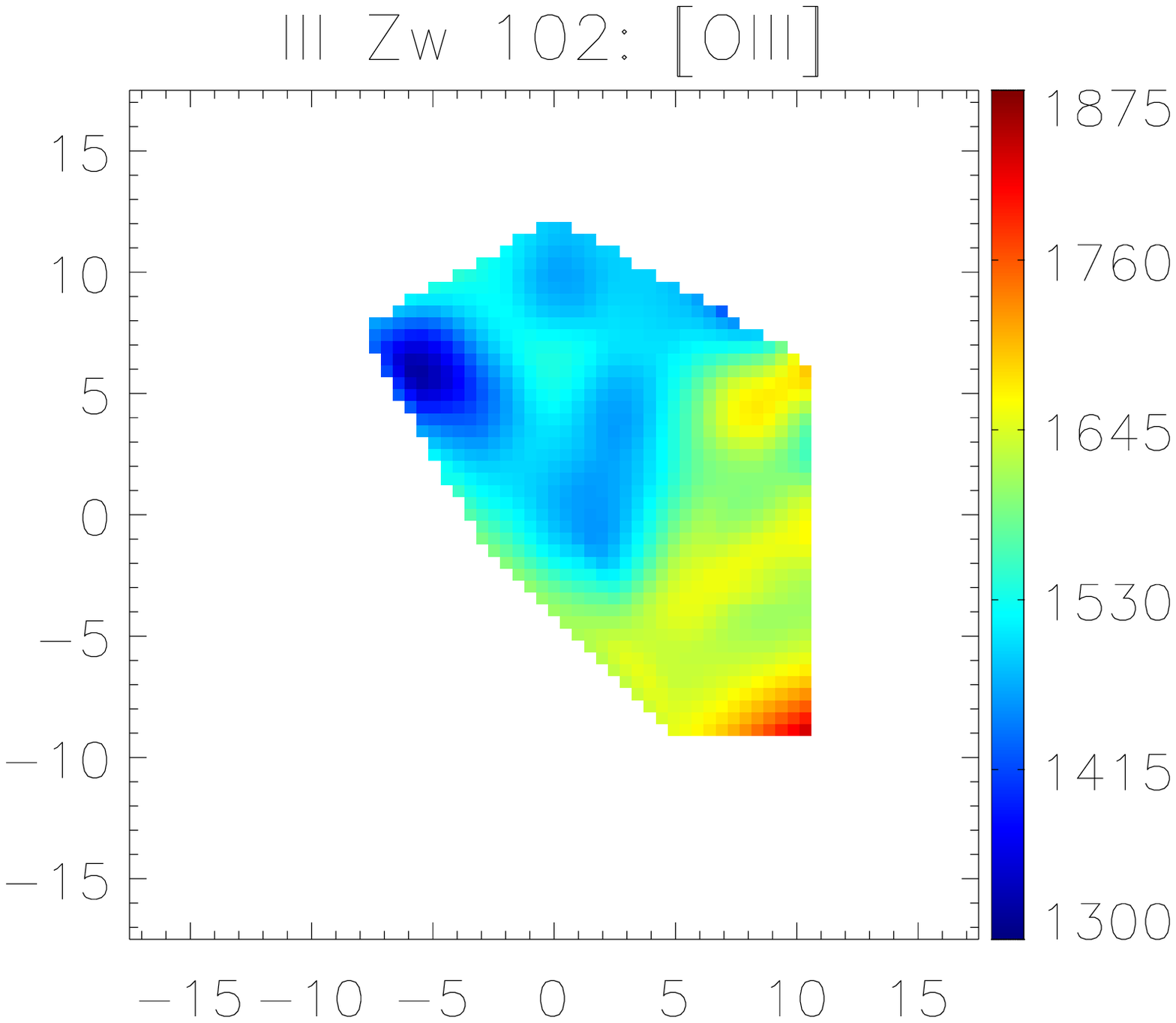}
\hspace*{0.0cm}\includegraphics[width=4.5cm]{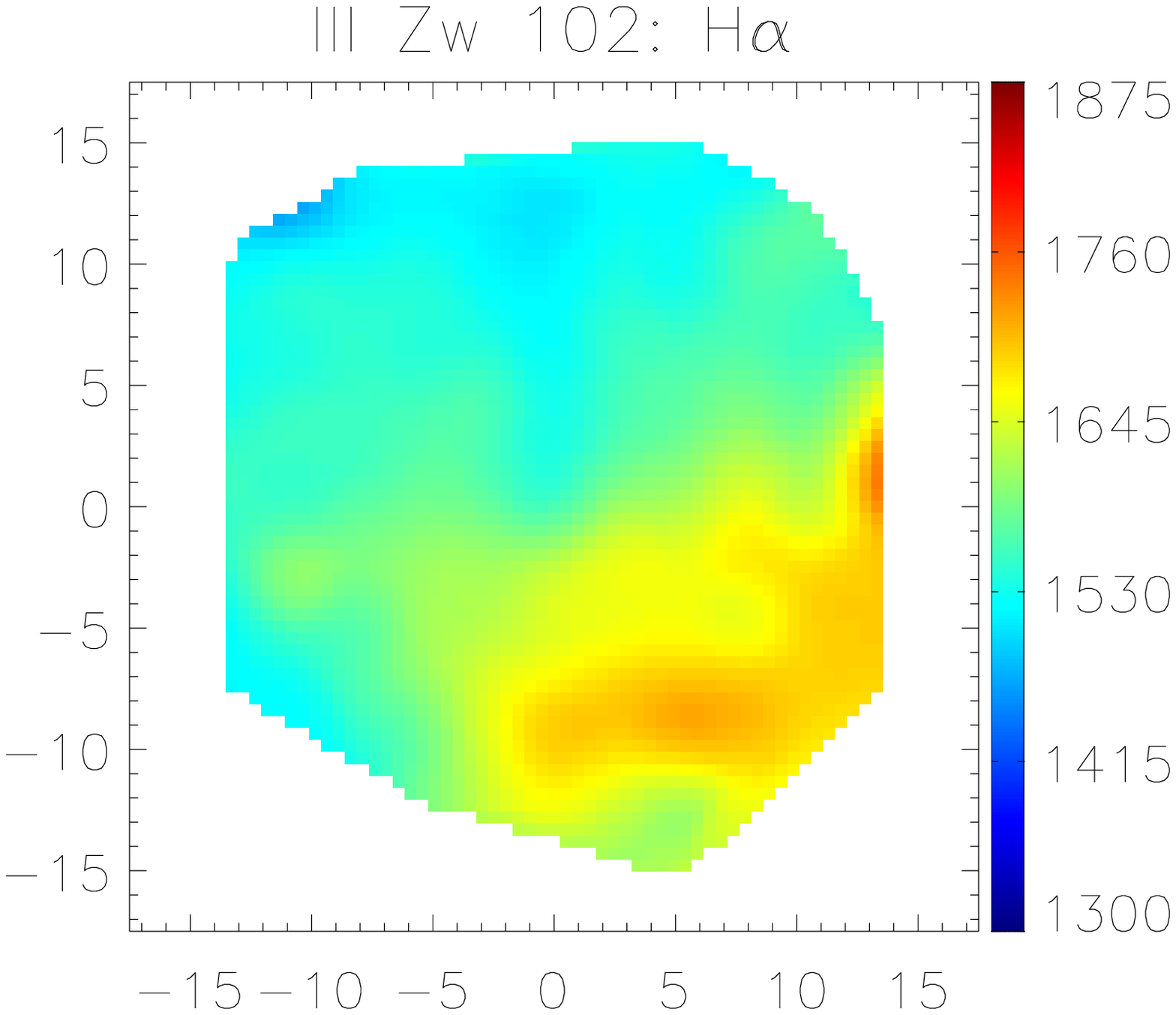} 
}}                      
\caption{Velocity field of the ionized gas in the central region of the
galaxies, for the [\ion{O}{3}]~$\lambda5007$ 
([\ion{S}{2}]~$\lambda\lambda6717,\;6731$ for Mrk~35) and the \Ha\ emission 
lines.  
Axis units are arcseconds; north is up, east to the left.} 
\label{Fig:velocity}
\end{figure*}

Figure~\ref{Fig:velocity} displays the velocity field of the ionized gas
derived from the [\ion{O}{3}]~$\lambda5007$ (column~1) and the \Ha\ (column~2)
emission lines (for Mrk~35 we used \Ha\ and [\ion{S}{2}]). In
Table~\ref{Table:sys_vel} we present the velocity at the position of the
optical nucleus derived from [\ion{O}{3}]~$\lambda5007$ and \Ha\ emission
lines, as well as the peak-to-peak velocity difference in the central 5 arcsec
of the galaxies. 

In general, the velocity maps are irregular, making their interpretation
difficult. In all five galaxies the velocity field of the inner regions are
inconsistent with ordered motions, albeit in three galaxies we also find a
large scale behavior of receding and approaching velocities. Although the
[\ion{O}{3}]~$\lambda5007$ and the \Ha\ velocity maps are in broad agreement,
there are some differences, which may reflect the fact that  low (\Ha) and
high-ionization ([\ion{O}{3}]~$\lambda5007$) lines trace  the kinematics of
two different gaseous components. Observational evidences 
\citep{Mulchaey92,Arribas97,Marquez98, GarciaLorenzo05} suggest a general
trend for the low-ionization gas to be a better tracer of the general dynamics
of the galaxies, while the high-ionization gas may present a more irregular
kinematical pattern  associated with outflows, inflows or tidal induced
motions. 

Mrk~370 and Mrk~35 display a general velocity pattern indicative of an overall
rotation; Mrk~370 shows similar velocities and amplitudes for both the high
and the low ionization gaseous components. III~Zw~102, while displaying the
largest differences between the [\ion{O}{3}]~$\lambda5007$ and the \Ha\
kinematics, also shows a velocity field resembling a rotational system.
Finally, Mrk~297 and Mrk~314 show small differences between high and
low-ionization gas: in these two systems, the velocity field is irregular in 
the whole mapped area.  
%
%


\section{Results and discussion on the individual galaxies}
\label{Sect:individual}

\subsection{Mrk~370}

This is the only galaxy in the sample that clearly fits into the BCD
luminosity range (with $M_B = -17.00$). Deep broad-band optical and
near-infrared (NIR) surface photometry was presented in C01a,b and
\cite{Noeske05} respectively, while C02 published a dedicated
spectrophotometric study which included \textsc{integral} data taken with the
\textsc{std2} bundle. The galaxy shows two bright SF knots, located in the
inner region, from which smaller knots emerge and form a structure resembling
spiral arms. These two central knots were cataloged as "double nucleus" in
\cite{MazBor93}. The outer isophotes are regular and elliptical.

The current {\sc ifu} data cover the central, brighter SF knots. We find
similar values of the electron densities (in the low density limit) in the
four selected regions, and constant oxygen abundances (within the
uncertainties), slightly smaller than those reported in C02 ($12+\log
\mbox{(O/H)}=8.6$ in \Rone\ and $12+\log \mbox{(O/H)}=8.7$ in \Rtwo).

Continuum and emission line maps show similar morphologies, with the two
central knots (\Rone\ and \Rtwo) dominating the galaxy emission; \Rtwo\ and
\Rthree\ merge together in the continuum frames. \Rone, the peak in line
emission, is not a strong continuum emitter. As already pointed out by C02 and
\cite{Noeske05}, the large equivalent \Ha\ width, the weak NIR flux, and the
blue optical colors, all indicate that \Rone\ is an \ion{H}{2} region, and not
the remnant of a captured galaxy, thus falsifying the double nucleus
hypothesis.

The [\ion{O}{3}]~$\lambda5007$/\Hb\ ratio reaches the highest values ($>4.0$)
in the emission peaks. The [\ion{O}{3}]~$\lambda5007$/\Hb\ and 
[\ion{N}{2}]~$\lambda6584$/\Ha\ maps have similar morphologies, with
[\ion{N}{2}]~$\lambda6584$/\Ha\ dropping in those regions where
[\ion{O}{3}]~$\lambda5007$/\Hb\ peaks. The \Ha/\Hb\ map shows  an
inhomogeneous pattern, in which the positions of the extinction peaks do not
coincide with those of the emission line peaks. 

Although the velocity fields derived from [\ion{O}{3}]~$\lambda5007$ and from 
\Ha\ are somewhat irregular, they suggest an overall rotation, with the
rotation axis oriented southeast northwest, broadly aligned with the
photometric minor axis of the \Ha\ distribution.

\subsection{Mrk~35}

Mrk~35 also fits, albeit by a small margin, into the luminosity range of
BCDs.  Deep optical and NIR surface photometry was presented in C01a,b and
C03, while a spectrophotometric analysis has been recently published by C07.
The starburst knots of Mrk~35 are distributed in a bar-like structure aligned
northeast-southwest. The brightest SF regions are located in the central part
of the galaxy, in a "heart-shaped" configuration; a tail departs from it
towards southwest, connecting with two moderately bright knots $\sim20\arcsec$
from the nucleus. The presence of Wolf-Rayet stars was reported by
\cite{Steel96} and C07. The starburst is placed atop a redder and regular
envelope of older stars.  

Our IFS data cover the central regions of the galaxy, the tail-like feature,
and part of the two knots detected at the end of the tail. We find noticeable
variations in the electron density, which peaks in the central, strongest \Ha\
knot (this knot contains about one half of the total \Ha\ emission of the
galaxy, C07). These findings, as well as the values found for the oxygen
abundances, are in good agreement with the study by C07 based on long-slit
spectra.

With the present spatial resolution both continuum and emission line maps show
a similar morphology. The individual star clusters seen by \cite{Johnson04}
are not resolved, and the emission peak in \Ha\ is located $\sim 3\arcsec$
northwest of the continuum peak. 

We could not compute the [\ion{O}{3}]~$\lambda5007$/\Hb\, nor the \Ha/\Hb\
ratio, as [\ion{O}{3}] and \Hb\ fall outside the spectral range. As expected
in objects photoionized by stars, the [\ion{N}{2}]~$\lambda6584$/\Ha\ map
traces the star-formation regions, with a minimum in the central knot,
\Rone. The small values of [\ion{N}{2}]~$\lambda6584$/\Ha\ in the whole field
of view indicate that shocks do not play a significant role.

The gas velocity field shows an overall regular rotation pattern, with the
rotation axis aligned north-south (close to the galaxy minor axis, see C07),
the west side receding and the east side approaching. However, in the central
regions the velocity field displays a S-shaped distortion. This distorsion,
also observed in the velocity profiles presented in C07, is suggestive of a
counter-rotating component, located in the depression in gas emission between
the central regions and the southwest knots. The velocity field is consistent
with the hypothesis of an on-going minor merger, as proposed by
\cite{Johnson04}.

\subsection{Mrk~297} 

Mrk~297 is the most luminous galaxy in the sample; with $M_{B}= -21.05$,  it
is far from being a dwarf, but since its inclusion in the pioneering paper of
\cite{ThuanMartin81}, it is included in most BCDs samples. Mrk~297 has turned
out to be the prototype of what are called today LBCG (see
\citealp{Kunth00,Cairos03,Garland04}). The galaxy is listed in the 
\textit{Atlas of Peculiar Galaxies} \citep{Arp66}. 

Optical and NIR surface photometry was presented in C01a,b and C03. The star
formation takes place in numerous regions spread over nearly the entire
system. \Ha\ and continuum maps reveal a very complex morphology, suggestive
of outflowing gas, with loops, tendrils  and numerous filaments (see C01a,b
and \citealp{MartinezDelgado07}). One remarkable feature is the double tail in
the east side of the galaxy, extending out in the north-south direction.
\cite{MartinezDelgado07} cataloged a total of 30 SF knots, for which they
provided \Ha\ photometry and colors. Ground-based \VK\ (C03) and HST \VR\
color maps \citep{Papaderos98} indicate a large-scale, inhomogeneous
absorption pattern. Mrk~297 has been studied extensively in the infrared, both
with the Infrared Space Observatory \citep{Metcalfe05} and with the Spitzer
Space Telescope \citep{Whelan07}. With a total IR luminosity of $1.0 \times
10^{11}L_{\sun}$, it is classified as a luminous infrared galaxy (LIRG). 
Mrk~297 is thought to have arisen from the merger of two gas-rich galaxies
\citep{Alloin79,Sage93}. \cite{Alloin79} identified the two peaks detected in
the continuum frames as two compact cores embedded in a common envelope.

Our \textsc{integral} data cover the whole starburst area, as well as a
substantial fraction of the LSB component. We measure density variations
across the galaxy, being the nucleus, \Rfour\ and \Rfive\ the regions with the
highest values ($\Ne \geq 200$ cm$^{-3}$). Such high densities could be
explained by collisions between molecular clouds of the interstellar medium,
or by gas inflow events. \Rfour\ and \Rfive\ have also the largest equivalent 
widths \citep{MartinezDelgado07} implying that they are very young starbursts,
probably ignited by the galaxy collision. Oxygen abundances are slightly
larger in the nucleus and in \Rthree; this abundance difference is consistent 
with the merger hypothesis. The values found for abundances ($8.46<
12+\log\mbox{(O/H)}< 8.64$) are typical of late-type spiral galaxies, and much
larger that those usually present in BCDs. The derived abundances are in
good agreement with the values reported by \cite{Calzetti94}. 

Continuum and emission line maps show different morphologies. In the continuum
maps we clearly distinguish the two peaks identified by \cite{Alloin79} as the
cores of the two merging systems, and a tail extending in the north-south
direction. The emission line maps display a more knotty pattern; several very
young regions are spread across the galaxy, the starburst having probably
been triggered by the collision of the two systems. The emission line maps
peak in \Rtwo, which includes one of the continuum maxima, while the continuum
peak (that is, the galaxy nucleus) is close to but not coincident with the
fainter  knot \Rthree.

Both the [\ion{O}{3}]~$\lambda5007$/\Hb\ and the
[\ion{N}{2}]~$\lambda6584$/\Ha\  excitation maps show an overall similar,
complex morphology. The [\ion{O}{3}]~$\lambda5007$/\Hb\ minima (peaks in
[\ion{N}{2}]~$\lambda6584$/\Ha) are found in the central region (nucleus and
\Rthree), where there is decreased \Ha\ emission. In the whole mapped area,
both line ratios are inside the \ion{H}{2}-region locus. The extinction
pattern is also highly inhomogeneous, in agreement with the optical-NIR color
map presented in C03. 

The [\ion{O}{3}]~$\lambda5007$ and \Ha\ kinematics are in good agreement. The
ionized gas velocity field shows chaotic rather than regular motions. The west
side of the galaxy is less disturbed than the east side; a sharp velocity
gradient in the EW direction is visible in the southern half of the map. This
velocity field could be the result of the superposition of two colliding
systems.

All in all, the morphological and kinematics results corroborate the
interaction scenario proposed in previous works
\citep{Alloin79,TaniguchiNoguchi91}. The double nucleus, the tail, the recent
star formation, the large velocity gradients in the extra-nuclear regions and
the perturbed gas in the whole field of view all are signs of a merger event.

\subsection{Mrk~314}

This is a LBCG, cataloged as a polar-ring galaxy candidate
\citep{Whitmore90,vanDriel00} and also included in the \cite{MazBor93} catalog
of multiple nuclei galaxies. Surface photometry in the optical and in the NIR
was presented in C01a,b, \cite{Caon05} and \cite{Noeske05}. Optical and NIR
broad-band images reveal three prominent peaks, aligned in the
northeast-southwest direction. The narrow band images show that the star
formation activity in the galaxy is distributed along the same direction, in a
bar-like structure, which extends out to about 5 kpc southwest of the nuclear
region (\citealt{Deeg97}; C01b). 

The selected \textsc{integral} configuration provides a higher spatial
resolution, which allows us to resolve all the central knots, but a smaller
field of view, which limits our study to the very central regions. The
electron density peaks in the nucleus and in \Rfive. The derived oxygen
abundances are constant within the uncertainties, and are in good agreement
with the values published in \cite{Shi05}. 

Emission line and continuum maps display slightly different morphologies. The
central knot, which is a moderate emitter in the continuum frames, is the peak
in the emission line maps, whereas the peak in the continuum peak, south  of
the central knot, is a weak emission line emitter: this indicates that they
have a different stellar content. Several SF knots, probably very young, are
only visible in the emission line maps.

The [\ion{O}{3}]~$\lambda5007$/\Hb\ ratio map has a morphology similar to the
emission line maps, and its peak coincides with the peak in \Ha\ emission. The
highest [\ion{O}{3}]~$\lambda5007$/\Hb\ value ($\geq 4$) is reached in \Rone.
Both the [\ion{O}{3}]~$\lambda5007$/\Hb\ and [\ion{N}{2}]~$\lambda6584$/\Ha\
maps display basically the same pattern, and have values typical of
\ion{H}{2}-like ionization in the whole field of view. The \Ha/\Hb\ ratio map
is irregular, and its features do not correlate with the ionized gas
distribution. The higher values in the southernmost region are suggestive of a
dust lane.

The velocity fields for [\ion{O}{3}]~$\lambda5007$ and \Ha\ have almost
identical pattern. The inner regions of Mrk~314 display a complex velocity
field, which is in broad agreement with the one published in
\cite{Shalyapina04}, which interpret it as the superposition of two kinematic
subsystems of ionized gas.

\subsection{III~Zw~102}

This is another LBCG which, like Mrk~297, is included in many BCDs samples 
after the \cite{ThuanMartin81} classification, but does not comply with the
luminosity criterion. The galaxy is listed in the Arp catalogue of Peculiar
Galaxies \citep{Arp66}, and is also classified as Ep or SAap. It is an
intriguing object, which shows regular external isophotes but a very clumpy
morphology in its inner parts. \cite{Whitmore90} considered III~Zw~102 to be
"related to polar-ring galaxies." Optical photometry was carried out by C01a;
high resolution \Ha\ maps were presented in C01b and \cite{MartinezDelgado07}.
The \Ha\ emission is concentrated in the central region, and is resolved into
numerous individual knots. There are plenty of holes, loops, and filamentary
structures. Especially interesting are the large curvilinear structures that
depart from the main body of the galaxy and extend out to 4 kpc from the
center. A strong dust lane is clearly visible in optical images and color maps
(C01b).

Our \textsc{integral} observations map the central starburst and part of the
LSB component. The selected spatial regions present variations in the electron
density, with the density peak being reached in the northern SF region, {\sc
r1}. As we pointed out in the case of Mrk~297, these density variations can be
due to collisions among molecular clouds, or to gas inflows. All the SF knots
have high oxygen abundances, with values typical of spiral galaxies.
Interesting in this galaxy is the weak [\ion{O}{3}]~$\lambda5007$ emission,
indicating that the ionizing stars are not very young; also, the spectral
shape is looks like that of a spiral galaxy, rather that resembling the flat
spectrum of a BCD. 

The continuum and the \Ha\ maps have different morphologies. The continuum map
shows a central maximum, that is the galaxy nucleus, while in line emission we
see a more complex morphology, with two peaks located slightly north of the
continuum peak, and a third, fainter peak further north.

The morphology of the [\ion{O}{3}]~$\lambda5007$/\Hb\ ratio resembles that of
the \Ha\ emission. [\ion{N}{2}]~$\lambda6584$/\Ha\ have values larger than
those typical for star-formation in the outer parts of the map; this could
indicate that shocks are playing a significant role. The extinction map is
inhomogeneous, and clearly shows the dust lanes  detected by C01b. 

The ionized gas velocity field shows an overall regular rotation pattern, 
around a kinematic axis roughly oriented ESE-WNW. The optical nucleus 
appears to be located on this axis.

\section{Summary and Conclusions}

We present results from what is, as far as we know, the first Integral Field
Spectroscopy analysis of a sample of galaxies catalogued as Blue Compact 
Dwarfs.
%
%

\begin{itemize}

\item With the help of high resolution optical images, we define several
regions of interest in each object, and for each of them we measure emission
line fluxes and compute the most relevant line ratios. These data show that:

\begin{itemize}

\item The strength of emission lines and absorption features, the shape of
the continuum, and the presence of other relevant lines significantly vary
across each galaxy. This indicates varying physical conditions and/or stellar
content.

\item All the regions studied have \ion{H}{2}-like ionization (i.e.
star-formation).

\item All the identified regions in the observed galaxies have low electron
densities, ranging from $\leq 100$ to 360 cm$^{-3}$, typical of classical
\ion{H}{2} regions.  The electron density also shows significant spatial
variations in four out of the five galaxies studied.

\item The derived oxygen abundances are relatively large in all the cases,
ranging from $Z_{\sun}\simeq0.3$ to $Z_{\sun}\simeq1.5$. No significant oxygen
abundances variations within a same object are found.

\end{itemize}

\item We build maps of the stellar continuum, emission line fluxes and
excitation ratios, finding that:

\begin{itemize}

\item Continuum and emission line morphologies are generally different.

\item The excitation ratios [\ion{O}{3}]~$\lambda5007$/\Hb\ and
[\ion{N}{2}]~$\lambda6584$/\Ha\ are typical of HII regions in the whole
observed fields. Only in the outer regions of III~Zw~102 the
[\ion{N}{2}]~$\lambda6584$/\Ha\ ratio may suggest the presence of shocks.

\item All the galaxies present a complex extinction pattern. Assuming that
the extinction coefficient is constant across the whole galaxy can lead to
considerable errors in the derivation of magnitudes and colors and in the
determination of the star formation rate and ages.

\item The \Ha/\Hb\ ratios are higher that the theoretical values, indicating 
the presence of significant amounts of dust in these galaxies.

\end{itemize}

\item In all the five galaxies the central regions display complex, distorted
ionized gas velocity fields, though large scale ordered motions are present in
three of them. With the current data we cannot determine whether these
perturbed velocity fields are the signature of interaction/mergers episodes or
they are the result of shocks, stellar winds or supernovas.

Further IFS observations are essential to investigate this issue: higher
resolution spectra, in order to measure the velocity dispersion of the gas, as
well as deeeper observations which allow to map the kinematics of the
stars, are indeed fundamental.

\item The galaxies studied here, although classified as BCDs, all show
characteristics very different from those of "genuine" BCD galaxies: they have
relatively high metallicities and significant amounts of dust; they show also
variations of such properties as electron densities, extinction and ionization
degrees, across the galaxies. These properties make them an 
especially attractive area, as they could be the local counterparts of the
blue, high metallicity, vigorously starbursting galaxies, detected at
intermediate redshift.

\end{itemize}

\acknowledgments

This paper is based on observations with the WHT, operated on the island of La
Palma by the Royal Greenwich Observatory in the Spanish Observatorio del Roque
de los Muchachos of the Instituto de Astrof\'\i sica de Canarias. We thank
J.~N. Gonz{\'a}lez-P{\'e}rez, J.~M. V{\'\i}lchez and P.~Papaderos for their
help in the initial stages of this project. We also thank  J.~Acosta-Pulido
for assistance with the software we used to analyze the \textsc{integral}
data. We are grateful to the anonymous referee whose detailed review and
criticism greatly helped to improve this paper. This research has made use of
the NASA/IPAC Extragalactic Database (NED), which is operated by the Jet
Propulsion Laboratory, Caltech, under contract with the National Aeronautics
and Space Administration. This work has been partially funded by the Spanish
``Ministerio de Ciencia y Tecnologia'' (grants PB97-0158, AYA2001-3939,
AYA2004-08260-C03-02, AyA2006-13682 and HA2006-0032). L.~M.~Cair{\'o}s
acknowledges support from the Alexander von Humboldt Foundation. N.~Caon and
C.~Kehrig are grateful for the hospitality of the Astrophysikalisches Institut
Potsdam, where part of this paper was written.

\appendix

\section{Atlas of Spectra}

Here for each galaxy we present a map with the spectrum, within a given
wavelength interval, of each fiber, placed on the corresponding spatial
position. The spectra have been rescaled to better show the line profile. 

\clearpage
\begin{figure*}                         
\resizebox{\hsize}{!}{\includegraphics{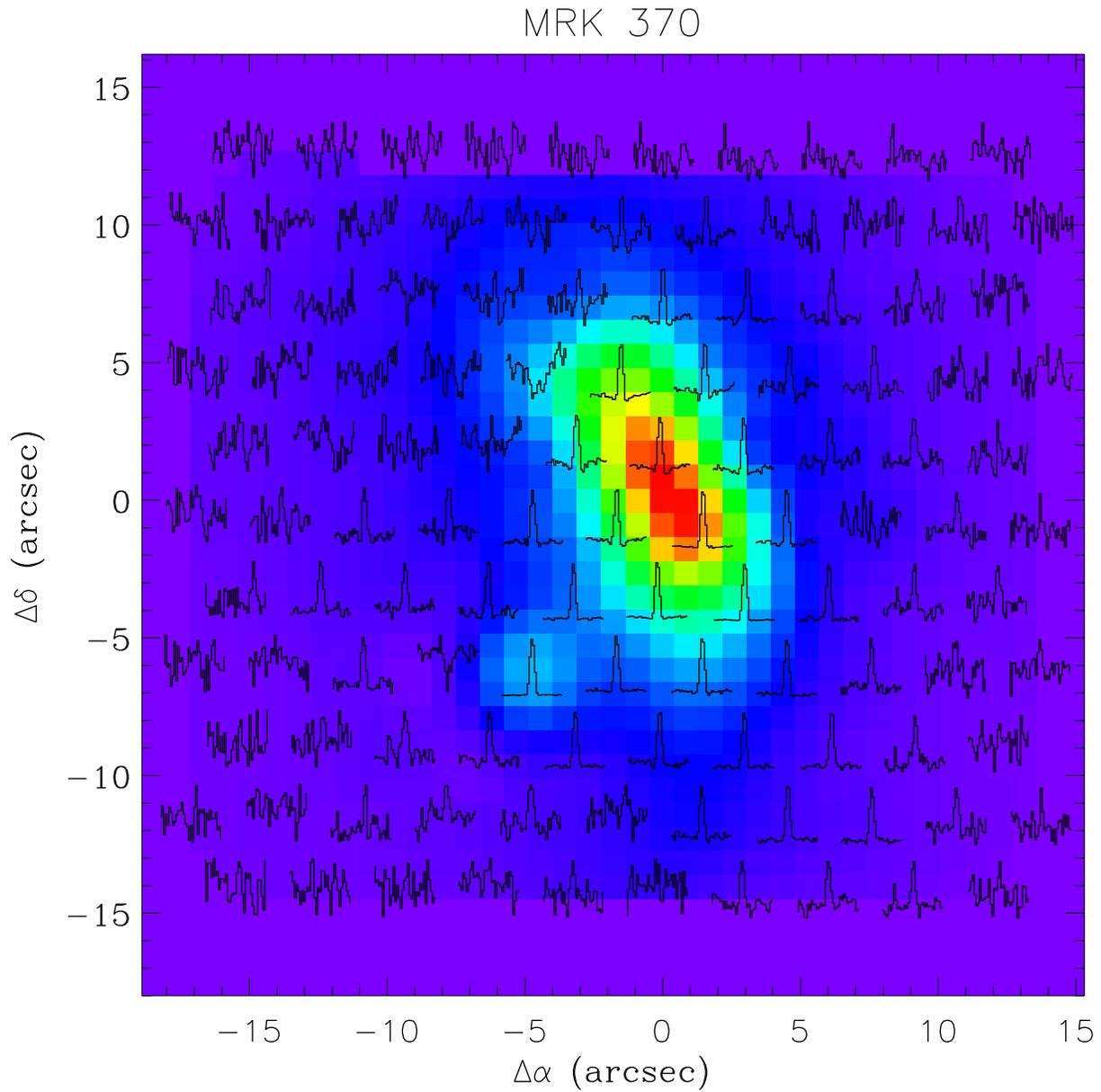}}
\caption{ Two-dimensional spectrum diagrams for Mrk~370 in the spectral range
4820--4925 \AA, which includes \Hb. The spectra are over-plotted on the 
recovered V filter image, obtained by integrating the spectrum of each fiber
within the wavelength interval of the Johnson-Cousin $V$-band filter.} 
\end{figure*}

\begin{figure*}                         
\resizebox{\hsize}{!}{\includegraphics{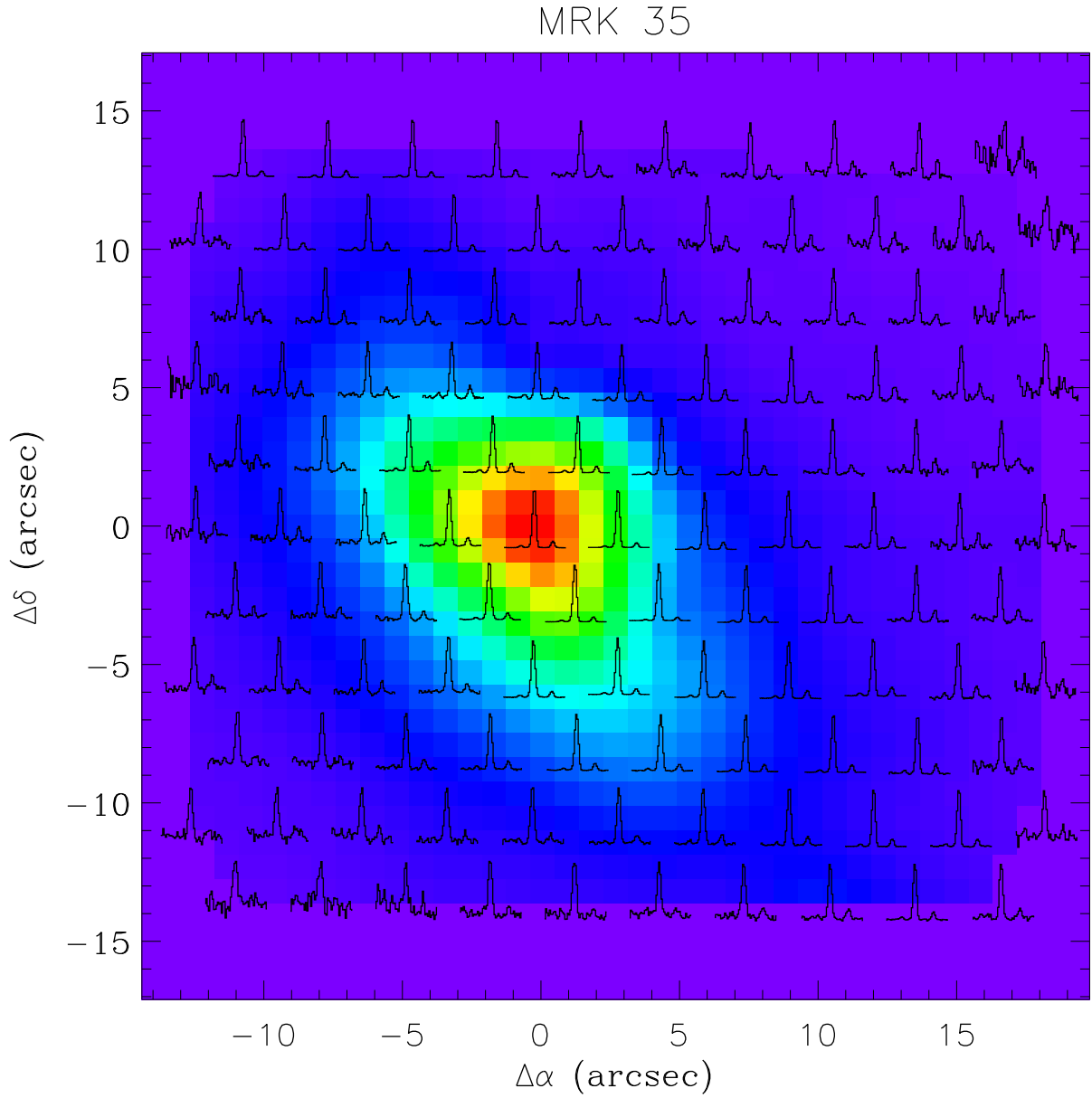}}
\caption{ Two-dimensional spectrum diagrams for Mrk~35 in the spectral range
6540--6630 \AA, which includes \Ha+[\ion{N}{2}]$\lambda\lambda6548, 6584$. 
The spectra are over-plotted on the recovered V filter image.}
\end{figure*}

\begin{figure*}                         
\resizebox{\hsize}{!}{\includegraphics{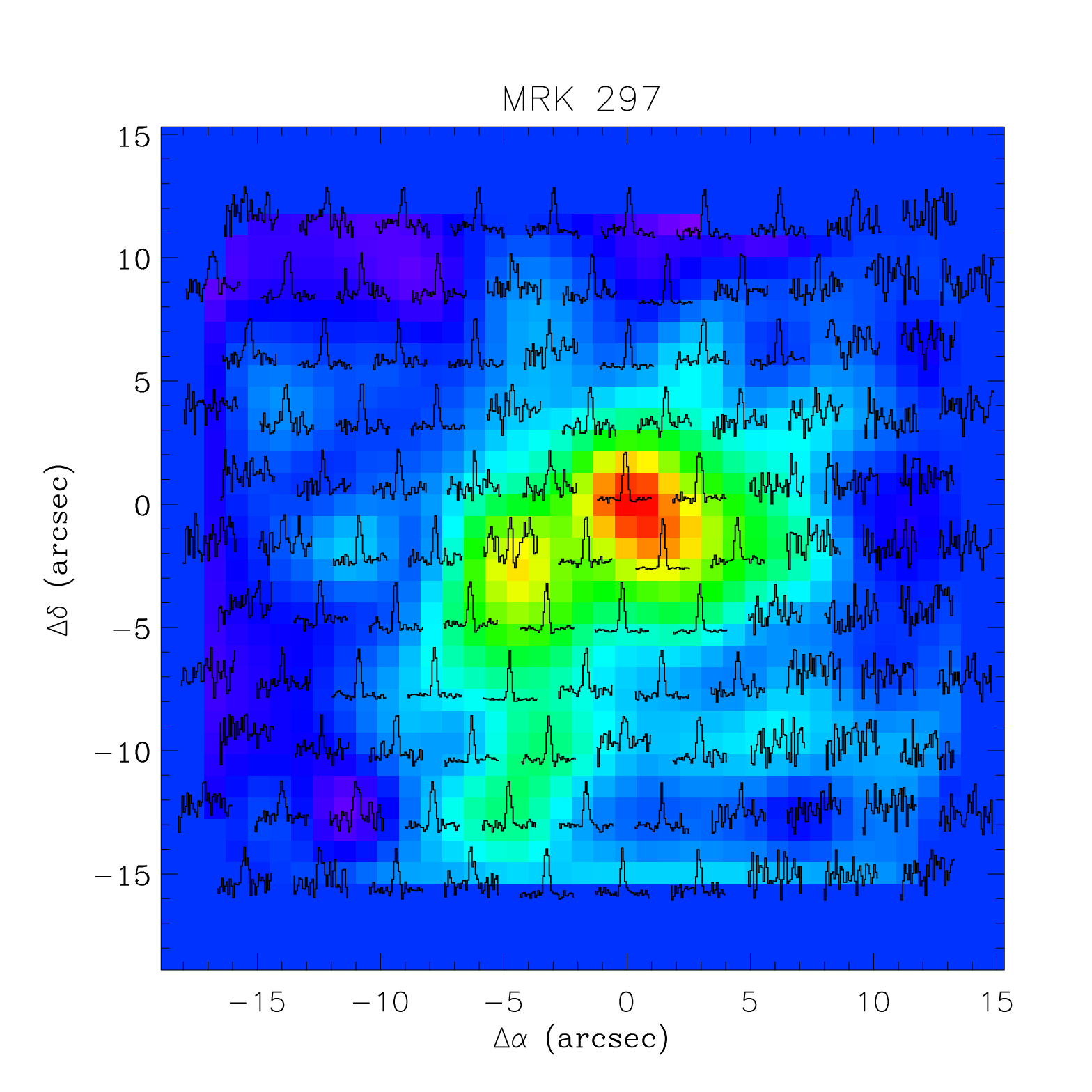}}
\caption{ Two-dimensional spectrum diagrams for Mrk~297 in the spectral 
range 4885--4990 \AA, which includes \Hb.
The spectra are over-plotted on the recovered V filter image.}
\end{figure*}

\begin{figure*}                         
\resizebox{\hsize}{!}{\includegraphics{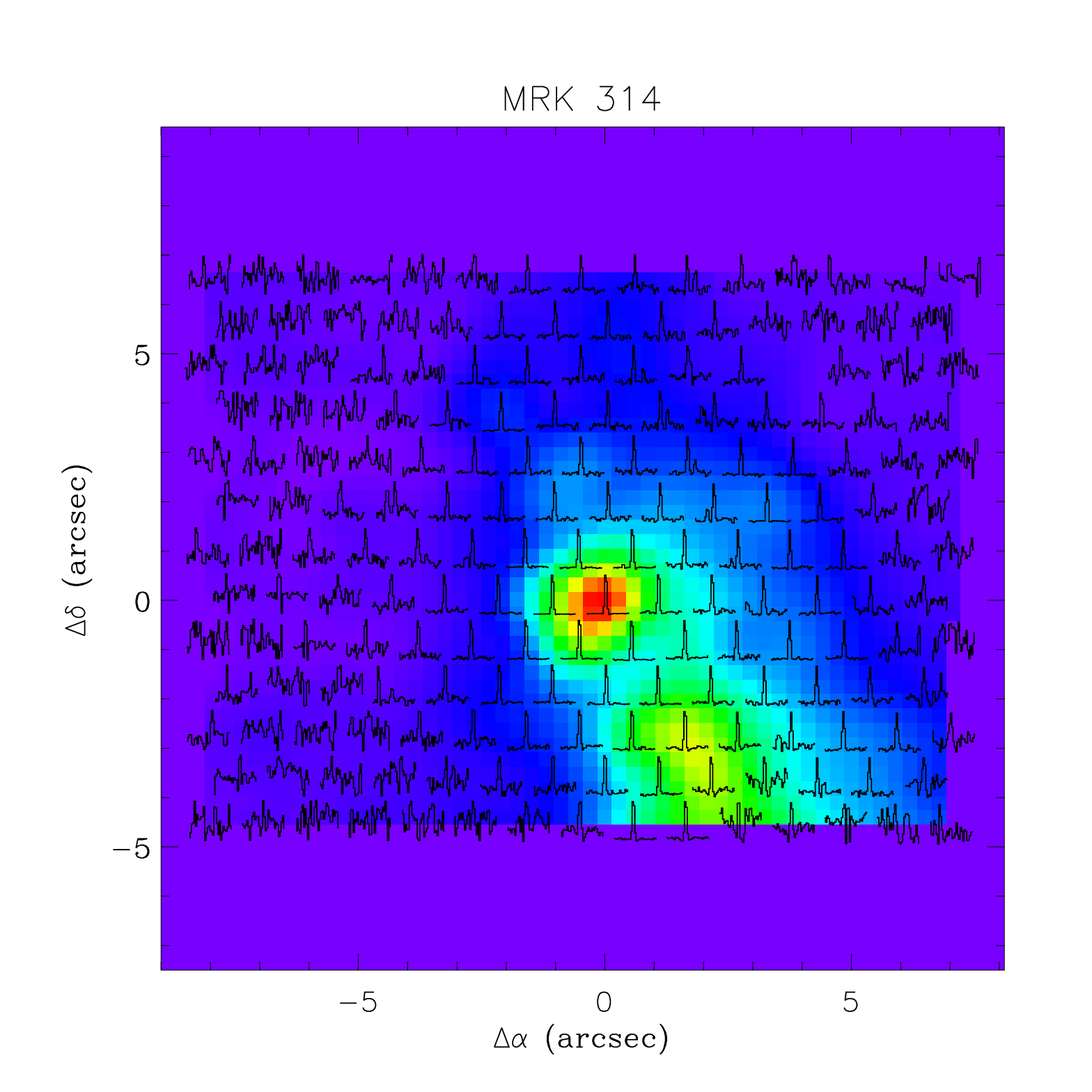}}
\caption{ Two-dimensional spectrum diagrams for Mrk~314 in the 
spectral range 4850--4950 \AA, which includes \Hb.
The spectra are over-plotted on the recovered V filter image.}
\end{figure*}

\begin{figure}   
\includegraphics[angle=0,width=\textwidth]{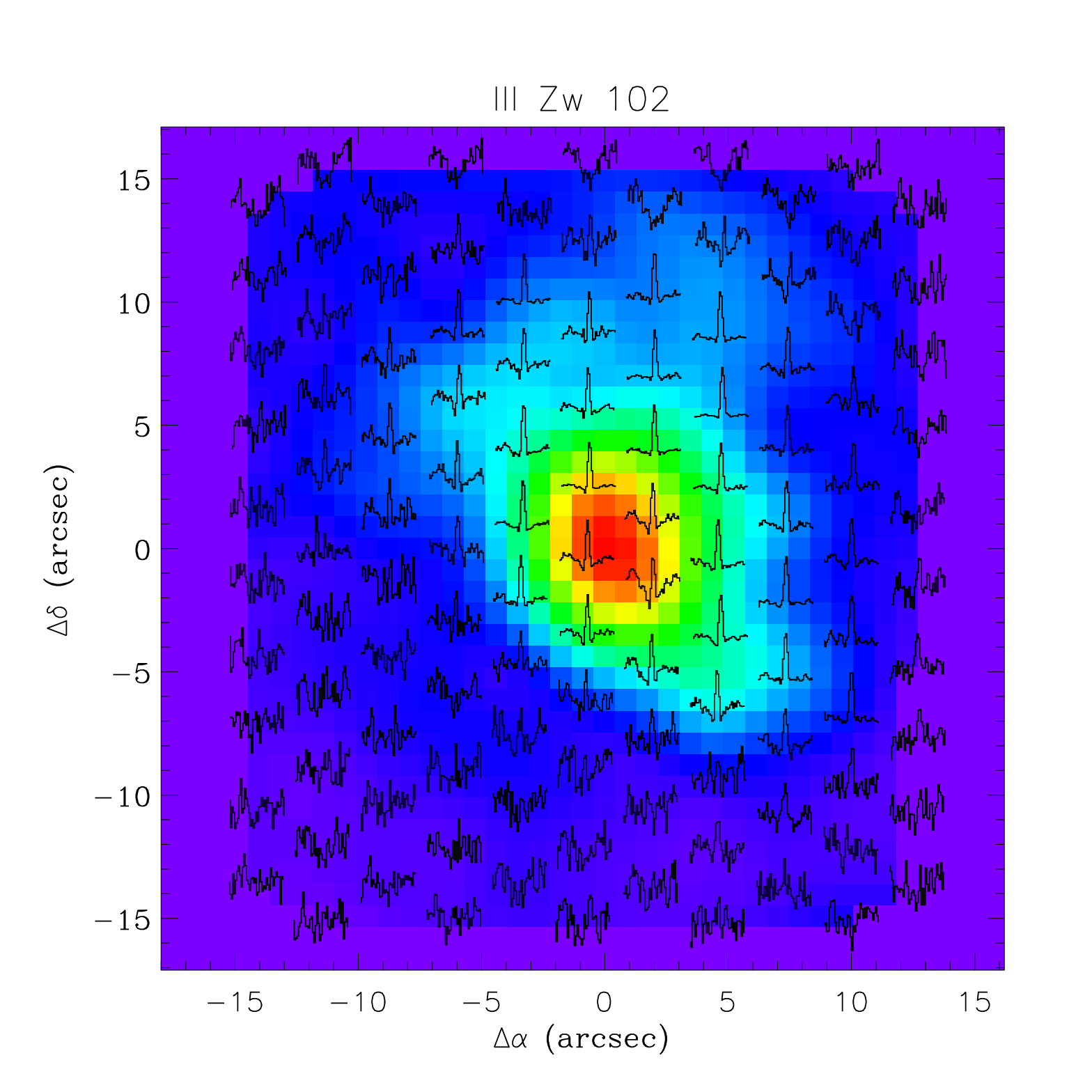}
\caption{Two-dimensional spectrum diagrams for III~Zw~102 in the spectral range
4830--4950 \AA, which includes \Hb.
The spectra are over-plotted on the recovered V filter image.}
\end{figure}

\clearpage
\begin{landscape}
\begin{deluxetable}{llcccccc}
\tabletypesize{\footnotesize}
\tablewidth{0pt}
\tablecaption{The galaxy sample}
\tablehead{
\colhead{Galaxy}              & 
\colhead {Other names}        & 
\colhead {R.A.}               &
\colhead {Decl.}              & 
\colhead {$D$}                & 
\colhead {$M_{B}$}            &
\colhead {Morph.}             & 
$A_B$                        \\
                              &                               
                              &  
(J2000)                       &
(J2000)                       & 
(Mpc)                         & 
(mag)                         & 
                              &      
(mag)                        \\
\multicolumn{1}{c}{(1)} & \multicolumn{1}{c}{(2)} & 
(3) & (4) & (5) & (6) & (7) & (8)}
\startdata
Mrk~370    & NGC~1036, IC~1828,UGC~02160               & 02 40 29.0 & +19 17 50 & 10.8 & $-17.00$ & Ch  & 0.399 \\
Mrk~35     & NGC~3353, UGC~05860, Haro~3, SBS~1042+562 & 10 45 22.4 & +55 57 37 & 15.6 & $-17.75$ & Ch  & 0.031 \\
Mrk~297    & NGC~6052/6064, UGC~10182, Arp~209         & 16 05 12.9 & +20 32 32 & 65.1 & $-21.05$ & Ext & 0.330 \\ 
Mrk~314    & NGC~7468, UGC~12329                       & 23 02 59.2 & +16 36 19 & 28.9 & $-18.53$ & Ch  & 0.368 \\ 
III~Zw~102 & NGC~7625, UGC~12529, Arp~212              & 23 20 30.1 & +17 13 32 & 22.7 & $-19.28$ & Ext & 0.109 \\  
\enddata
\label{Table:data}
\tablecomments{Columns (3) and (4): units of right ascension are hours, 
minutes, and seconds, and units of declination are degrees, arcminutes, and
arcseconds.
Col.~(5): distance computed assuming a Hubble flow, with a Hubble
constant $H_0 = 75$ km  sec$^{-1}$ Mpc$^{-1}$, and applying the correction 
for the Local Group infall into Virgo.
Col.~(6): asymptotic photometry obtained by extrapolating the growth curves, 
and corrected for galactic extinction (C01b).
Col.~(7): morphological classification following C01b.
Col.~(8): galactic extinction, from \cite{Schlegel98}.}
\end{deluxetable}
\clearpage
\end{landscape}

\clearpage
\begin{landscape}
\begin{deluxetable}{lcccccccc}
\tabletypesize{\footnotesize}
\tablewidth{0pt}
\tablecaption{Log of the Observations}
\tablehead{
\colhead{Galaxy}         & 
\colhead {Date}          & 
\colhead{Bundle}         & 
\colhead{Spectral Range} & 
\colhead{Fiber diameter} & 
\colhead{Dispersion}     & 
\colhead{texp}           & 
\colhead{Air mass}       & 
\colhead{Spatial scale} \\
                      &     
                      &  
                      & 
(\AA )                & 
(arcsec)              & 
(\AA \ pixel$^{-1}$)  & 
(sec)                 &     
                      & 
(pc/$\arcsec$)\\
\multicolumn{1}{c}{(1)} & (2) & (3) & (4) & (5) & (6) & (7) & (8) & (9)}
\startdata
Mrk~370    & 1999 Aug 22 & STD3 & 4324-7304 & 2.7 &2.99 & $3\times    1200$ & 1.01--1.04 & \phn53 \\
Mrk~35     & 2003 Feb 25 & STD3 & 5481-6937 & 2.7 &1.43 & $3\times \phn900$ & 1.16--1.19 & \phn76 \\
Mrk~297    & 1999 Aug 22 & STD3 & 4324-7304 & 2.7 &2.99 & $2\times    1200$ & 1.10--1.48 &    313 \\ 
Mrk~314    & 1999 Aug 22 & STD2 & 4414-7452 & 0.9 &2.99 & $3\times    1800$ & 1.02--1.04 &    140 \\  
III~Zw~102 & 1999 Aug 22 & STD3 & 4324-7304 & 2.7 &2.99 & $3\times    1800$ & 1.07--1.21 &    110 \\  
\enddata
\label{Table:logobs}
\end{deluxetable}
\clearpage
\end{landscape}

\clearpage
\begin{landscape}
\begin{deluxetable}{lccccccc}
\tabletypesize{\footnotesize}
\tabletypesize{\scriptsize}
\tablewidth{0pt}
\tablecaption{The computed emission line ratios in the different regions}
\tablehead{
\colhead{Galaxy} & 
\colhead{Region} & 
\colhead{$\frac{\Ha}{\Hb}$} & 
\colhead{$\log(\frac{[\mbox{\ion{O}{3}}]\;\lambda5007}{\Hb})$} & 
\colhead{$\log(\frac{[\mbox{\ion{N}{2}}]\;\lambda6584}{\Ha})$} & 
\colhead{$\log(\frac{[\mbox{\ion{S}{2}}]\;\lambda\lambda6717,\;6731}{\Ha})$} &  
\colhead{$\log(\frac{[\mbox{\ion{O}{1}}]\;\lambda6300}{\Ha})$} & 
\colhead{$\frac{[\mbox{\ion{S}{2}}]\;\lambda6717}{[\mbox{\ion{S}{2}}]\;\lambda6731}$} 
}
\startdata
Mrk~370    &  N   & $2.87\pm0.35$ & $ \phn0.26\pm0.02$ & $-0.84\pm0.02$ & $-0.65\pm0.02$ & $-2.06\pm0.07$ & $1.39\pm0.06$ \\
           &  R1  & $2.66\pm0.33$ & $ \phn0.32\pm0.02$ & $-0.96\pm0.03$ & $-0.79\pm0.02$ & \nodata        & $1.33\pm0.04$ \\
           &  R2  & $2.86\pm0.36$ & $ \phn0.30\pm0.02$ & $-0.88\pm0.03$ & $-0.68\pm0.02$ & $-2.07\pm0.09$ & $1.39\pm0.05$ \\
           &  R3  & $2.93\pm0.38$ & $ \phn0.25\pm0.02$ & $-0.81\pm0.02$ & $-0.51\pm0.02$ & \nodata        & $1.41\pm0.05$ \\ \hline
Mrk~35     &  N   & \nodata       &     \nodata        & $-1.22\pm0.02$ & $-1.02\pm0.02$ & $-2.13\pm0.07$ & $1.23\pm0.03$ \\
           &  R1  & \nodata       &     \nodata        & $-1.09\pm0.02$ & $-0.87\pm0.02$ & $-1.97\pm0.05$ & $1.29\pm0.03$ \\
           &  R2  & \nodata       &     \nodata        & $-0.87\pm0.02$ & $-0.49\pm0.02$ & $-1.46\pm0.03$ & $1.39\pm0.04$\\
           &  R3  & \nodata       &     \nodata        & $-1.00\pm0.01$ & $-0.72\pm0.02$ & $-1.78\pm0.04$ & $1.39\pm0.03$\\ \hline
Mrk~297    &  N   & $3.91\pm1.50$ & $    -0.11\pm0.19$ & $-0.67\pm0.03$ & $-0.40\pm0.03$ & $-1.39\pm0.11$ & $1.20\pm0.17$\\
           &  R1  & $2.59\pm0.37$ & $ \phn0.27\pm0.03$ & $-0.91\pm0.03$ & $-0.67\pm0.02$ & $-1.67\pm0.09$ & $1.43\pm0.07$ \\
           &  R2  & $3.52\pm0.44$ & $ \phn0.31\pm0.02$ & $-0.87\pm0.01$ & $-0.65\pm0.02$ & $-1.50\pm0.04$ & $1.30\pm0.04$\\
           &  R3  & $2.99\pm0.39$ & $    -0.22\pm0.03$ & $-0.66\pm0.01$ & $-0.54\pm0.02$ & $-1.43\pm0.05$ & $1.30\pm0.07$\\
	   &  R4  & $5.12\pm0.77$ & $ \phn0.36\pm0.04$ & $-0.74\pm0.03$ & $-0.69\pm0.02$ & $-1.57\pm0.06$ & $1.23\pm0.10$\\
           &  R5  & $3.06\pm0.62$ & $ \phn0.22\pm0.07$ & $-0.86\pm0.03$ & $-0.61\pm0.04$ & $-1.38\pm0.07$ & $1.12\pm0.21$\\
           &  R6  & $3.38\pm0.47$ & $ \phn0.07\pm0.03$ & $-0.75\pm0.02$ & $-0.58\pm0.02$ & $-1.42\pm0.05$ & $1.26\pm0.07$\\
           &  R7  & $3.46\pm0.57$ & $ \phn0.26\pm0.05$ & $-0.88\pm0.02$ & $-0.66\pm0.02$ & \nodata        & $1.32\pm0.10$\\ \hline
Mrk~314    &  N   & $3.25\pm0.39$ & $ \phn0.65\pm0.01$ & $-1.34\pm0.03$ & $-0.94\pm0.02$ & $-2.02\pm0.05$ & $1.27\pm0.07$\\
           &  R1  & $3.11\pm0.37$ & $ \phn0.61\pm0.01$ & $-1.29\pm0.03$ & $-0.85\pm0.02$ & $-1.89\pm0.05$ & $1.33\pm0.05$\\
           &  R2  & $3.29\pm0.40$ & $ \phn0.54\pm0.02$ & $-1.23\pm0.02$ & $-0.74\pm0.02$ & $-1.80\pm0.05$ & $1.44\pm0.05$\\
           &  R3  & $2.90\pm0.38$ & $ \phn0.60\pm0.03$ & $-1.34\pm0.04$ & $-0.84\pm0.02$ & \nodata        & $1.31\pm0.10$ \\
           &  R4  & $3.16\pm0.39$ & $ \phn0.46\pm0.02$ & $-1.12\pm0.03$ & $-0.69\pm0.02$ & $-1.63\pm0.08$ & $1.39\pm0.09$\\ 
           &  R5  & $3.03\pm0.38$ & $ \phn0.47\pm0.02$ & $-1.06\pm0.03$ & $-0.62\pm0.02$ & $-1.67\pm0.07$ & $1.25\pm0.09$\\ \hline
III~Zw~102 &  N   & $6.30\pm0.82$ & $    -0.78\pm0.11$ & $-0.37\pm0.01$ & $-0.59\pm0.02$ & \nodata        & $1.23\pm0.06$ \\ 
           &  R1  & $5.10\pm0.72$ & $    -0.61\pm0.08$ & $-0.37\pm0.01$ & $-0.38\pm0.02$ & $-1.53\pm0.06$ & $1.13\pm0.08$\\
           &  R2  & $5.21\pm0.65$ & $    -0.71\pm0.06$ & $-0.42\pm0.01$ & $-0.54\pm0.02$ & \nodata        & $1.29\pm0.06$ \\
           &  R3  & $5.03\pm0.63$ & $    -0.65\pm0.04$ & $-0.39\pm0.01$ & $-0.55\pm0.02$ & $-1.77\pm0.06$ & $1.34\pm0.05$\\
           &  R4  & $5.50\pm0.71$ & $    -0.57\pm0.05$ & $-0.43\pm0.01$ & $-0.54\pm0.02$ & \nodata        & $1.36\pm0.06$
\enddata
\tablecomments{The line ratios have been corrected for galactic extinction.
The quoted uncertainties include both measurement errors and the uncertainty 
on the calibration factor, which is about 12\% for the \Ha/\Hb\ ratio, 
5\% for [\ion{O}{3}]~$\lambda~5007$/\Hb, and $\leq 3$\% for the other ratios.}
\label{Table:emissionlines}
\end{deluxetable}
\clearpage
\end{landscape}

\clearpage

\begin{landscape}
\begin{deluxetable*}{lccc}
\tabletypesize{\footnotesize}
\tablewidth{0pt}
\tablecaption{Electron densities and oxygen abundances}
\tablehead{
\colhead{Galaxy}  & \colhead {Region} & \colhead {\Ne} &
\colhead{12 +$\log$(O/H)}}
\startdata
Mrk~370    &  N    &  $\leq100$  &  8.51   \\
           &  R1   &  $\leq100$  &  8.42   \\
           &  R2   &  $\leq100$  &  8.48   \\
           &  R3   &  $\leq100$  &  8.53   \\ \hline
Mrk~35     &  N    &  200        &  8.23   \\
           &  R1   &  131        &  8.32   \\
           &  R2   &  $\leq100$  &  8.48   \\
           &  R3   &  $\leq100$  &  8.40   \\ \hline
Mrk~297    &  N    &  239        &  8.63   \\
           &  R1   &  $\leq100$  &  8.46   \\
           &  R2   &  120        &  8.48   \\
           &  R3   &  120        &  8.64   \\
           &  R4   &  200        &  8.58   \\
           &  R5   &  360        &  8.49   \\
           &  R6   &  165        &  8.57   \\
           &  R7   &  100        &  8.48   \\ \hline
Mrk~314    &  N    &  153        &  8.14   \\
           &  R1   &  $\leq100$  &  8.18   \\
           &  R2   &  \nodata    &  8.22   \\
           &  R3   &  110        &  8.14   \\
           &  R4   &  $\leq100$  &  8.30   \\  
           &  R5   &  176        &  8.35   \\ \hline
III~Zw~102 &  N    &  200        &  8.85   \\ 
           &  R1   &  343        &  8.84   \\
           &  R2   &  130        &  8.81   \\
           &  R3   &  $\leq100$  &  8.83   \\
           &  R4   &  $\leq100$  &  8.81   \\
\enddata
\label{Table:physicalpara}
\tablecomments{Abundances were computed using the empirical calibrations
published in \cite{Denicolo02}, whose typical uncertainty is $\pm 0.2$ 
dex.}
\end{deluxetable*}
\end{landscape}

\clearpage

\begin{landscape}
\begin{deluxetable}{lccccc}
\tabletypesize{\footnotesize}
\tablewidth{0pt}
\tablecaption{Systemic velocity of the sample galaxies at the position 
of the optical nucleus}
\tablehead{
\colhead{Galaxy}  & \colhead{$V_{\rm sys}$ (km s$^{-1}$)} & 
\colhead{$V_{[\mbox{\ion{O}{3}}]~\lambda5007}$ (km s$^{-1}$)} & 
\colhead{$\Delta V_{[\mbox{\ion{O}{3}}]}$ (km s$^{-1}$)} & 
\colhead{$V_{\Ha}$ (km s$^{-1}$)} &  
\colhead{$\Delta V_{\Ha}$ (km s$^{-1}$)} \\
\multicolumn{1}{c}{(1)} & (2) & (3) & (4) & (5) & (6)
} 
\startdata
Mrk~370    & \phn790 & $\phn715\pm\phn6$ & $\phn35\pm\phn9$ & $\phn735\pm5$ & $\phn28\pm5$ \\
Mrk~35     & \phn944 & \nodata           & \nodata          & $\phn958\pm5$ & $\phn31\pm7$ \\
Mrk~297    &    4739 & $4720\pm\phn9$    &    $125\pm14$    &    $4714\pm5$ &    $118\pm7$ \\ 
Mrk~314    &    2081 & $2117\pm\phn6$    & $\phn58\pm\phn8$ &    $2104\pm5$ & $\phn52\pm7$ \\  
III~Zw~102 &    1633 & $1463\pm37$       &    $254\pm50$    &    $1588\pm5$ & $\phn66\pm7$ \\  
\hline 
\enddata
\label{Table:sys_vel}
\tablecomments{Column (2) lists the systemic velocity from NASA Extragalactic 
database (NED) (http://nedwww.ipac.caltech.edu/).
Cols. (4) and (6) list the half velocity amplitude in the central 5 
arcsec of the galaxy.}
\end{deluxetable}

\clearpage
\end{landscape}

\end{document}